\documentclass[
superscriptaddress,aps,preprintnumbers,amsmath,amssymb,prd,nofootinbib,preprint]{revtex4}

\usepackage[dvipdfm]{graphicx}
\usepackage{epstopdf}
\usepackage{dcolumn}
\usepackage{bm}
\usepackage{hyperref}
\usepackage{cancel}
\usepackage{amsmath, amssymb, slashed, epsf, color}
\usepackage{ascmac}
\begin{document}

\def\o{\over}
\def\beq{\begin{eqnarray}}
\def\eeq{\end{eqnarray}}
\newcommand{\gsim}{ \mathop{}_{\textstyle \sim}^{\textstyle >} }
\newcommand{\lsim}{ \mathop{}_{\textstyle \sim}^{\textstyle <} }
\newcommand{\vev}[1]{ \left\langle {#1} \right\rangle }
\newcommand{\bra}[1]{ \langle {#1} | }
\newcommand{\ket}[1]{ | {#1} \rangle }
\newcommand{\EV}{ {\rm eV} }
\newcommand{\KEV}{ {\rm keV} }
\newcommand{\MEV}{ {\rm MeV} }
\newcommand{\GEV}{ {\rm GeV} }
\newcommand{\TEV}{ {\rm TeV} }
\def\diag{\mathop{\rm diag}\nolimits}
\def\Spin{\mathop{\rm Spin}}
\def\SO{\mathop{\rm SO}}
\def\O{\mathop{\rm O}}
\def\SU{\mathop{\rm SU}}
\def\U{\mathop{\rm U}}
\def\Sp{\mathop{\rm Sp}}
\def\SL{\mathop{\rm SL}}
\def\tr{\mathop{\rm tr}}
\def\Mpl{M_{\rm pl}}


\title{
Gaugino mass in heavy sfermion scenario
}

\author{Keisuke Harigaya}
\affiliation{Kavli IPMU (WPI), UTIAS, The University of Tokyo, Kashiwa, 277-8583, Japan}
\affiliation{ICRR, University of Tokyo, Kashiwa, Chiba 277-8582, Japan}
\begin{abstract}
The heavy sfermion scenario is naturally realized when supersymmetry breaking fields are
charged under some symmetry or are composite fields.
There, scalar partners of standard model fermions and the gravitino are as heavy as $O(10\mathchar`-1000)$ TeV while
gauginos are as heavy as $O(1)$ TeV.
The scenario is not only consistent with the observed higgs mass, but also is free from cosmological problems such as the Polonyi problem and the gravitino problem.
In the scenario, gauginos are primary targets of experimental searches.
In this thesis, we discuss gaugino masses in the heavy sfermion scenario.
First, we derive the so-called anomaly mediated gaugino mass
in the superspace formalism of supergravity with a Wilsonian effective action.
Then we calculate gaugino masses generated through other possible one-loop corrections by extra light matter fields and the QCD axion.
Finally, we consider the case where some gauginos are degenerated in their masses with each other,
because the thermal relic abundance of the lightest supersymmetric particle as well as the the strategy to search gauginos drastically change in this case.
After calculating the thermal relic abundance of the lightest supersymmetric particle for the degenerated case, we
discuss the phenomenology of
gauginos at the Large Hadron Collider and cosmic ray experiments.

\end{abstract}

\date{\today}
\maketitle
\preprint{IPMU 15-0137}

\newpage
\tableofcontents

\newpage
\section{Introduction}

Supersymmetry (SUSY) has been intensively studied as a fundamental law of nature for following reasons:
SUSY reduces degrees of divergences in quantum field theories~\cite{Grisaru:1979wc}, and hence stabilizes the vast separation between the electroweak scale and the Planck scale or the grand unification scale~\cite{MaianiLecture,Veltman:1980mj,Witten:1981nf,Kaul:1981wp}.
If SUSY is broken by some strong gauge dynamics,
the smallness of the electroweak scale is explained by dimensional transmutation.
In the minimal SUSY extension of the standard model (MSSM), three gauge coupling constants unify, which is consistent with the prediction of the simplest grand unified theory (GUT), namely the $SU(5)$ GUT~\cite{Georgi:1974sy}.
With an $R$ parity preserved, the lightest SUSY particle (LSP) is a good candidate of dark matter~\cite{Goldberg:1983nd,Ellis:1983ew}.

The discovery of the standard-model-like higgs as heavy as $125$ GeV~\cite{Aad:2012tfa,Chatrchyan:2012ufa} strongly constrains the mass spectrum of MSSM particles.
In the MSSM, quartic couplings of higgs fields are given by $D$ term potentials of standard model gauge interactions in the SUSY limit.
Thus the mass of the standard-model-like higgs is lighter than the mass of the $Z$ boson at the tree level.%
\footnote{We do not consider the case where the discovered higgs is not the lightest higgs.}
The observed large higgs mass indicates that quantum corrections from SUSY breaking effects to the higgs mass are significant~\cite{Okada:1990gg,Okada:1990vk,Ellis:1990nz,Haber:1990aw}.
To explain the observed higgs mass,
soft scalar masses of scalar top quarks are required to be larger than about $3\mathchar`-5$ TeV~\cite{Hahn:2013ria}.

We note that gravity mediated soft scalar masses larger than $O(10)$ TeV are compatible with physics in the early universe.
Planck scale-suppressed mediation, where the gravitino mass is as large as soft scalar masses, with the gravitino mass larger than $O(10)$ TeV, is free of the constraint from the Big-Bang-Nucleosynthesis even for a high reheating temperature~\cite{Weinberg:1982zq,Pagels:1981ke,Khlopov:1984pf,Kawasaki:1994af,Kawasaki:2004qu,Jedamzik:2006xz,Kawasaki:2008qe}.%

With scalar masses and the gravitino mass larger than $O(10)$ TeV, one may naively expect that the LSP mass is also of the same order, which leads to the overproduction of the LSP from thermal bath in the early universe.%
\footnote{If the reheating temperature of the universe is smaller than the mass of the LSP, a correct abundance of the LSP can be obtained even if the LSP mass is $O(10)$ TeV~\cite{Chung:1998rq,Giudice:2000ex,Allahverdi:2002nb,Allahverdi:2002pu,Harigaya:2013vwa,Harigaya:2014waa,Harigaya:2014tla}.}
However, if SUSY breaking fields are charged or composite, gaugino masses are far smaller than soft scalar masses due to an approximate classical super-Weyl symmetry.
Actually, this is the case with many dynamical SUSY breaking models~(see e.g. Refs.~\cite{Affleck:1983vc,Affleck:1984mf,Affleck:1984xz,Luty:1997nq}).%
\footnote{For a dynamical SUSY breaking model with a singlet SUSY breaking field, see Ref.~\cite{Harigaya:2013ns} for example.}
It should be noted that SUSY breaking models with charged or composite SUSY breaking fields are free from the Polonyi problem~\cite{Coughlan:1983ci,Ibe:2006am}.

In this thesis, we consider the SUSY breaking scenario where the gravitino mass is larger than $O(10)$ TeV and SUSY breaking fields are charged or composite, which we refer to as the ``heavy sfermion scenario".
The PeV SUSY~\cite{Wells:2003tf,Wells:2004di}, the pure gravity mediation model~\cite{Ibe:2006de,Ibe:2011aa,Ibe:2012hu}, the minimal split SUSY model~\cite{ArkaniHamed:2012gw} and the spread SUSY model~\cite{Hall:2011jd} belong to this scenario.
In Sec.~\ref{sec:PGM}, we review the theoretical framework of the heavy sfermion scenario, including its cosmology.

Further, we assume that the Dirac mass term of the higgsino, so-called the $\mu$ term, is also as large as the gravitino mass.
Indeed, the $\mu$ term as large as the gravitino mass is naturally explained if any charges of the up and the down type higgs chiral multiplets add up to zero~\cite{Inoue:1991rk,Casas:1992mk} (see also Ref.~\cite{Giudice:1988yz}).
The heavy sfermion scenario with this origin of the $\mu$ term is called the pure gravity mediation model~\cite{Ibe:2006de,Ibe:2011aa,Ibe:2012hu}.

Assuming that the higgsino is heavy, we pay attention to gaugino masses in the heavy sfermion scenario.
In the heavy sfermion scenario, gaugino masses are at least given by the anomaly mediation~\cite{Randall:1998uk,Giudice:1998xp}.
The anomaly mediation yields gaugino masses proportional to the gravitino mass and the beta functions of the corresponding gauge coupling constants.
The LSP is the neutral wino with a mass of $O(0.1\mathchar`-1)$ TeV for the gravitino mass of $O(100\mathchar`-1000)$ TeV.%
\footnote{If the higgsino threshold correction is large, the bino can also be the LSP.}
Due to its large (co)annihilation cross-section, the thermal abundance of the neutral wino is smaller than the observed dark matter abundance as long as the wino mass is smaller than $3$ TeV~\cite{Hisano:2006nn}.

Sec.~\ref{sec:AMSB} is devoted to the derivation of the anomaly mediated gaugino mass in the superspace formulation of supergravity with a Wilsonian effective action~\cite{Harigaya:2014sfa}.
In the superspace formulation of supergravity~\cite{Wess:1992cp}, the anomaly mediated gaugino mass in the Wilsonian effective action invariant under the super-diffeomorphism was not known~\cite{deAlwis:2008aq,DEramo:2013mya}.
This is because the graivitino mass, which is the origin of the anomaly mediated gaugino mass, is the vacuum expectation value (VEV) of the scalar component of the supergravity multiplet.
Couplings of the supergraivity multiplet to the gauge multiplet are strongly constrained by the super-diffeomorphism invariance.
Thus, it is difficult (actually impossible) to write down the anomaly mediated gaugino mass in the Wilsonian effective action
in a manifestly super-diffeomorphism invariant way.
We give a super-diffeomorphism invariant expression of the anomaly mediated gaugino mass by taking the path-integral measure into account. 

In Sec.~\ref{sec:correction}, we discuss
the deviation of gaugino masses
from the prediction of the anomaly mediation in the MSSM.
In the presence of a flat direction coupling to gauge charged matters, gaugino masses receive corrections as large as the anomaly mediation~\cite{Pomarol:1999ie}.
Actually,  this is the case with a KSVZ-type QCD axion~\cite{Kim:1979if,Shifman:1979if}. The KSVZ-type QCD axion couples to vector-like matter fields charged under the standard model gauge interaction.
Thus, gaugino masses in general deviate from the prediction of the anomaly mediation in the MSSM if the KSVZ-type QCD axion exists~\cite{Nakayama:2013uta}.

Gaugino masses also receive corrections as large as the anomaly mediation if there are vector-like matter fields whose masses are smaller than the gravitino mass~\cite{Nelson:2002sa,Hsieh:2006ig,Gupta:2012gu}.
Actually, vector-like matter fields as heavy as the gravitino are predicted in models with an anomaly free discrete $R$ symmetry~\cite{Kurosawa:2001iq,Harigaya:2013vja}.

With the above two corrections, the gaugino mass spectrum drastically changes.
For example, the gluino mass can be lighter than the prediction of the anomaly mediation in the MSSM, which enhances the detectability of the gluino at the Large Hadron Collider (LHC).
We pay attention to the case where some gauginos are degenerated in their masses with each other, because the thermal relic abundance of the LSP as well as the the strategy to search gauginos drastically change in this case.
We refer to this region of gaugino masses as the ``gaugino coannihilation region".
In Sec.~\ref{sec:pheno},
we calculate the thermal relic abundance of the LSP in the gaugino coannihilation region, and discuss the phenomenology of gaugino searches at the Large Hadron Collider (LHC) and cosmic ray experiments.

\newpage
\section{Heavy sfermion scenario}
\label{sec:PGM}

In this section, we review the heavy sfermion scenario.
We first review the relation between sfermion masses and the observed higgs mass, and show that heavy sfermion masses are suggested.
Then we discuss the mass spectrum of SUSY particles in the heavy sfermion scenario, where SUSY breaking fields are
charged or composite fields.
As we will see, gaugino masses are one-loop suppressed in comparison with soft scalar masses.
Finally, we investigate the compatibility of the heavy sfermion scenario with cosmology.
We show the upper bound on the wino mass from the thermal relic abundance of the wino.
We also show that the gravitino problem
 and the saxion/axino problem are
is considerably relaxed in the heavy sfermion scenario.

\subsection{Higgs mass and scalar mass}
\label{sec:higgs mass}

\subsubsection{Tree level higgs mass}
We first calculate the tree level higgs mass.
At the tree level in the MSSM, the potential of the up-type higgs $H_u = (h_u^+, h_u^0)^T$ and the down-type higgs $H_d = (h_d^0, h_d^-)^T$ is given by
\begin{eqnarray}
\label{eq:higgs potential}
V (H_u,H_d) &=&
   \left(|\mu|^2 + m_{H_u}^2\right) \left(|h_u^0|^2 + |h_u^+|^2\right)
+ \left(|\mu|^2 + m_{H_d}^2\right) \left(|h_d^0|^2 + |h_d^-|^2\right)\nonumber\\
&& + \left[ b_H \left( h_u^+ h_d^- - h_u^0 h_d^0 \right) + {\rm h.c.}\right] \nonumber \\
&& + \frac{1}{8} (g^2 + g'^{2})\left(|h_u^0|^2 + |h_u^+|^2 - |h_d^0|^2 -|h_d^-|^2  \right)^2 
 + \frac{1}{2}g^2 | h_u^+ h_d^{0*} + h_u^0 h_d^{-*}  |^2,
\end{eqnarray}
where $\mu$, $m_{H_u}^2$, $m_{H_d}^2$, $b_H$, $g$ and $g'$ are the supersymmetric higgsino mass, the soft scalar squared masses of the up-type and the down-type higgs, the holomorphic quadratic soft mass, the $SU(2)_L$ and $U(1)_Y$ gauge coupling constants, respectively.

The minimum of the potential in Eq.~(\ref{eq:higgs potential}) is calculated in Appendix~\ref{sec:higgs}.
In this section we assume the decoupling limit, where higgs bosons except for the standard-model like one are far heavier that the $Z$ boson. Then the standard-model like higgs $h$ is given by
\begin{eqnarray}
h_u^0 \rightarrow \frac{1}{\sqrt{2}} {\rm sin}\beta ~h,~~ h_d^0 \rightarrow \frac{1}{\sqrt{2}} {\rm cos}\beta~ h,
\end{eqnarray}
where ${\rm tan}\beta = \vev{h_u^0}/ \vev{h_d^0}$.
The potential of $h$ is given by
\begin{eqnarray}
V(h) = \frac{\lambda}{8} (h^2 -v^2)^2,\\
\lambda \equiv \frac{1}{4} {\rm cos}^2 (2 \beta) (g^2 + g'^{2}),
\end{eqnarray}
where $v \simeq 246$ GeV is the VEV of the higgs $h$ determined by quadratic mass terms.
Remembering that the mass of the $Z$ boson is given by $m_Z^2 = \frac{1}{4} (g^2 + g'^{2}) v^2 $, the mass of $h$, $m_h$, is given by
\begin{eqnarray}
m_h^2 = \lambda v^2 =  {\rm cos}^2(2\beta) m_Z^2~.
\end{eqnarray}
At the tree level, the mass of the standard-model like higgs is lighter than the mass of the $Z$ boson, $m_Z=91.2$ GeV.

\subsubsection{Quantum correction by SUSY breaking}

\begin{figure}[t]
\begin{center}
\includegraphics[width=0.9\linewidth]{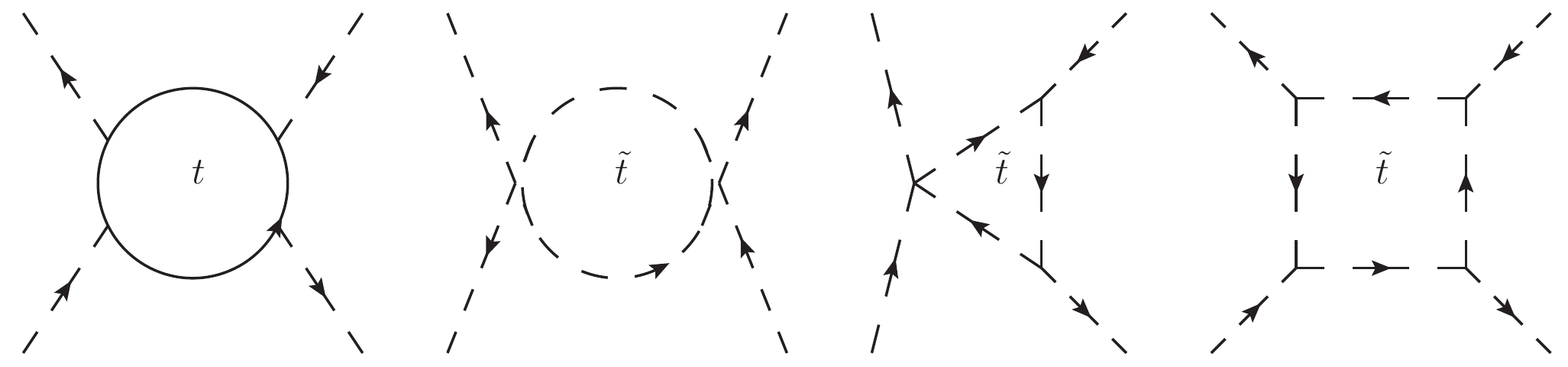}
\caption{\sl \small Threshold correction to the quartic coupling.}
\label{fig:threshold}
\end{center}
\end{figure}

The tree level mass of the higgs is bounded from above because the quartic coupling of higgs $\lambda$ is given by the $D$ term potential, whose value is restricted by SUSY.
SUSY breaking effects can raise the quartic coupling through quantum corrections.
At the one-loop level, corrections are dominated by the following two effects.

One is the finite threshold correction involving stops, whose one-loop diagrams are shown in Fig.~\ref{fig:threshold}.
They correct the quartic coupling as~\cite{Okada:1990gg}
\begin{eqnarray}
\label{eq:lambda threshold}
\Delta \lambda = \frac{6}{16\pi^2} y_t^4 \left(
\frac{1}{4}\frac{|X_t|^2}{\bar{m}_{\tilde{t}}^2} - \frac{1}{192} \frac{|X_t|^4}{\bar{m}_{\tilde{t}}^4}
\right) ,\\
X_t \equiv A_t - \mu {\rm cot} \beta,~~ \bar{m}_{\tilde{t}}^2 \equiv \frac{m_{\tilde t_L}^2 + m_{\tilde t_R}^2}{2},
\end{eqnarray}
where $y_t$, $A_t$, $m_{\tilde{t}_L}^2$ and $m_{\tilde{t}_R}^2$ are the top yukawa, the soft trilinear coupling of higgses and stops, the soft squared masses of the left-hand and right-hand stops, respectively.
Another is the running of the quartic coupling.
For simplicity, we assume that SUSY particles have the same mass and hence the running is determined by
the standard model interaction from the soft mass scale down to the electroweak scale,
\begin{eqnarray}
\label{eq:lambda running}
\frac{{\rm d}} {{\rm dln}\mu } \lambda = \frac{1}{16\pi^2} \left[
12 \left(\lambda^2 - y_t^4 + \lambda y_t^2 \right) - \lambda \left( 3 g'^{2} + 9 g^2 \right)
+ \frac{3}{4} \left(
g'^{4} + 2 g'^{2} g^2 + 3 g^4
\right)
\right],
\end{eqnarray}
where $\mu$ is the renormalization scale.
The correction in Eq.~(\ref{eq:lambda threshold}) should be added at the soft mass scale and the renormalization equation (\ref{eq:lambda running}) should be solved with the corrected boundary condition at the soft mass scale.

\begin{figure}[t]
\begin{center}
\includegraphics[width=0.7\linewidth]{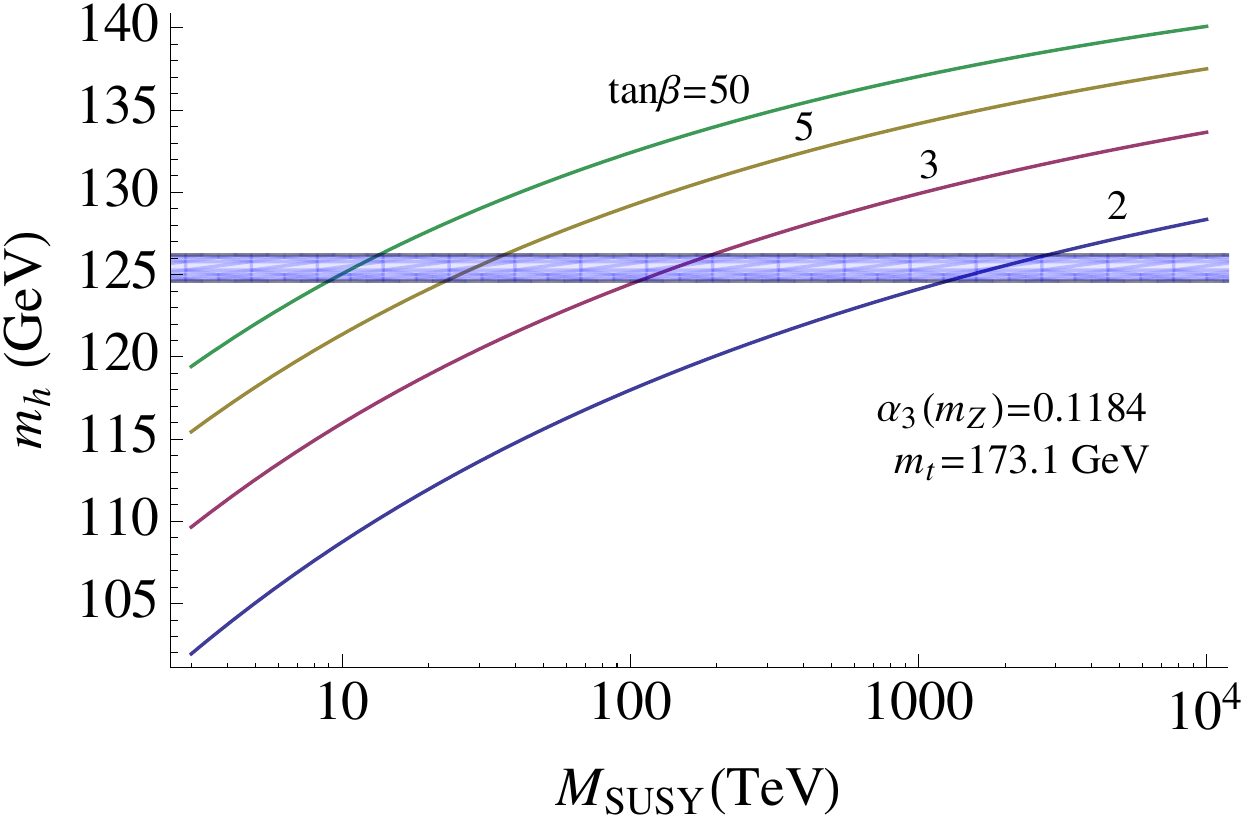}
\caption{\sl \small Higgs mass for given soft mass scales and ${\rm tan}\beta$.
The The blue band shows the observed higgs mass, $m_h = 125.36\pm 0.8$ GeV~\cite{Aad:2014aba}.
}
\label{fig:higgs mass}
\end{center}
\end{figure}

In Fig.~\ref{fig:higgs mass}, we show the higgs mass obtained from Eqs.~(\ref{eq:lambda threshold}) and (\ref{eq:lambda running})
with $m_{\tilde{t}_L}^2 = m_{\tilde{t}_R}^2 = |\mu|^2 \equiv M_{\rm SUSY}^2$ and $|A_t| \ll M_{\rm SUSY}$.
Here, we have also included the threshold correction at the electroweak scale summarized in Ref.~\cite{Degrassi:2012ry}.
The blue band shows the observed higgs mass, $m_h = 125.36\pm 0.8$ GeV~\cite{Aad:2014aba}.
It can be seen that $m_h \simeq 125.36$ GeV is obtained for sufficiently large $M_{\rm SUSY}$.
For example, for ${\rm tan}\beta = O(1)$, the observed higgs mass requires $M_{\rm SUSY}>O(100)$ TeV.
For the recent accurate calculation, see Ref.~\cite{Hahn:2013ria}.

\subsection{SUSY breaking without singlet}

In most of dynamical SUSY breaking models~(see e.g. Refs.~\cite{Affleck:1983vc,Affleck:1984mf,Affleck:1984xz,Luty:1997nq}),
SUSY breaking fields are charged under some symmetry or are composite fields.
We refer to this type of SUSY breaking as ``SUSY breaking without singlet".
In this subsection, we discuss the mass spectrum of SUSY particles expected in SUSY breaking without singlet.

We denote the SUSY breaking field as $Z$ and assume the following simplest effective superpotential,
\begin{eqnarray}
W_{\rm eff} = \Lambda^2 Z + W_0,
\end{eqnarray}
where $\Lambda$ is the SUSY breaking scale and $W_0$ is a constant term.%
\footnote{When the SUSY breaking field $Z$ have a charge under some symmetry, $\Lambda$ has a charge so that the superpotential is invariant under the symmetry.
That is, the SUSY breaking is associated with the breaking of the symmetry.
}
The gravitino mass is given by $m_{3/2}^* = W_0 / \Mpl^2$, where $\Mpl$ is the reduced Planck mass.
Vanishing of the cosmological constant requires that  $|\Lambda|^2  = \sqrt{3} |m_{3/2}|\Mpl$, where we have normalized $Z$ so that it is canonical.

\subsubsection{Soft squared scalar mass}
Let us first discuss soft squared scalar mass terms of chiral multiplets $Q^i$ in the MSSM.
The SUSY breaking field $Z$ in general couples to $Q^i$ through the Kahler potential,
\begin{eqnarray}
K = Q^iQ^{\bar{i}\dag} + Z Z^\dag + \frac{c_{i\bar{j}}}{\Mpl^2} Q^iQ^{\bar{j}\dag} Z Z^\dag.
\end{eqnarray}
We assume that the SUSY breaking sector couples to the standard model sector only through Planck scale suppressed interactions and hence $c_{i\bar{j}}$ is at largest $O(1)$.
The soft squared mass term of $Q^i$ is given by
\begin{eqnarray}
m_{i\bar{j}}^2 = \left(\delta_{i\bar{j}} - 3 c_{i\bar{j}}  \right)|m_{3/2}|^2.
\end{eqnarray}
Thus, in general, the soft squared scalar mass term is as large as the gravitino mass.
Among MSSM higgses, only the standard model-like higgs is light while other higgses have the same mass as large as  the gravitino mass.

We note that generic $c_{i\bar{j}}$ without any structures induce flavor changing neutral currents, CP violations and lepton flavor violations.
For discussions on these issues in the heavy sfermion scenario, we refer to Refs.~\cite{Sato:2012xf,Moroi:2013sfa,Moroi:2013vya,Sato:2013bta,Altmannshofer:2013lfa,Fuyuto:2013gla,Tanimoto:2014eva}.

\subsubsection{$\mu$ term and $b_H$ term}

In order for the electroweak symmetry to be broken by VEVs of MSSM higgses,
the $\mu$ term must be as large as or smaller than soft scalar mass terms of higgses (see Appendix~\ref{sec:higgs}).
The origin of the correct magnitude of the $\mu$ term is one of the key issues in SUSY model-building, which is dubbed as the ``$\mu$ problem".

A trivial way to obtain the $\mu$ term is to assume that the combination $H_u H_d$ has an $R$ charge of 2
 while is neutral under other symmetries.%
 \footnote{We use a normalization of the $R$ charge where the superpotential has an $R$ charge of 2.}
Then the $\mu$ term of any magnitude is allowed.
The natural value would be as large as the fundamental scale such as the Plack scale and the GUT scale.
Indeed, this is the case with the minimal $SU(5)$ GUT model~\cite{Witten:1981nf,Dimopoulos:1981yj,Dimopoulos:1981zb,Sakai:1981gr}.
In the landscape point of view~\cite{Bousso:2000xa,Kachru:2003aw,Susskind:2003kw,Denef:2004ze},
the small $\mu$ term may be selected from the landscape by the anthropic principle~\cite{Weinberg:1987dv}.
$b_H=\mu m_{3/2}$ is obtained by the supergravity effect. 

Here, instead of readily adopting the anthropic principle, we show two ways to naturally obtain a small $\mu$ term.
In both cases, the $\mu$ term is forbidden by a symmetry and is given by the breaking of the symmetry.%
\footnote{
A solution to the $\mu$ problem in this way requires the GUT group to be a product one~\cite{Witten:2001bf,Harigaya:2015zea}.
}
We do not discuss the origin of the smallness of the breaking scale in detail here.
The smallness is naturally explained, for example, if the breaking scale is generated by dimensional transmutation.

The first model assume the Peccei-Quinn (PQ) symmetry~\cite{Peccei:1977hh,Peccei:1977ur}.
The PQ charge of MSSM fields is given in Tab.~\ref{tab:PQ}.
With this charge assignment, yukawa couplings are allowed while the $\mu$ term is forbidden.
The $\mu$ term is provided by the following coupling to a PQ breaking field $P$ with a PQ charge $1/n$~\cite{Kim:1983dt},
\begin{eqnarray}
W\supset \frac{P^n}{\Mpl^{n-1}} H_u H_d.
\end{eqnarray}
For $n=2$, the $\mu$ term is of $O(0.1\mathchar`-1000)$ TeV  for the PQ breaking scale of $O(10^{10}\mathchar`-10^{12})$ GeV.
This range is consistent with the lower bound from the burst duration of SN1987A (Ref.~\cite{Raffelt:2006cw} and references therein) and the ``upper bound" from the cosmic abundance of the axion by an initial misalignment angle~\cite{Preskill:1982cy,Abbott:1982af,Dine:1981rt}. 

The $b_H$ term is given by the VEV of the scalar component of the supergravity multiplet and the $F$ term of $P$ as large as $\vev{P}m_{3/2}$ (see Sec.~\ref{sec:axion}). Both contributions result in the $b_H$ term of $O(m_{3/2} \mu)$.
If $\mu \ll m_{3/2}$, $b_H$ is smaller than the soft squared masses of sfermions, and hence ${\rm tan}\beta$ is large unless cancellation occurs (see Eq.~(\ref{eq:tanb})).
Then the gravitino mass must be less than $O(10)$ TeV to explain the observed higgs mass. In this case, however, gaugino masses discussed later are too small that it is inconsistent with constraints from the LHC. Thus, the $\mu$ term must be as large as the gravitino mass.

\begin{table}[tb]
\begin{center}
\begin{tabular}{|c||c|c|c|c|}
& $H_u$ & $H_d$ & $\bar{u},Q, \bar{e}$ & $\bar{d},L$ \\ \hline
$U(1)_{\rm PQ}$ & $0$ & $-1$ & $0$ & $1$ 
\end{tabular}
\end{center}
\caption{PQ charge assignment of MSSM fields}
\label{tab:PQ}
\end{table}%

The second model assumes the $R$ symmetry. We assume that the combination $H_u H_d$ has vanishing charges under any symmetries.%
\footnote{If $Z$ is a fundamental field with an $R$ charge of $2$ and without any other charges, an order one yukawa coupling between $Z$ and $H_u H_d$ in the superpotential is possible. Then higgs fields obtain large vacuum expectation values. We assume that $Z$ is a composite field, has an $R$ charge other than $2$, or charged under non-$R$ symmetries.}
Then following terms are allowed in the Kahler potential,
\begin{eqnarray}
K \supset c_1 H_u H_d + c_2 H_u H_d\frac{ZZ^\dag}{\Mpl^2} + {\rm h.c.}.
\end{eqnarray}
Then the $\mu$ and $b_H$ terms are given by~\cite{Ibe:2006de}
\begin{eqnarray}
\mu = c_1 m_{3/2}^*,~~ b_H = (2c_1- 3c_2)|m_{3/2}|^2.
\end{eqnarray}
The $\mu$ term of $O(m_{3/2})$ can be understood by observing that the chiral higgsino pair has an $R$ charge of $-2$ while the gravitino mass $m_{3/2}$ has an $R$ charge of $-2$.
Since $b_H$ is as large as the soft squared mass term, the natural value of ${\rm tan}\beta$ is $O(1)$ (see Eq.~(\ref{eq:tanb})).
Then the observed higgs mass requires $m_{3/2}>O(100)$ TeV.
The $\mu$ term can be far smaller than the gravitino mass by tuning $c_1$, but we assume a natural value, $\mu\sim m_{3/2}$.

\subsubsection{Gaugino mass}

If $Z$ is charged under some symmetry, the coupling between $Z$ and a gauge multiplet through the gauge kinetic function is forbidden because the gauge kinetic function is neutral under any symmetries.
If $Z$ is a singlet composite field in a dynamical SUSY breaking model, the following coupling may be possible,
\begin{eqnarray}
\int {\rm d}^2 \theta \frac{Z \Lambda_{\rm dyn}^n }{4\pi \Mpl^{n+1}} W^\alpha W_\alpha,
\end{eqnarray}
where $\Lambda_{\rm din}\sim (4\pi)^{1/2} \Lambda $ is the dynamical scale of the SUSY breaking model.
Here, we assume that $Z$ corresponds to a composite field composed of $n+1$ chiral multiplets.
In both cases, the tree-level gaugino mass is far smaller than the gravitino mass.

It was pointed out in Refs.~\cite{Randall:1998uk,Giudice:1998xp} that the following gaugino mass is generated through the conformal anomaly,
\begin{eqnarray}
M_\lambda^{({\rm AM})} = - \frac{\beta(g^2)}{2g^2} m_{3/2},
\end{eqnarray}
where $g$ and $\beta (g^2)$ are the gauge coupling constant and the beta function of $g^2$, respectively.
The anomaly mediated gaugino mass is derived in Sec.~\ref{sec:AMSB}.

At the one-loop level in the MSSM, bino, wino, and gluino masses ($M_1^{\rm (AM)}$, $M_2^{\rm (AM)}$, and $M_3^{\rm (AM)}$) are given by
\begin{eqnarray}
M_1^{\rm (AM)} = \frac{g_1^2}{16\pi^2} \frac{33}{5} m_{3/2},
\qquad
M_2^{\rm (AM)} = \frac{g_2^2}{16\pi^2} m_{3/2},
\qquad
M_3^{\rm (AM)} = -\frac{g_3^2}{16\pi^2} 3 \, m_{3/2},
\label{eq: AM}
\end{eqnarray}
where $g_1$, $g_2$ and $g_3$ are the gauge coupling constants of $U(1)_Y$, $SU(2)_L$ and $SU(3)_c$, respectively.
Here, we use the GUT normalization for the $U(1)_Y$ gauge coupling constant, $g_1^2 = g'^2 (5/3)$.

\begin{figure}[b]
\begin{center}
\includegraphics[width=0.5\linewidth]{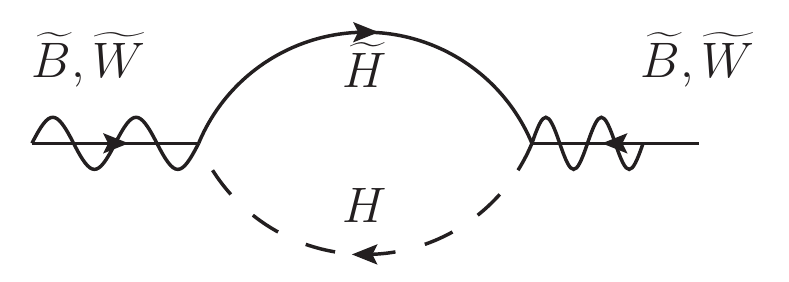}
\caption{\sl \small higgsino threshold corrections to masses of electroweak gauginos.}
\label{fig: higgsinoth}
\end{center}
\end{figure}

In addition to the anomaly mediation, 
electroweak gauginos (bino and wino) receive threshold corrections from the higgsino via the diagram in Fig.~\ref{fig: higgsinoth}. The corrections are evaluated as~\cite{Giudice:1998xp} (the calculation is the essentially same as the one in Sec.~\ref{sec:correction vector}),
\begin{eqnarray}
\Delta M_1^{\rm (HT)} = \frac{g_1^2}{16\pi^2} \frac{3}{5} L,
\qquad
\Delta M_2^{\rm (HT)} = \frac{g_2^2}{16\pi^2} L,
\qquad
L \equiv \frac{\mu \, m_A^2 \, \sin 2\beta}{|\mu|^2 - m_A^2}
\ln \frac{|\mu|^2}{m_A^2},
\label{eq: HT}
\end{eqnarray}
where $m_A$ is 
the mass of heavy higgses. The contributions are comparable to those of anomaly mediated contributions when $\mu = {\cal O}(m_{3/2})$ and $\tan \beta = {\cal O}(1)$.

\begin{figure}[t]
\begin{center}
\includegraphics[width=0.47\linewidth]{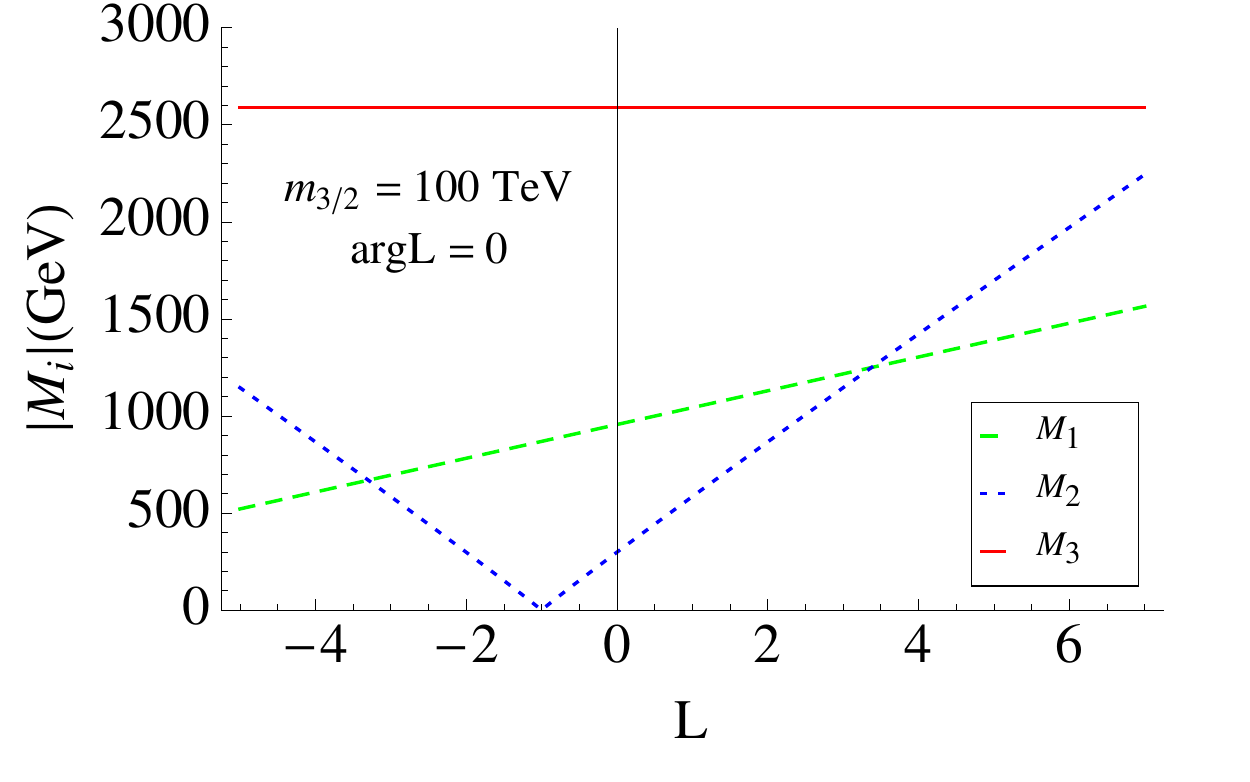}
\caption{\sl \small
Gaugino (bino, wino, and gluino) masses in the high-scale SUSY breaking scenario of the MSSM.
}
\label{fig: gaugino masses}
\end{center}
\end{figure}

In the MSSM, physical masses of the gauginos $M_i$ are obtained by adding the contributions in Eqs.~(\ref{eq: AM}) and (\ref{eq: HT}), and also considering the effect of renormalization group running of the masses down to those scales from $M_{\rm SUSY}$,
\begin{eqnarray}
\label{eq:renormalization eq}
 \frac{{\rm dln}M_i(\mu)}{{\rm dln}\mu} &=&
  -\frac{g_i^2(\mu)}{8\pi^2}b_i,~~~(b_1,b_2,b_3)= (0,6,9),\nonumber\\
M_i(M_{i,{\rm phys}}) &=& M_{i,{\rm phys}}. 
\end{eqnarray}
In Fig.~\ref{fig: gaugino masses}, the gaugino masses are shown as a function of $L$ assuming the phase of the higgsino threshold corrections to be zero (${\rm arg}\,L = 0$), and $M_{\rm SUSY} = m_{3/2} = 100$\,TeV.
Unless the higgsino threshold correction is large, the wino is the LSP.

The wino LSP is constrained by the disappearing track search at the LHC as $M_2 >  270$ GeV~\cite{Aad:2013yna}.
Also, the search for jets with missing energy at the LHC put the constraint on the gluino mass, $M_3 \gsim 1.4$ TeV~\cite{Chatrchyan:2014lfa} unless the gluino is degenerated with the LSP.
Thus, the gravitino mass larger than $O(100)$ TeV is required, which is consistent with the observed higgs mass for ${\rm tan}\beta = O(1)$.

\subsubsection{$A$ term}

Let us briefly mention the $A$ term.
As is the case with gaugino masses, $A$ term is suppressed in SUSY breaking without singlet.
However, the $A$ term is generated through the conformal anomaly. For a superpotential $W = y Q_1Q_2 Q_3$, the corresponding $A$ term is given by the wave function normalization as~\cite{Randall:1998uk,Giudice:1998xp}
\begin{eqnarray}
{\cal L}_{A~{\rm term}} &=& y A(\mu) Q_1 Q_2 Q_3 + {\rm h.c.},~~
A(\mu) = -\frac{1}{2} \sum_i \gamma_i (\mu) m_{3/2},\nonumber\\
{\cal L}_{\rm kin,renormalized} &= & \int {\rm d}^2\theta {\rm d}^2\theta^\dag Z_i (\mu) Q^i Q^{\bar{i}\dag} ,~~   \gamma_i(\mu)\equiv \frac{{\rm dln}Z_i(\mu)}{{\rm dln}\mu},
\end{eqnarray}
which is one-loop suppressed in comparison with the gravitino mass.

\subsection{Compatibility with cosmology}

We have shown that  the gravitino mass of $O(100)$ TeV is consistent with the observed higgs mass and the constraint from the LHC in the heavy sfermion scenario.
Here, we investigate the compatibility of the large gravitino mass with cosmology.
We discuss the thermal relic abundance of the LSP,
the gravitino problem,
and the saxion/axino problem.
For the compatibility of the large gravitino mass with inflation models, see Refs.~\cite{Buchmuller:2012wn,Buchmuller:2012bt,Nakayama:2012dw,Nakayama:2012hy,Harigaya:2012hn,Takahashi:2013cxa,Harigaya:2013pla,Harigaya:2014pqa,Nakayama:2014xca,Harigaya:2014roa}.

\subsubsection{Thermal abundance of the wino LSP}
Unless the higgsino threshold correction is large, the wino is the LSP.
The neutral and the charged wino masses degenerate with each other at the tree-level.
Through quantum corrections by electroweak interactions, the neutral wino becomes lighter than the charged wino by $\sim 170$ MeV~\cite{Cheng:1998hc,Feng:1999fu,Gherghetta:1999sw,Yamada:2009ve,Ibe:2012sx}.

\begin{figure}[t]
\begin{center}
\includegraphics[width=0.5\linewidth]{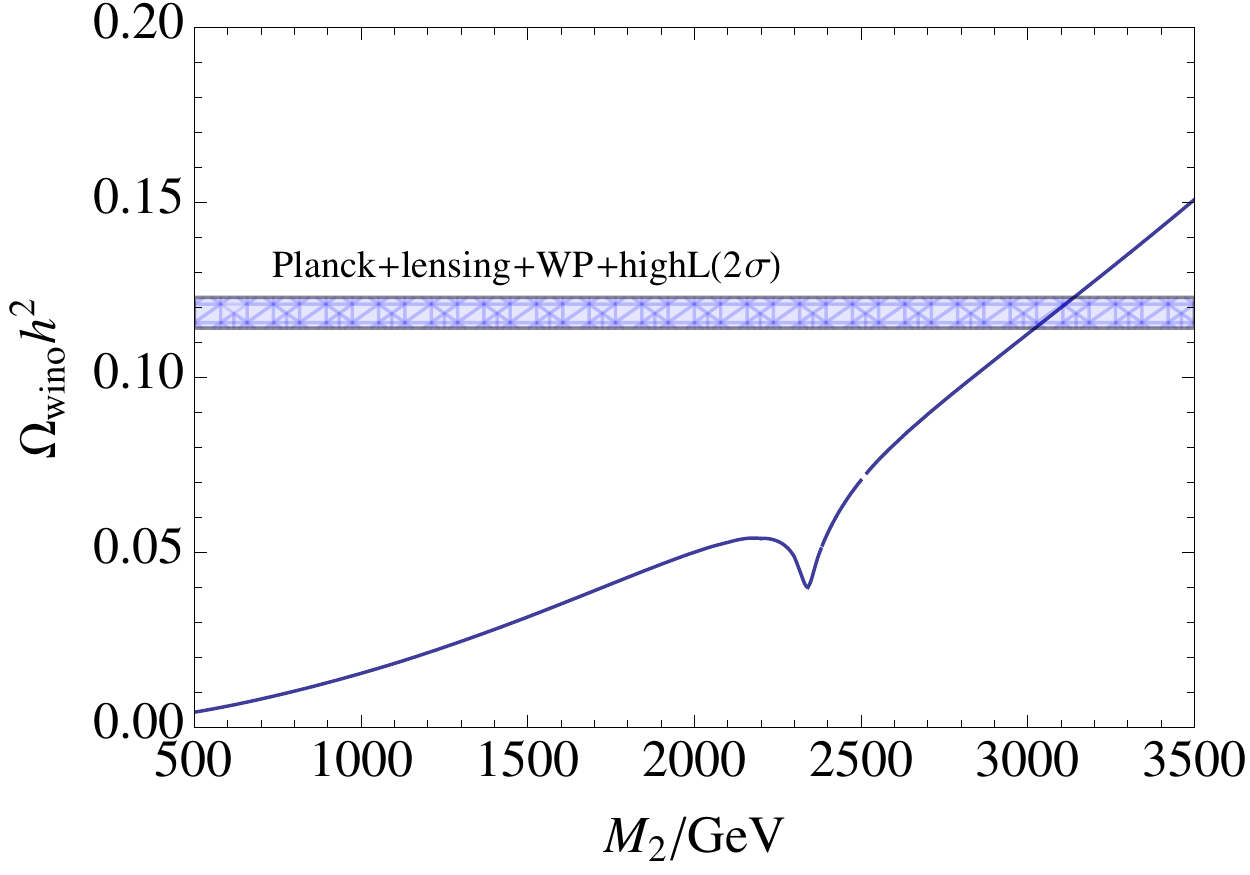}
\caption{\sl \small
Thermal abundance of the wino for a given wino mass (see Sec.~\ref{sec:pheno}).
The blue band shows the observed dark matter abundance by the Planck experiment.
}
\label{fig:wino abundance}
\end{center}
\end{figure}

The thermal abundance of the neutral wino LSP has been calculated in Ref.~\cite{Hisano:2006nn}, including the coannihilation between the charged and the neutral wino as well as the Sommerfeld effect (see Sec.~\ref{sec:pheno}).
In Fig.~\ref{fig:wino abundance}, we show the thermal abundance of the wino for a given wino mass.
The blue band shows the observed dark matter abundance by the Planck experiment~\cite{Ade:2013zuv}.
For $M_2\simeq 3.1$ TeV, the thermal abundance of the wino is consistent with the observed dark matter abundance.

In the present universe, the neutral wino annihilates into a pair of $W$ bosons at the tree level and a pair of photons at the one-loop level.
The former mode yields gamma-rays, positrons and light elements with spread spectra while the latter yields gamma-rays with a line spectrum.
For the wino search with cosmic-rays, see Refs.~\cite{Moroi:1999zb,Ibe:2012hu,Bhattacherjee:2014dya}.

\subsubsection{Gravitino problem}
Let us consider the production of the gravitino from thermal bath in the early universe.
Since the gravitino interacts with MSSM particles through higher dimensional interactions, the production of the gravitino from thermal bath is more efficient for higher temperature. Therefore, the number density of the gravitino is determined by the reheating temperature as~\cite{Kawasaki:1994af,Kawasaki:2008qe},
\begin{eqnarray}
\frac{n_{3/2}}{s}\simeq 2.3 \times
  10^{-13} \times\frac{T_{\rm RH}}{10^{9}~{\rm GeV}} ,
\label{eq: gravitino abundance}
\end{eqnarray}
where $s$ it the entropy density and we have used the approximation that gaugino masses are far smaller than the gravitino mass.

The produced gravitino eventually decays into the LSP. The abundance of the LSP produced in this way is given by
\begin{eqnarray}
\Omega_{\rm LSP}^{({\rm grav)}}h^2 = \frac{m_{\rm LSP}}{3.6\times 10^{-9}~{\rm GeV}}\frac{n_{3/2}}{s}
= 0.12 \times \frac{m_{\rm LSP}}{900~{\rm GeV}} \frac{T_{\rm RH}}{2\times10^{9}~{\rm GeV}}.
\end{eqnarray}
It is known that the thermal leptogenesis~\cite{Fukugita:1986hr} requires the reheating temperature larger than about $2\times 10^9$ GeV~\cite{Giudice:2003jh,Buchmuller:2004nz}.
Then, the thermal leptogenesis puts an upper bound on the mass of the LSP, $m_{\rm LSP} \lsim 1 $ TeV.

Next, let us discuss the constraint from the Big Bang Nucleosynthesis (BBN).
The gravitino decays with a rate $\Gamma_{3/2} \simeq 12 m_{3/2}^3/ (32\pi \Mpl^2)$, where we have assumed that sfermions and the higgsino are heavier than the gravitino.%
\footnote{Inclusion of the decay into sfermions and higgsino does not change the following discussion.}
The lifetime of the gravitino is given by
\begin{eqnarray}
\tau_{3/2} = \Gamma_{3/2}^{-1} = 0.06~{\rm sec} \left(\frac{m_{3/2}}{ 80~{\rm TeV}}\right)^{-3}.
\end{eqnarray}
Note that the gravitino dominantly decays into the gluino and the gluon.
In this case, if the lifetime of the gravitino is longer than about $0.1$ sec, decay products of the gravitino hadronically interact with light elements and hence spoil the success of the BBN~\cite{Kawasaki:2004qu}.
For $m_{3/2}\gsim 100$ TeV, the constraint from the BBN is absent.

\subsubsection{Saxion/axino problem}

In the heavy sfermion scenario, the saxion/axino problem is relaxed.
In general PQ mechanisms, there is a pseudo-Nambu-Goldstone boson $a$ associated with the spontaneous breaking of the PQ symmetry, which is called the axion. The axion couples to the Pontryagin density of the QCD through the anomaly of the PQ symmetry.
The strong dynamics of the QCD generate the potential of the axion. At the minimum of the potential, the $\theta$ angle of the QCD vanishes, which solves the strong CP problem~\cite{Peccei:1977hh,Peccei:1977ur,Weinberg:1977ma,Wilczek:1977pj}.

In SUSY models, the axion is associated with a light scalar $s$ called the saxion and a light fermion $\tilde{a}$ called the axino.
$a$, $s$ and $\tilde{a}$ form a chiral multiplet, which we refer to as the axion multiplet ${\cal A}$.
The saxion and the axino are as light as the gravitino. Then their dynamics in the early universe cause various problems.
Here, we discuss two problems which are important for $m_{3/2} \gsim $ TeV.
For a rigorous discussion on generic gravitino masses, see Ref.~\cite{Kawasaki:2007mk} and references therein.

\vspace{1cm}

\noindent
{\tt Dark radiation from saxion}

The saxion produced in the early universe in general decays into the axion at the tree level through the Kahler potential,
\begin{eqnarray}
K \supset ({\cal A} + {\cal A}^\dag)^3 / f_a,
\end{eqnarray}
where $f_a$ is the decay constant of the axion.
Here, we identify the imaginary part of the lowest component of ${\cal A}$ as the axion.
The axion produced in this way has a large momentum and hence works as radiation.
The extra radiation component, which is called the dark radiation, changes the expansion history of the universe and hence is constrained e.g.~by
the BBN,
the cosmic microwave background and
the structure formation.

There are two sources of the saxion production in the early universe.
One is the coherent oscillation of the saxion.
The mass of the saxion is given by SUSY breaking, which is dominated by an inflaton sector and the SUSY breaking sector during and after inflation, respectively.
Thus, the minima of the saxion potential during and after inflation are in general different and
typically separated by the decay constant of the axion.
The saxion starts its oscillation with an initial amplitude $s_i \sim f_a$ as the Hubble scale of the universe becomes comparable to the mass of the saxion $m_s$. Assuming that the inflaton decays after the saxion starts its oscillation, the energy density of the saxion before it decays is
\begin{eqnarray}
\left(\frac{\rho_s}{s}\right)_{\rm osc} = \frac{1}{8} T_{\rm RH} \frac{s_i^2}{\Mpl^2} = 2.2 \times 10^{-9} {\rm GeV} \frac{T_{\rm RH}}{10^9~{\rm GeV}}
\left(\frac{f_a}{10^{10}~{\rm GeV}}\right)^2 \left(\frac{s_i}{f_a}\right)^2.
\end{eqnarray}

The other source is the thermal scattering.
The saxion interacts with thermal bath with a rate
\begin{eqnarray}
\Gamma_{s,{\rm th}} \simeq \frac{\alpha_3^3}{128\pi^2 f_a^2} \times (n_g  + n_q )\simeq 4\times 10^{-6} \frac{T^3}{f_a^2},
\end{eqnarray}
where $n_g$ and $n_q$ are the thermal number density of the gluon and the quark.
The saxion is in thermal bath if the interaction rate is sufficiently larger than the Hubble rate, that is,
\begin{eqnarray}
T >  5\times 10^{9}~{\rm GeV} \left(\frac{f_a}{10^{11}~{\rm GeV}}\right)^2 \equiv T_D.
\end{eqnarray}
The energy density of the saxion before it decays is
\begin{eqnarray}
\left(\frac{\rho_s}{s}\right)_{\rm th} = 
\left\{
\begin{array}{ll}
1~{\rm GeV} \left(\frac{m_s}{{\rm TeV}}\right) & (T_{\rm RH} >  T_D ), \\
 2\times 10^{-4}~{\rm GeV}  \frac{m_s}{{\rm TeV}} \frac{T_{\rm RH}}{10^8~{\rm GeV}} \left(\frac{f_a}{10^{12}~{\rm GeV}}\right)^{-2}   & (T_{\rm RH} <  T_D ).
\end{array}
\right.
\end{eqnarray}

The constraint on the dark radiation is conventionally expressed by the effective neutrino number $N_{\rm eff}$ defined by
\begin{eqnarray}
\rho_{\rm rad}(T_\gamma) = \left[ 1 + \frac{7}{8} N_{\rm eff} \left(\frac{T_\nu}{T_\gamma}\right)^4  \right] \rho_\gamma (T_\gamma),
\end{eqnarray}
where $\rho_{\rm rad}$, $\rho_\gamma$, $T_\mu$ and $T_\gamma$ are the energy density of the total radiation, the photon, the temperature of the neutrino and the photon, respectively.
The standard cosmology predicts $N_{\rm eff} \simeq 3$ and the deviation $\Delta N_{\rm eff}$ is constrained.

As the saxion dominantly decays into the axion, $\Delta N_{\rm eff}$ is given by
\begin{eqnarray}
\frac{\rho_s}{s} \simeq 0.3 g_{*s}(T_s)^{-1} \Delta N_{\rm eff} T_s,
\end{eqnarray}
where $g_{*s}$ is the effective degree of freedom of the entropy density and $T_s$ is the temperature at which the saxion decays,
\begin{eqnarray}
T_s \simeq 0.3 \times \sqrt{\Gamma_s \Mpl}, ~~ \Gamma_s \sim \frac{1}{64\pi}\frac{m_s^3}{f_a^2}. 
\end{eqnarray}
We adopt the limit $\Delta N_{\rm eff} <1$ by the Planck experiment~\cite{Ade:2013zuv}.

\vspace{1cm}

\noindent
{\tt LSP overproduction from axino}

The axino is produced from thermal bath and eventually decays into the LSP.
If the temperature at which the axino decay $T_{\tilde{a}}$,
\begin{eqnarray}
T_{\tilde{a}} \simeq 0.3 \times \sqrt{\Gamma_a \Mpl}, ~~ \Gamma_s \sim \frac{\alpha_3^2}{256\pi^2}\frac{m_{\tilde{a}}^3}{f_a^2},
\end{eqnarray}
 is smaller than the freeze-out temperature of the LSP, $\sim m_{\rm LSP} /20$, 
the LSP produced in this way contributes to the dark matter density.

The number density of the axino produced from thermal bath is given by~\cite{Covi:2001nw,Brandenburg:2004du}
\begin{eqnarray}
\label{eq:axino abundance}
\frac{n_{\tilde{a}}}{s} = 2\times 10^{-5} \frac{T_{\rm RH}}{10^8~{\rm GeV}} \left(\frac{f_a}{10^{12}~{\rm GeV}}\right)^{-2}.
\end{eqnarray}
Here, we assume that the axino is not in thermal bath ($T_{\rm RH} \lsim T_D$). Otherwise, the LSP produced from the axino over-closes the universe.
From Eq.~(\ref{eq:axino abundance}), the abundance of the LSP is given by
\begin{eqnarray}
\label{eq:LSP axino}
\Omega_{\rm LSP}^{({\rm axino)}}h^2 = \frac{m_{\rm LSP}}{3.6\times 10^{-9}~{\rm GeV}}\frac{n_{\tilde{a}}}{s}.
\end{eqnarray}

\vspace{1cm}

\begin{figure}[tb]
\begin{tabular}{cc}
\begin{minipage}{0.5\hsize}
\begin{center}
  \includegraphics[width=0.9\linewidth]{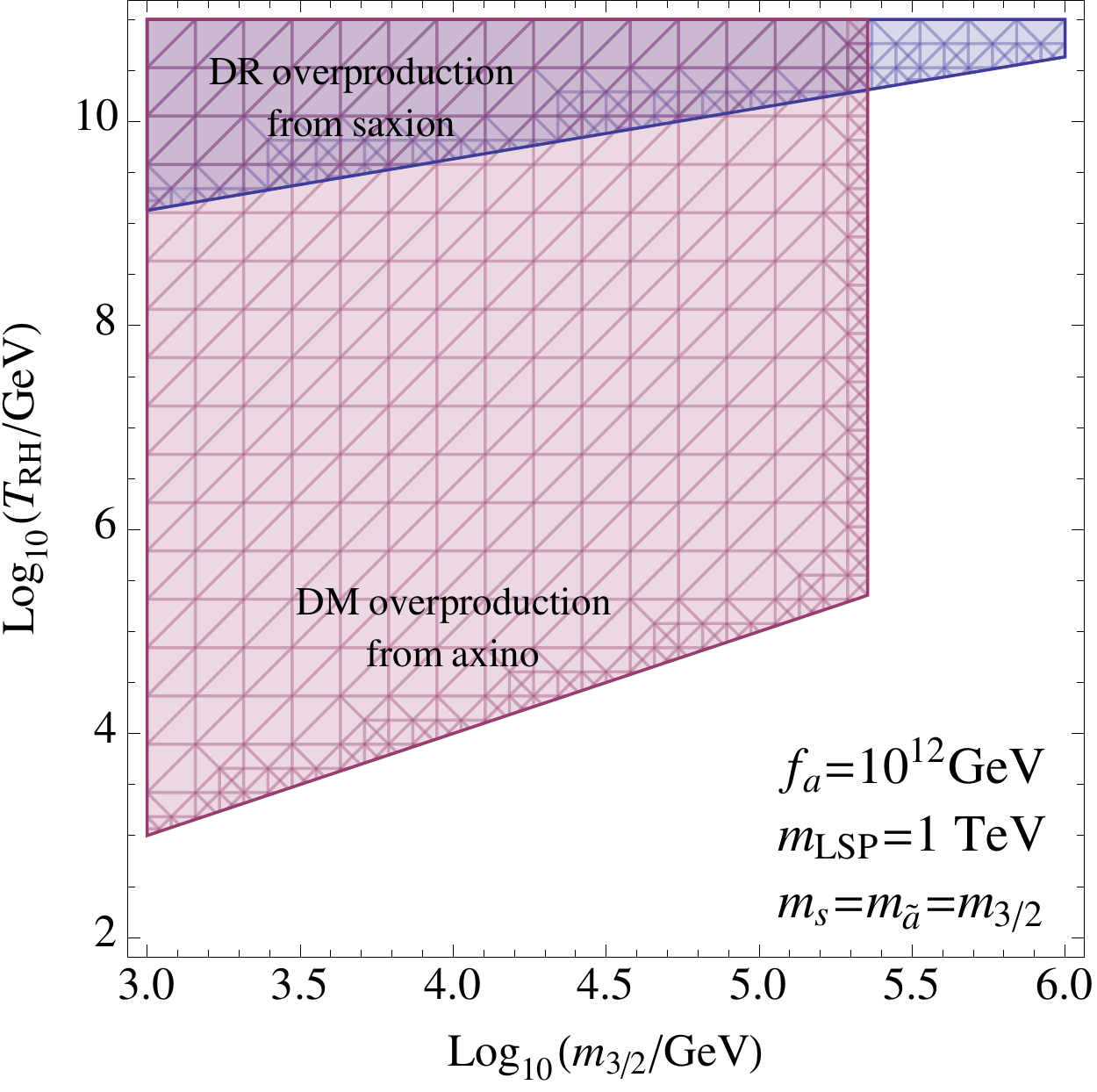}
\end{center}
\end{minipage}
 &
\begin{minipage}{0.5\hsize}
\begin{center}
  \includegraphics[width=0.9\linewidth]{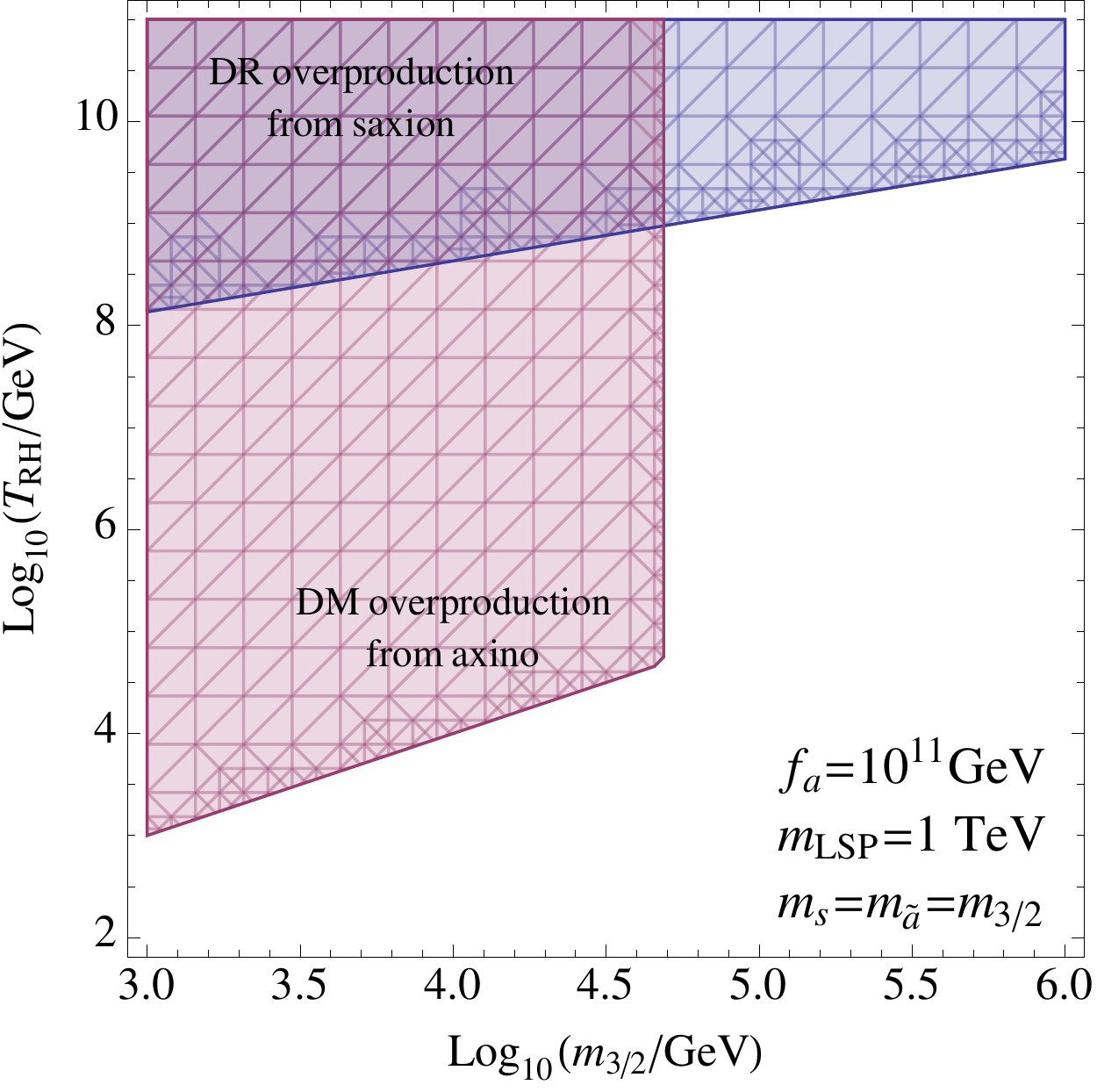}
\end{center}
\end{minipage}
\\
\begin{minipage}{0.5\hsize}
\begin{center}
  \includegraphics[width=0.9\linewidth]{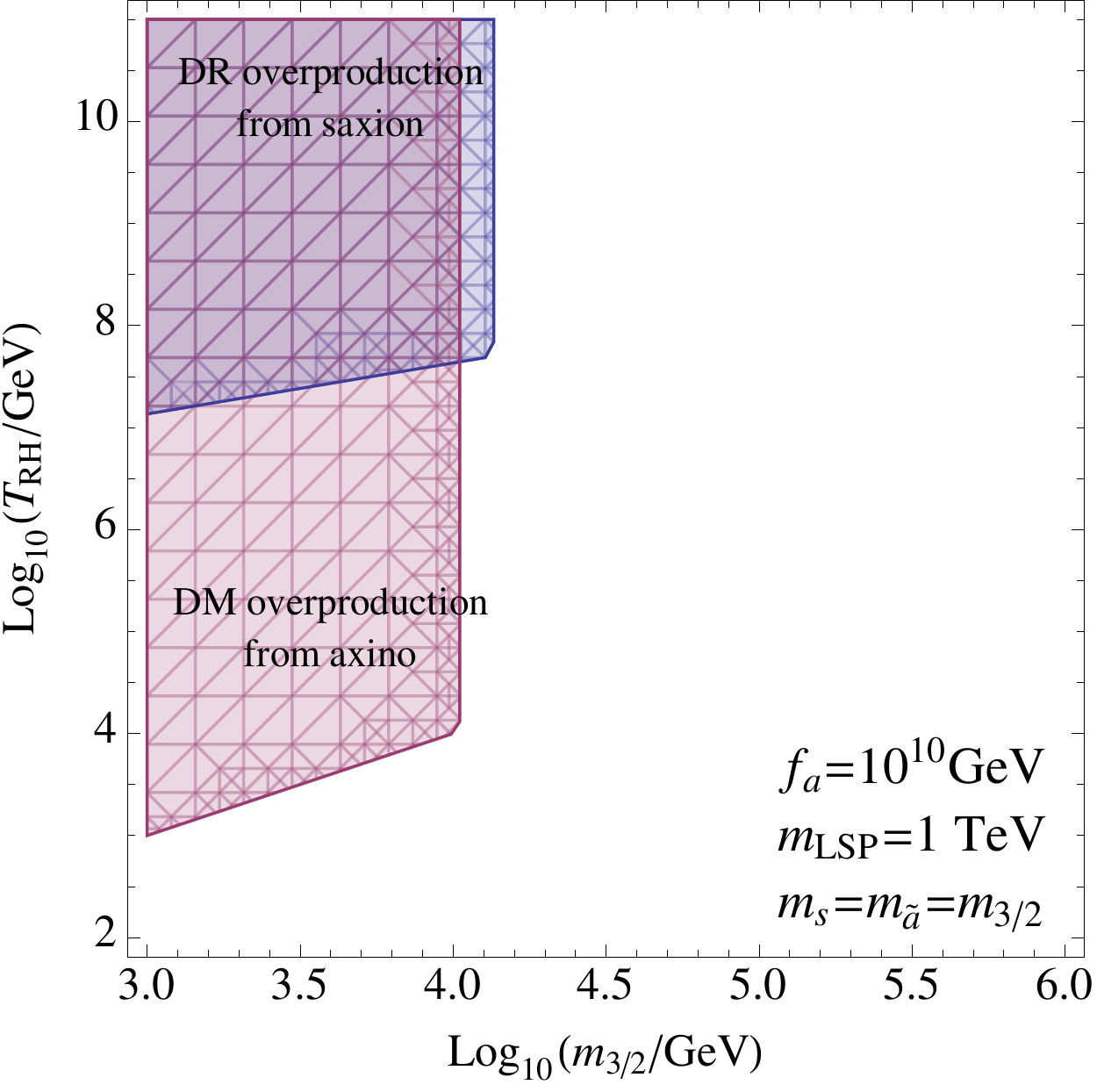}
\end{center}
\end{minipage}
 &
\begin{minipage}{0.5\hsize}
\begin{center}
  \includegraphics[width=0.9\linewidth]{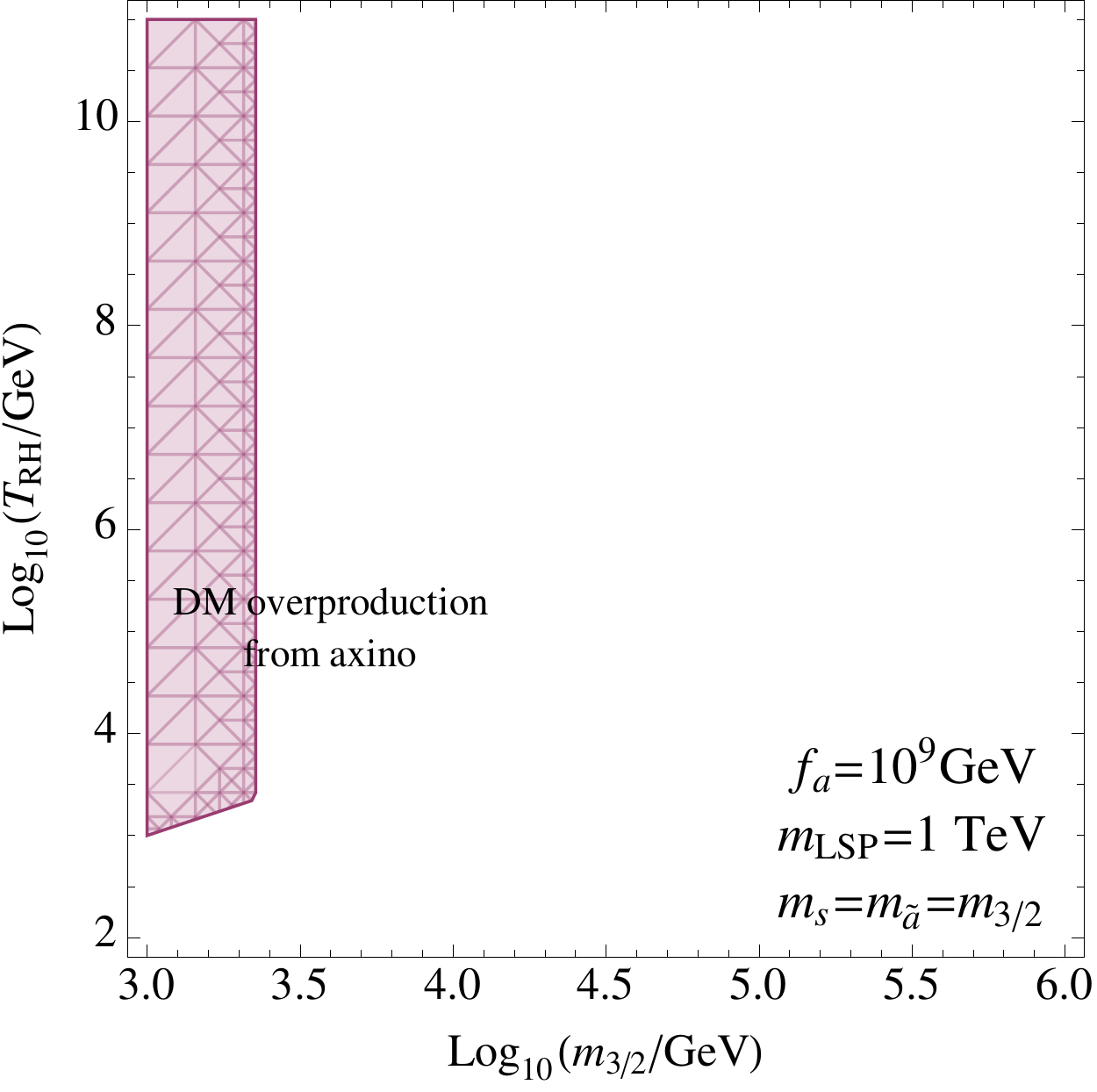}
\end{center}
\end{minipage}
\end{tabular}
\caption{\sl \small
The upper bound on the reheating temperature for a given gravitino mass and the axion decay constant.
In the blue-shaded region, the saxion produces too much dark radiation.
In the red-shaded region, the axino over-produces the LSP. 
}
\label{fig:axion constraint}
\end{figure}

In Fig.~\ref{fig:axion constraint}, we show the upper bound on the reheating temperature for a given gravitino mass and the axion decay constant.
Here, we assume that $m_s = m_{\tilde{a}} =m_{3/2}$ and $m_{\rm LSP} = 1$ TeV.
In the blue-shaded region, the saxion produces too much dark radiation.
In the red-shaded region, the axino over-produces the LSP.  
It can be seen that the constraint is weaker for larger gravitino masses.
This is because the saxion and the axino decay faster for larger masses.
For example, for $f_a \sim 10^{10}$ GeV and $m_{3/2} \sim 100$ TeV, the saxion/axino problem is absent.

\clearpage
\section{Anomaly mediated gaugino mass}
\label{sec:AMSB}

\begin{screen}
This section is based on Ref.~\cite{Harigaya:2014sfa};\\
{\it \underline{Keisuke Harigaya} and Masahiro Ibe, 
``Anomaly Mediated Gaugino Mass and Path-Integral Measure,''
Phys.\ Rev.\ D {\bf 90}, 043510 (2014),
Copyright (2014) by the American Physical Society.}
\end{screen}

Anomaly mediated gaugino mass~\cite{Randall:1998uk,Giudice:1998xp} is the essential ingredient of the heavy sfermion scenario.
In this section, we derive the anomaly mediated gaugino mass in the superspace formalism of supergravity with a Wilsonian effective action.
For instructive discussions on the anomaly mediation, we refer to Refs.~\cite{Bagger:1999rd,Boyda:2001nh,Dine:2007me,Jung:2009dg,Sanford:2010hc,DEramo:2012qd,DEramo:2013mya,Dine:2013nka,DiPietro:2014moa}.

\subsection{Approximate super-Weyl symmetry in classical action}
\label{sec:classical}
Before discussing the anomaly mediated gaugino mass, let us first clarify the
gaugino mass expected in the local supergravity action at the classical level.%
\footnote{Here, we assume that the classical action consists of local interactions.
If the classical action is allowed to be non-local, an arbitrary gaugino mass of $O(m_{3/2})$ 
can be introduced by using the non-local term in Eq.~(\ref{eq:non-local}) without 
conflicting with the super-diffeomorphism invariance.
}
In our discussion, we concentrate ourselves on a situation where 
SUSY is dominantly broken by some charged fields under some symmetries or by some composite fields.
Otherwise direct interactions between the SUSY breaking fields and gauge multiplets 
lead to the ``tree-level" gaugino mass of the order of the gravitino mass, $m_{3/2}$.
Under this assumption, direct interactions between the SUSY 
breaking fields and the gauge supermultiplets are suppressed at least by a second power of 
the Planck scale, and hence resultant gaugino masses 
from those interactions are negligible.
By the same reason, we also assume that no SUSY breaking field obtains a VEV of the order of the Planck scale.%
\footnote{These assumptions also eliminate contributions to 
gaugino masses from the Kahler and sigma-model anomalies (see Refs.~\cite{Bagger:1999rd,DEramo:2012qd} and Appendix \ref{sec:AMSB singlet}).
}

Once we assume that gaugino masses from couplings to the SUSY breaking 
sector are highly suppressed, remaining sources of gaugino masses are couplings
to the supergravity multiplets.
As is well known, however, gaugino masses from tree level interactions 
to the supergravity multiplets are also suppressed in spite of the apparent $F$-term VEVs of $O(m_{3/2})$ 
in the supergravity multiplets. 
As we shortly discuss, the absence of $O(m_{3/2})$ gaugino masses from the supergravity multiplets is due to an approximate super-Weyl symmetry,
which is the key to understand the origin of the anomaly mediated gaugino mass in the next subsection.
For the time being, we restrict ourselves to the gaugino mass generation in a $U(1)$ gauge theory 
with a pair of vector-like matter fields.

\subsubsection{Classical supergravity action}
In this thesis, we follow the notation and the formulation in Ref.~\cite{Wess:1992cp} (see Appendix~\ref{sec:SUGRA review}), 
except for the notation of complex conjugate (we use $\dagger$) and for
the normalization of gauge supermultiplets that we adopt in Ref.~\cite{ArkaniHamed:1997mj}.
For a simple model with charged chiral multiplets $Q$ and $\bar{Q}$, and a $U(1)$ gauge multiplet $V$, 
the classical supergravity action is given by,
\begin{eqnarray}
\label{eq:classical action}
{\cal L} =  \Mpl^2 \int {\rm d}^2 \Theta\, 2 {\cal E} \, \frac{3}{8} \left({\cal D}^{\dag2} - 8R\right){\rm exp}\left[-\frac{K}
{3\Mpl^2}\right]
+ 
\frac{1}{16g^2}\int {\rm d}^2 \Theta\, 2{\cal E} \,W^\alpha W_\alpha + {\rm h.c.},\nonumber \\
K
= Q^\dag e^{2V} Q + \bar{Q}^\dag e^{-2V} \bar{Q} +\cdots,~~
W_\alpha \equiv -\frac{1}{4} \left({\cal D}^{\dag 2}- 8R\right) \left(
e^{-2V} {\cal D}_\alpha e^{2V}
\right),
\end{eqnarray}
where $\Theta^\alpha$, ${\cal E}$, ${\cal D}_{\alpha}$, $R$, $K$, and $g$ are the chiral fermionic coordinate, the chiral density, the superspace covariant derivative, the superspace curvature, the Kahler potential, and the gauge coupling constant, respectively.
Here, we have assumed that the chiral multiplets $Q$ and $\bar{Q}$ are massless.
By expanding with respect to the chiral multiplets, we can extract relevant  interactions,
\begin{eqnarray}
\label{eq:matter kin}
{\cal L}_{\rm kin,matter} &=&  -\frac{1}{8} \int {\rm d}^2 \Theta\,2 {\cal E} \left({\cal D}^{\dag2} - 8R\right) \left( Q^\dag e^{2V} Q + \bar{Q}^\dag e^{-2V} \bar{Q} \right) + {\rm h.c} \ ,\\
\label{eq:gauge kin}
{\cal L}_{\rm kin,gauge} &=& \frac{1}{16g^2}\int {\rm d}^2 \Theta \,2{\cal E}\, W^\alpha W_\alpha + {\rm h.c.}\ ,
\end{eqnarray}
from which we can extract gauge interactions and kinetic terms.
Other interactions are suppressed by the Planck scale.

Now, let us expand $W^\alpha$, ${\cal E}$, and $R$ in terms of component fields;
\begin{eqnarray}
\label{eq:expansion}
W^\alpha &=& - 2 i \lambda^\alpha + \cdots \nonumber\ , \\
2 {\cal E} & = & e( 1-M^* \Theta^2) + \cdots\nonumber\ ,\\
R & = & -\frac{1}{6} M - \frac{1}{9} |M|^2\Theta^2 + \cdots .
\end{eqnarray}
Here, $\lambda^\alpha$, $e$, and $M$ are the gaugino, the determinant of the vielbein, 
and the auxiliary scalar component of the supergravity multiplet, respectively. 
Ellipses denote terms which are irrelevant for our discussion on the gaugino mass.
The auxiliary field $M$ is fixed by the equation of motion as
\begin{eqnarray}
 M^*  = - 3 m_{3/2}\ ,
 \label{eq:auxM}
\end{eqnarray}
where we have omitted contributions from $F$ terms of SUSY breaking fields which
are negligible under the assumption we have made at the beginning of this section.

Since the chiral density ${\cal E}$ has a non-vanishing $\Theta^2$ term,
it might look non-trivial why a gaugino mass of $O(m_{3/2})$ does not appear
from the interaction in Eq.~(\ref{eq:gauge kin}).%
\footnote{Technically speaking, the contribution from the $\Theta^2$ term cancels with the contribution from $M$ in $W_\alpha$; $W_\alpha$ is given by covariant derivatives of $V$ and hence the supergravity multiplet is included in $W_\alpha$.}
In the rest of this section, we show that the absence of the gaugino mass in the classical action is understood by an approximate super-Weyl symmetry.

\subsubsection{Approximate super-Weyl symmetry}
Let us consider the super-Weyl transformation parameterized by a chiral scalar $\Sigma$ (see Ref.~\cite{Wess:1992cp} and Appendix \ref{sec:SW}),%
\footnote{In this thesis we define an infinitesimal transformation of a superfield $\cal X$ by
${\cal X}' ={\cal X} -\delta {\cal X} $.}
\begin{eqnarray}
\label{eq:SW}
\delta_{\rm SW} {\cal E} &=& 6 \Sigma {\cal E} + \frac{\partial}{\partial \Theta^\alpha}\left(S^\alpha {\cal E}\right)\ , \nonumber\\
\delta_{\rm SW} R &=&  -4 \Sigma R  - \frac{1}{4}\left( {\cal D}^{\dag 2} - 8 R\right) \Sigma^\dag - S^\alpha \frac{\partial}{\partial \Theta^\alpha}R\ , \nonumber\\
\delta_{\rm SW} W^\alpha &=& - 3 \Sigma W^\alpha + \cdots\ ,\nonumber\\
\delta_{\rm SW} Q &=& w\Sigma Q - S^\alpha \frac{\partial}{\partial \Theta^\alpha}Q\ , \nonumber\\
S^\alpha &\equiv& \Theta^\alpha \left( 2 \Sigma^\dag - \Sigma \right)| +\Theta^2 {\cal D}^\alpha \Sigma|\ ,
\end{eqnarray}
where ellipses denote terms which are irrelevant for our discussion.
${\cal X}|$ denotes the lowest component of a superfield $\cal{X}$.

A parameter $w$ is the Weyl weight of $Q$.%
\footnote{If ${Q}$ is not a chiral scalar but a chiral density with a density weight $\tilde w$, 
the super-Weyl transformation is given by,
\begin{eqnarray}
\delta_{\rm SW} {Q} &=& w\Sigma {Q} - S^\alpha \frac{\partial}{\partial \Theta^\alpha}{Q}
+ \tilde{w} {Q}\frac{\partial}{\partial \Theta^\alpha}S^\alpha \ .
\end{eqnarray}
}
From Eqs.\,(\ref{eq:expansion}) and (\ref{eq:SW}),  the transformation laws of $e$, $M$ and $\lambda^\alpha$ are 
given by
\begin{eqnarray}
\label{eq:Mlambda}
\delta_{\rm SW} e &=& 4 \left(\Sigma + \Sigma^\dag\right)| e, \nonumber \\
\delta_{\rm SW} M &=& -2 (2 \Sigma - \Sigma^\dag)|M + \frac{3}{2} {\cal D}^{\dag 2}\Sigma^\dag|, \nonumber\\
\delta_{\rm SW} \lambda^\alpha &=& - 3 \Sigma| \lambda^\alpha.
\end{eqnarray}

From the transformation laws of the component fields in Eq.\,(\ref{eq:Mlambda}), it is clear that the possible
origin of a gaugino mass of $O(m_{3/2})$, 
\begin{eqnarray}
\int {\rm d}^4 x\, e M^{(*)} \lambda \lambda, 
\label{eq:MLL}
\end{eqnarray}
is not invariant under the super-Weyl transformation.
This shows that the gaugino mass is generated only through terms which break the super-Weyl symmetry.

As we immediately see, the kinetic term of the gauge multiplet in Eq.\,(\ref{eq:gauge kin}) 
is invariant under the super-Weyl transformation, and hence, does not contribute to the gaugino mass.
Higher dimensional terms omitted in Eq.\,(\ref{eq:classical action}) are, on the other hand,
not invariant under the super-Weyl transformation.
Contributions from such terms to the gaugino mass are, however, at the largest of $O(m_{3/2}^2/\Mpl)$, 
and hence are negligible.
Altogether, we find that there is no gaugino mass of $O(m_{3/2})$ originated from couplings to the supergravity multiplets
due to the approximate super-Weyl symmetry.%
\footnote{The term in Eq.\,(\ref{eq:MLL}) is invariant under the $R$-symmetry and the dilatational symmetry parts of the super-Weyl symmetry, which are parameterized by the lowest component of $\Sigma$.
Thus, the gaugino mass from the couplings to the supergravity multiplets 
cannot be forbidden by the $R$-symmetry nor the dilataional symmetry.}

For later convenience, let us also note that the terms of massless matter fields in Eq.~(\ref{eq:matter kin})
are also invariant under the super-Weyl symmetry.
That is, for $w = -2$, it can be shown that
\begin{eqnarray}
\label{eq:SW kin}
\delta_{\rm SW} \left( \left(\bar{{\cal D}}^2 - 8R\right) \left(Q^\dag  Q\right) \right) = -6 \Sigma  \left(\bar{{\cal D}}^2 - 8R\right) \left( Q^\dag  Q \right) -S^\alpha \frac{\partial}{\partial \Theta^\alpha} \left( \left(\bar{{\cal D}}^2 - 8R\right) \left(Q^\dag  Q\right) \right) \ .
\end{eqnarray}
From Eqs. (\ref{eq:SW}) and (\ref{eq:SW kin}),
terms in Eq.~(\ref{eq:matter kin}) are invariant under the super-Weyl transformation.

Finally, let us stress that interaction terms of the gauge supermultiplets which are unsuppressed 
by the Planck scale is uniquely determined to the form of Eqs.~(\ref{eq:matter kin}) and~(\ref{eq:gauge kin}) by 
the super-diffeomorphism invariance and by the  gauge invariance. 
Thus, one may regard the approximate super-Weyl symmetry as an accidental one.
Due to this accidental symmetry, the gaugino mass of $O(m_{3/2})$ is suppressed at the classical level.

\subsection{Anomaly of the super-Weyl symmetry and Gaugino Mass}
\label{sec:measure}
In the last subsection, we have shown that no gaugino mass of $O(m_{3/2})$ is generated through
couplings to the supergravity multiplets even
though the chiral density has a non-zero $F$ term,
due to the approximate super-Weyl symmetry.
However, the approximate super-Weyl symmetry is in general broken by quantum effects.
In this subsection, we investigate effects of quantum violation of the approximate super-Weyl symmetry 
by Fujikawa's method\,\cite{Fujikawa:1979ay} in a Wilsonian effective action.

\subsubsection{Wilsonian effective action}
To discuss quantum effects on the super-Weyl symmetry, 
we take the local classical action in the previous section 
(Eq.\,(\ref{eq:classical action})) as a Wilsonian effective action with a cutoff at the Planck scale. 
Here, let us remind ourselves that effective quantum field theories suffer 
from ultraviolet divergences, and hence, they are well-defined only after the divergences are properly regularized.
In our arguments, we presume an ultraviolet regularization such that the ``tree-level" action
at the cutoff scale is manifestly invariant under the super-diffeomorphism and the gauge transformations.
We refer to this super-diffeomorphism invariant tree-level action at the cutoff scale as the Wilsonian effective action.%
\footnote{Although we fix the cutoff scale to the Planck scale for a while,
the following discussion is essentially unchanged as long as the cutoff scale 
is far larger than the gravitino mass. 
We also discuss effects of the change of the cutoff scale later.}

The Wilsonian effective action in general includes higher dimensional interactions than those 
in Eq.\,(\ref{eq:classical action}) suppressed by the cut off scale.
As we have discussed, however, contributions from those terms to the gaugino mass are highly suppressed
by the cutoff scale and hence negligible. 
One concern is whether non-local interaction terms appear in the Wilsonian effective action at the cutoff scale, 
which could lead to the gaugino mass of $O(m_{3/2})$.
In our argument, we presume that such non-local interactions do not show up in the Wilsonian effective action,
which is reasonable because we are dealing with effective field theories after integrating out only ultraviolet modes.

\subsubsection{Super-diffeomorphism invariance}
In the above definition of the super-diffeomorphism invariant theory,
there is a missing ingredient, the measure of the path-integral. 
As elucidated in Ref.~\cite{Fujikawa:1979ay}, the path-integral measure plays a crucial role in treating quantum
violations of symmetries.
Moreover, the definition of the ``tree-level" interactions in the Wilsonian effective action depends on the choice 
of the path-integral measure, which we will encounter shortly.
To clarify these issues, let us first discuss which path-integral measure we should use in conjunction with the ``tree-level" 
Wilsonian action.

Under the infinitesimal (chiral) super-diffeomorphism transformation,%
\footnote{
In the formulation given in Ref.~\cite{Wess:1992cp},
gauge symmetries are fixed except for the diffeomorphism invariance, the local Lorentz symmetry and the supergravity symmetry. Thus, $\eta^M$ is restricted so that it parameterizes only the diffeomorphism and the supergravity transformation.
We retain the word ``super-diffeomorphism" to simplify our expression.
The word ``super-diffeomorphism" used in Appendix~\ref{sec:SUGRA review}, on the other hand, refers to the full super-diffeomorphism.
}
$Q$ and ${\cal E}$ transform as 
\begin{eqnarray}
\label{eq:Sdiff}
Q \rightarrow Q'  &=& Q - \eta^M(x,\Theta)\partial _M Q\ ,\nonumber \\
{\cal E} \rightarrow {\cal E}' &=& {\cal E}- \eta^M(x,\Theta)\partial _M {\cal E} - (-)^M \left(\partial_M \eta^M\left(x, \Theta\right)\right){\cal E}\ ,
\end{eqnarray}
where $M = (m,\alpha)$ denotes the indices of the chiral super coordinate $(x^m,\Theta^\alpha)$,
$\eta^M (x, \Theta)$ parameterizes the super-diffeomorphism, and
$(-)^M = (1, -1)$ for $M = (m,\alpha)$. 
As is shown in Appendix~\ref{sec:Sdiff measure}, naive path-integral measures of chiral fields
are not invariant under the super-diffeomorphism due to the anomaly of the gauge interactions,
i.e.
\begin{eqnarray}
 [D Q]  \to [D Q'] \neq  [D Q] \ , \quad
  [D \bar {Q}]  \to [D \bar{Q}'] \neq  [D \bar{Q}] \ .
 \label{eq:chiral measure}
\end{eqnarray}
Instead, anomaly free measures are given by
\begin{eqnarray}
\label{eq:SD measure}
[D\left(2{\cal E}\right)^{1/2}Q]
\ , \quad [D\left(2{\cal E}\right)^{1/2} \bar {Q}]\ .
\end{eqnarray}
For a later purpose, we define weighted chiral fields $Q_{\rm diff} = \left(2{\cal E}\right)^{1/2}Q$ ($\bar{Q}_{\rm diff} = \left(2{\cal E}\right)^{1/2}\bar{Q}$) which are no more chiral scalar fields but chiral density fields 
with  density weights $1/2$.

In our discussion, we take the super-diffeomorphism invariant Wilsonian 
effective action. 
Therefore, in order to obtain a super-diffeomporphism invariant quantum theory,
we inevitably use the super-diffeomorphism invariant path-integral measure in Eq.\,(\ref{eq:SD measure}).
If we use different measures instead, we need to add appropriate super-diffeomorphism variant counter terms 
to the tree-level Wilsonian action so that the super-diffeomorphism is restored in the quantum theory.

\subsubsection{Anomaly of the super-Weyl symmetry}
Once we choose appropriate path-integral measures for the charged fields, 
we can now discuss quantum violation of the super-Weyl symmetry.
Here, since we are interested in the gaugino mass, we only look at the breaking of the super-Weyl
symmetry by the anomaly of the corresponding gauge interaction.

Before proceeding further, let us comment on a technical point.
As   in Eq.\,(\ref{eq:SW}), the super-Weyl transformation 
is accompanied by a super-diffeomorphism parameterised by $S^\alpha$, so that 
the super-Weyl transformation is expressed in terms of the component fields
defined in the chiral superspace spanned by $(x,\Theta)$.
The accompanied super-diffeomorphism, however, makes it complicated to
discuss the quantum violation of the super-Weyl symmetry.
To avoid such a complication, we only consider a subset of the super-Weyl transformation
where $\Sigma$ has only an $F$-term, i.e.
\begin{eqnarray}
\Sigma(x,\Theta) = f(x)\Theta^2\ .
\end{eqnarray}
Here, $f$ is an arbitrary function of the space-time.
Under this restricted super-Weyl transformation, we find $S^\alpha = 0$, and hence,
no super-diffeomorphism is accompanied.
We refer this type of the super-Weyl transformation  as an ``$F$-type" super-Weyl transformation.
It should be noted that the $F$-type super-Weyl transformation is
sufficient to forbid the gaugino mass
in the discussion of Sec.\,\ref{sec:classical}.
In the followings, we concentrate on the anomalous breaking of the $F$-type super-Weyl symmetry.

Now let us examine the invariance of the path-integral measures in Eq.\,(\ref{eq:SD measure})
under the $F$-type super-Weyl transformation.
Under the transformation, $Q_{\rm diff}$ and $\bar{Q}_{\rm diff}$ are not invariant but transform by
\begin{eqnarray}
Q_{\rm diff} = \left(2{\cal E}\right)^{1/2}Q \to Q_{\rm diff}' = e^{-\Sigma} Q_{\rm diff}\ , 
\quad
\bar{Q}_{\rm diff} = \left(2{\cal E}\right)^{1/2}\bar{Q} \to \bar{Q}_{\rm diff}' = e^{-\Sigma} \bar{Q}_{\rm diff}\ . 
\end{eqnarray}
Here, we have used the fact that the super-Weyl weight of the massless chiral fields are $-2$
so that the kinetic term of the chiral fields in Eq.\,(\ref{eq:matter kin}) is invariant under the super-Weyl symmetry.
Thus, due to the Konishi-Shizuya anomaly~\cite{Konishi:1985tu}, we find that the super-diffeomorphism 
invariant measure is not invariant under the $F$-type super-Weyl transformation.
Instead, the $F$-type super-Weyl invariant measures are given by
\begin{eqnarray}
\label{eq:SW measure}
[D Q_{\rm SW}]& \equiv& [D  \left(2 {\cal E}\right)^{1/3} Q ] = [D  \left(2 {\cal E}\right)^{-1/6} Q_{\rm diff} ]\ ,\\
\left[D \bar{Q}_{\rm SW}\right]&\equiv & [D  \left(2 {\cal E}\right)^{1/3} \bar{Q} ] = [D  \left(2 {\cal E}\right)^{-1/6} \bar{Q}_{\rm diff} ]\ ,
\end{eqnarray}
where $Q_{\rm SW}$ and $\bar{Q}_{\rm SW}$ are invariant under the the $F$-type super-Weyl transformation.
Here, the weighted chiral superfields  $Q_{\rm SW}$ and $\bar{Q}_{\rm SW}$ have  density weights $1/3$.

It should be commented that the component fields of $Q_{\rm SW}$ ($\bar{Q}_{\rm SW}$)  defined by
\begin{eqnarray}
Q_{\rm SW} &=& e^{1/3}[A_{Q_{\rm SW}} + \sqrt{2} \Theta \chi_{Q_{\rm SW}} + \Theta^2 F_{Q_{\rm SW}}]\ ,
\label{eq:SW component}
\end{eqnarray}
have canonical kinetic terms,
in a sense that kinetic terms does not contain the supergravity multiplets
in the flat limit:
For a generic chiral scalar superfield, $X = A + \sqrt{2} \Theta \chi + \Theta^2 F$, 
the chiral projection of its complex conjugate is given by
\begin{eqnarray}
\label{eq:chiral projection}
\left({\cal D}^{\dag2}-8R\right)X^\dag & = & - 4 F^* + \frac{4}{3} M A^* +
\Theta^\alpha \left[
-4 i \sqrt{2} \sigma^m \partial_m \chi^\dag
\right]\nonumber\\
&&
+\Theta^2\left[
-4 \partial^2 A^*
-\frac{8}{3} M^* F^* +\frac{8}{9} A^* |M|^2 
\right] + \cdots\ ,
\end{eqnarray}
where the ellipses denote higher dimensional terms.
Then, by remembering that the component fields of $Q_{\rm SW}$ are related to those of $Q$ via
\begin{eqnarray}
Q = \left( 1 + \frac{1}{3}M^* \Theta^2\right)
\left(A_{Q_{\rm SW}} + \sqrt{2} \Theta \chi_{Q_{\rm SW}} + \Theta^2 F_{Q_{\rm SW}}\right) + \cdots\ ,
\label{eq:SW component2}
\end{eqnarray}
we find that the kinetic terms of the component fields of $Q_{SW}$ are canonical and decouple from $M$.%
\footnote{In terms of the component fields of $Q$, $M$ does not decouple 
from the kinetic term and mixes with the scalar fields via, $M^* F_Q^* A_Q$
as well as $|A_Q|^2|M|^2$ terms.}
Therefore, it is appropriate to identify the component fields of $Q_{SW}$ as the component fields 
of the corresponding chiral field in the rigid SUSY,%
\footnote{Here, we have neglected higher dimensional terms.
If we take them into account, we need to perform a Kahler-Weyl transformation to achieve 
the canonical normalisation in the Einstein frame.
}
\begin{eqnarray}
Q_{\rm rigid\,\,SUSY}  = A_{Q_{\rm SW}} + \sqrt{2} \theta \chi_{Q_{\rm SW}} + \theta^2 F_{Q_{\rm SW}}\ ,
\end{eqnarray}
with $\theta$ being the fermionic coordinate of the rigid superspace.

\subsubsection{Gaugino mass in the Wilsonian effective action}
\label{sec:mass}
As we have discussed in the previous section, 
the gaugino mass vanishes if the $F$-type super-Weyl symmetry is preserved,
and it is generated only through violations of the $F$-type super-Weyl symmetry.
As relevant terms of the gauge supermultiplet
preserve the super-Weyl symmetry, the gaugino mass appearing in the super-diffeomorphism 
invariant ``tree-level" Wilsonian action is highly suppressed.

The approximate $F$-type super-Weyl symmetry is, however, 
anomalously broken by the super-diffeomorphism invariant measure $[D Q_{\rm diff}]$.
To read off the gaugino mass from this violation, it is transparent to change 
the path-integral measure to the $F$-type super-Weyl invariant measure, $[D Q_{\rm SW}]$,
so that the super-Weyl variance is apparent in the corrected ``tree-level" Wilsonian action. 
In fact, the change of the measures from $[DQ_{\rm diff}]$ to $[DQ_{SW}]$
is accompanied by the Konishi-Shizuya anomaly~\cite{Konishi:1985tu},%
\footnote{The identity in Eq.~(\ref{eq:translation}) is not quite correct.
In general, ${\mit \Delta} S$ involves higher dimensional terms suppressed by the cut off of the Wilsonian effective action.
However, such higher-dimensional terms are negligible.}
\begin{eqnarray}
\label{eq:translation}
[D Q_{\rm diff}] [D\bar{Q}_{\rm diff}] [D Q^\dag_{\rm diff}] [D\bar{Q}^\dag_{\rm diff}] &=&
[D Q_{\rm SW}] [D\bar{Q}_{\rm SW}] [D Q^\dag_{\rm SW}] [D\bar{Q}^\dag_{\rm SW}]
\times {\rm exp}
\left[
i {\mit\Delta} S
\right],\nonumber\\
{\mit\Delta} S &=& \frac{1}{16} \frac{1}{2\pi^2}\times \int {\rm d}^4x\, {\rm d}^2\Theta\, 2{\cal E}\, {\ln}(2{\cal E})^{1/6}\, W^\alpha W_\alpha + {\rm h.c.}\ .
\end{eqnarray}
Accordingly, the ``tree-level" Wilsonian effective action which should be taken in conjunction with  $[D Q_{\rm SW}]$
is given by,
\begin{eqnarray}
\label{eq:SSW}
S = S_{SD} + {\mit \Delta}S\ .
\end{eqnarray}
Here, $S_{SD}$ denotes the super-diffeomorphism invariant local Wilsonian effective action discussed above.
Without surprise, ${\mit\Delta} S$ is not invariant under the super-diffeomorphism, 
which cancels the anomalous breaking of the super-diffeomorphism invariance 
by $[D Q_{\rm SW}]$.
We summarize properties of the measures in Table.~\ref{tab:measures}.%
\footnote{Throughout this thesis, we presume the regularization scheme of the path-integral measure
which reproduce the Konishi-Shizuya anomaly in the form in Eq.\,(\ref{eq:translation}).
In the dimensional regularization/reduction, on the other hand, 
the change of the path-integral measures is not accompanied by the rescaling anomaly,
while the approximate super-Weyl symmetry is explicitly broken by the relevant interactions 
which eventually leads to a consistent gaugino mass\,\cite{Boyda:2001nh}.
}
\begin{table}[tb]
\begin{center}
\begin{tabular}{c||c|c|c|}
    & measure & action & gaugino mass\\ \hline
\,$[D Q_{\rm diff}]$ \,&\, $SD$,\,  $\cancel{SW}$ \, & \,$SD$,\, $SW$ \,& \, hidden in the measure\, \\
\,$[D Q_{\rm SW}]$ \,& \,$\cancel{SD}$,\, $SW$ \, & $\,\cancel{SD}$,\, $\cancel{SW}$\,  &\, apparent in the action\,
\end{tabular}
\end{center}
\caption{Properties of two path-integral measures.
Here, $SD$ and $SW$ denote the super-diffeomorphism and the  
$F$-type super-Weyl invariances, respectively.
Cancel lines denote non-invariances.
}
\label{tab:measures}
\end{table}%

Armed with a correct ``tree-level" Wilsonian action along with the super-Weyl invariant measure, 
we can now read off the gaugino mass directly from the local term in the action, ${\mit \Delta}S$,
which leads to
\begin{eqnarray}
\label{eq:gaugino mass}
M_\lambda/g^2= + \frac{1}{2} \frac{1}{2\pi^2} {\ln}(2{\cal E})^{1/6}|_{\Theta^2} =  -\frac{1}{24\pi^2} M^* = +\frac{1}{16\pi^2}\times 2m_{3/2}\ ,
\end{eqnarray}
where ${\cal X}|_{\Theta^2}$ denotes the $\Theta^2$ component of a superfield ${\cal X}$.
This gaugino mass reproduces the anomaly mediated gaugino mass given in Refs.~\cite{Randall:1998uk,Giudice:1998xp}.
In this way, we find that the anomaly mediated gaugino mass can be read off from 
the super-diffeomorphism non-invariant term ${\mit\Delta} S$ in the superspace formalism of supergravity.%
\footnote{In this thesis, we concentrate on the anomaly mediated gaugino mass at one-loop level.}

\subsubsection{Radiative corrections from path-integration}
So far, we have fixed the Wilsonian scale to $\Mpl$ and have not performed any path-integration. 
Here, let us discuss effects of the path-integration.
After integrating out modes above a scale $\Lambda (<\Mpl)$,
the Wilsonian effective action at $\Lambda$ is again given by the form of Eq.\,(\ref{eq:SSW}),
with renormalized coefficients and higher dimensional operators suppressed not only by $\Mpl$ but also by $\Lambda$.
Due to the presence of cutoff scales, the super-Weyl symmetry in the Wilsonian action at the scale $\Lambda$
is hardly preserved.
As we have discussed, however, the relevant terms of the matter and the gauge supermultiplets 
have an approximate super-Weyl symmetry accidentally due to the super-diffeomorphism invariance.
Therefore, radiative corrections
do not generate the gaugino mass term 
beyond the one in Eq.~(\ref{eq:gaugino mass}) up to $\Lambda$ or $\Mpl$ suppressed corrections.

It should be also noted that, among various corrections, the ones 
from diagrams which involve  Planck suppressed interactions
lead to higher dimensional operators suppressed at least by a single power of $\Mpl$ in the effective action 
at $\Lambda$.%
\footnote{
If there are ultraviolet divergences which are cancelled only by non-local terms, 
$\Mpl$ suppressed interactions could lead to higher dimensional operators suppressed not 
by $\Mpl$ but only by $\Lambda$ at the cutoff scale $\Lambda$.
The Bogoliubov-Parasiuk-Hepp-Zimmermann prescription~\cite{Bogoliubov:1957gp,Hepp:1966eg,Zimmermann:1969jj} shows that ultraviolet divergences in general can be renormalized away by local terms.
}
Effects to lower dimensional operators through ultra-violet divergences are renormalized 
by the shifts of the corresponding operators~\cite{Polchinski:1983gv}.
Visible effects of higher dimensional operators only show up 
through higher dimensional operators even in the effective action at $\Lambda$.

Concretely, radiative corrections from loop diagrams involving gravity supermultiplets (in particular gravitinos with small momenta to flip the chirality)
may lead to higher dimensional operators such as $|M|^n M^* \lambda\lambda$ ($n\ge 0$)
suppressed only by $\Mpl^2 \Lambda^{n-2}$.
Such diagrams involving the gravitinos however damp for $\Lambda \ll m_{3/2}$.
Therefore, they contribute to the gaugino mass at most of $O(m_{3/2}^3/\Mpl^2)$.

From these arguments, we see that higher dimensional operators which are suppressed by not $\Mpl$ 
but only by $\Lambda$ in the Wilsonian effective action at the cutoff scale $\Lambda$ 
are generated only from relevant interactions of the matter and gauge supermultiplets.
Such effects can be properly taken care of within the renormaizable effective theory
of the matter and the gauge supermultiplets  with softly broken SUSY.

Let us emphasize again that the super-diffeomorphism violation is not arbitrary in the Wilsonian effective action at $\Lambda$, although the super-diffeomorphism invariance is broken by $[DQ_{\rm SW}]$.
The super-diffeomorphism violation in the Wilsonian action is uniquely given  by ${\mit\Delta S}$ 
at each Wilsonian scale, so that the super-diffeomorphism is preserved in the quantum theory.
Thus, the accidental approximate super-Weyl symmetry which is the outcome of the super-diffeomorphism
invariance is justified even after performing path-integration.

Putting all together, we  find that the anomaly mediated gaugino mass can be extracted
from the super-diffeomorphism non-invariant local term in the Wilsonian effective action
at the scale $\Lambda \gg m_{3/2}$ in the superspace formalism of the supergravity.
Radiative corrections to the gaugino mass operator are dominantly given by relevant interactions 
of the matter and the gauge supermultiplets.
Therefore, once we extract a gaugino mass at some high cutoff scale, we can use the gaugino mass
as the boundary condition of the renormalization group equation at $\Lambda$ 
in the low-energy effective renormalizable supersymmetric theory with soft SUSY breaking.

\subsubsection{Decoupling effects of massive matter}
Before closing this section, let us consider the contribution to the gaugino mass 
from charged matter multiplets with a supersymmetric mass $m$ far larger than $m_{3/2}$,
\begin{eqnarray}
{\cal L}_{\rm mass } = \int {\rm d}^2\Theta\, 2{\cal E} \,m Q \bar{Q} + {\rm h.c.} \ .
\end{eqnarray}
If the cutoff scale of the Wilsonian effective action is far above $m$, the mass $m$ is negligible 
in comparison with the kinetic term and hence the above discussion holds. 
When the cutoff scale is below $m$, the mass term dominates over the kinetic term.
In that situation, the approximate super-Weyl symmetry is such that the mass term is invariant.%
\footnote{In the Pauli-Villars regularization, the anomaly mediated gaugino mass is understood by the difference of super-Weyl invariant measures between massive Pauli-Villars fields and massless matter fields (see Sec.~\ref{sec:Pauli-Villars}).}
This observation leads to the Weyl weights of $-3$ for $Q$ and $\bar{Q}$, i.e. 
$\delta _{\rm SW,massive}Q = -3 \Sigma Q + \cdots$, and hence,
the super-Weyl invariant measures of the massive matter are given by
\begin{eqnarray}
\label{eq:SW mass measure}
[D Q_{\rm SW,massive}] \equiv [D  \left(2 {\cal E}\right)^{1/2} Q ]\ , 
\quad
[D \bar{Q}_{\rm SW,massive}] \equiv [D  \left(2 {\cal E}\right)^{1/2} \bar{Q} ]\ ,
\end{eqnarray}
which coincide with the super-diffeomorphism invariant measures in Eq.\,(\ref{eq:SD measure}).
Thus, below the scale $m$, the approximate super-Weyl symmetry is well described by 
the super-diffeomorphism invariant Wilsonian effective action, i.e. ${\mit \Delta} S = 0$, 
and hence, no anomaly mediated gaugino mass term appears up to $O(m_{3/2}^2/m)$ contributions.
This argument reconfirms the insensitivity of the anomaly mediated gaugino mass to ultraviolet physics~\cite{Giudice:1998xp}.

If $m$ is close to $m_{3/2}$, the decoupling does not hold in general.
The Wilsonian effective action below the mass threshold of $Q$ and $\bar{Q}$ includes terms suppressed only by $m$, which might contribute to the gaugino mass as large as $O(m_{3/2}^2/m)$.
Integration of $Q$ and $\bar{Q}$ should be performed explicitly, as is the case with the higgsino threshold correction in the MSSM~\cite{Giudice:1998xp}.

\subsubsection{Anomaly mediation a la Pauli-Villars regulalization}
\label{sec:Pauli-Villars}
Here, we show how our method to extract the gaugino mass works in the Pauli-Villars regularization~\cite{Pauli:1949zm}.
In the Pauli-Villars regularization scheme, 
we introduce Pauli-Villars fields, a pair-of fermonic chiral scalar multiplets $P$ and $\bar{P}$ with a unit charge, 
and give them a supersymmetric mass term $\Lambda$ which corresponds to the cutoff scale;
\begin{eqnarray}
{\cal L} = \int {\rm d}^2\Theta\, 2{\cal E}\, \Lambda P \bar{P} + {\rm h.c.}\ .
\end{eqnarray}
As we discussed in Sec.\,\ref{sec:measure}, it is convenient to use the  $F$-type super-Weyl invariant measure, 
$[DQ_{SW}]$, to extract the gaugino mass from the Wilsonian action.
If we also take the measure of the Pauli-Villars fields to be $[DP_{SW}]$, however,
the counter terms associated with the change the measures are cancelled due to the 
opposite statistic of the  Pauli-Villars fields.
Thus, in this case, the $F$-type super-Weyl invariant measure does not invoke the counter term in Eq.~(\ref{eq:translation}), ${\mit\Delta} S$.

In the absence of ${\mit\Delta} S$, what is the origin of the gaugino mass?
As we have shown, the gaugino mass is generated only from violations of the approximate $F$-type super-Weyl symmetry.
For a energy scale well below $\Lambda$, the approximate $F$-type super-Weyl symmetry is 
explicitly broken by the mass term of the Pauli-Villars fields.
Thus, the integration of the Pauli-Villars fields generates the gaugino mass, as is discussed in Ref.~\cite{Giudice:1998xp}.%
\footnote{More explicitly, the masses of the fermions and the scalars in the Paulli-Villars multiplets
are split by the coupling to $M$ through $\int {\rm d}^2\Theta (2 {\cal E})^{1/3} \Lambda P_{SW} \bar{P}_{SW}$.
}

We can also extract the gaugino mass without explicitly performing the integration of the Pauli-Villars fields.
Well below the mass scale $\Lambda$, a good approximate super-Weyl symmetry is 
the one which is consistent with the mass term of the Pauli-Villars fields.
Thus, the appropriate measures to read off the gaugino mass from the action is 
the combination of $[DQ_{SW}]$ and $[DP_{\rm diff}]$.
With these measures, the counter term is again given by ${\mit\Delta} S$ in Eq.~(\ref{eq:translation}),
from which we can directly read off the anomaly mediated gaugino mass.

\subsection{Fictitious super-Weyl gauge symmetric formulation}
\label{sec:fictitious SW}
In Refs.~\cite{deAlwis:2008aq,DEramo:2013mya}, 
the origin of the gaugino mass has been discussed in the superspace formalism of supergravity
with the help of a fictitious (and exact) super-Weyl gauge symmetry 
by introducing a chiral super-Weyl compensator field, $C$, in the track of Ref.~\cite{Kaplunovsky:1994fg}. 
We call this super-Weyl symmetry as the fictitious super-Weyl gauge symmetry throughout this thesis
in order to distinguish it from the approximate super-Weyl symmetry we have discussed so far.
One of the key to settle the puzzle in the discussion of Refs.~\cite{deAlwis:2008aq,DEramo:2013mya},
which we explain later,
is how to write down the anomaly mediated gaugino mass term
in a gauge independent way of the fictitious super-Weyl gauge symmetry.
In this section, we show how to write down the gauge independent gaugino mass term,
where the knowledge on the super-diffeomorphism invariant path-integral measure plays a crucial role.

\subsubsection{Fictitious super-Weyl gauge symmetry}
The fictitious (and exact) super-Weyl gauge symmetry is introduced to the action in Eq.~(\ref{eq:classical action}) 
by performing a finite super-Weyl transformation in Eq.~(\ref{eq:SW}) with $\Sigma = {\ln}\,C/2$ and $w=0$~\cite{Kaplunovsky:1994fg}.
The resulting classical acton is given by
\begin{eqnarray}
\label{eq:SW classical action}
{\cal L} &=&  \Mpl^2 \int {\rm d}^2 \Theta\,2 {\cal E}'  \frac{3}{8} \left({\cal D}'^{\dag2} - 8R'\right)C C^\dag{\exp}\left[-\frac{K'}
{3\Mpl^2}\right] \nonumber\\
&& + 
\frac{1}{16g^2}\int {\rm d}^2 \Theta\, 2{\cal E}'\, W'^\alpha W'_\alpha + {\rm h.c.}\ ,
\end{eqnarray}
where primes denote fields after the transformation.
Now, the action is exactly invariant under the super-Weyl symmetry in Eq.\,(\ref{eq:SW})
in terms of ${\cal E}'$, $W'^\alpha$, $Q'$ and $\bar{Q}'$ with $w = 0$,
while giving a Weyl weight $-2$ to the ``super-Weyl compensator" $C$, 
\begin{eqnarray}
\delta_{\rm SW,fic} C &=& -2 \Sigma C -S^\alpha \frac{\partial}{\partial \Theta^\alpha}C\,,\nonumber \\
\delta_{\rm SW,fic} {\cal E}' &=& 6 \Sigma {\cal E}' + \frac{\partial}{\partial \Theta^\alpha}\left(S^\alpha {\cal E}'\right)\ , \nonumber\\
\delta_{\rm SW,fic} R' &=&  -4 \Sigma R'  - \frac{1}{4}\left( {\cal D}^{'\dag 2} - 8 R'\right) \Sigma^\dag - S^\alpha \frac{\partial}{\partial \Theta^\alpha}R'\ , \nonumber\\
\delta_{\rm SW,fic} W^{'\alpha} &=& - 3 \Sigma W^{'\alpha} + \cdots\ ,\nonumber\\
\delta_{\rm SW,fic} Q' &=&  - S^\alpha \frac{\partial}{\partial \Theta^\alpha}Q'\ , \nonumber\\
S^\alpha &\equiv& \Theta^\alpha \left( 2 \Sigma^\dag - \Sigma \right)| +\Theta^2 {\cal D}^\alpha \Sigma|\ .
\end{eqnarray}
It should be noted that the compensator $C$ is a gauge degree of freedom, which can be completely eliminated 
by performing the fictitious super-Weyl transformation.
In other words, one may take any $C$ so that a calculation one performs is as simple as possible.%
\footnote{A singular transformation leading to $C = 0$ should be avoided.}
In particular, in the presence of the compensator, the equation of the motion of $M'$ is changed from Eq.\,(\ref{eq:auxM})
to 
\begin{eqnarray}
F^C - \frac{1}{3}M'^* = m_{3/2} \ ,
\end{eqnarray}
where we have taken $C = 1 + F^C \Theta^2$.
Thus, for example, it is convenient to take the gauge where $M' = 0$, which is adopted
in Ref.~\cite{DEramo:2013mya} up to higher dimensional terms.

\subsubsection{Gaugino mass}
The super-Weyl transformation performed to introduce $C$ is anomalous
where the measure is transformed from $[D Q_{\rm diff}]$ to $[D Q'_{\rm diff}]$.%
\footnote{The weighted chiral field $Q_{\rm diff}$ has a Weyl weight $3$ for $w = 0$.}
The transformation invokes the following term in the Wilsonian effective action,
\begin{eqnarray}
\label{eq:counter' SW}
\Delta S'_C =  + \frac{1}{16} \frac{3}{4\pi^2} \int {\rm d}^4 x\, {\rm d}^2\Theta \,2{\cal E}'\, {\ln}\,C \,W'^\alpha W'_\alpha + {\rm h.c.}\ .
\end{eqnarray}
This term can be also derived from the condition that the fictitious super-Weyl symmetry is free from 
the gauge anomaly~\cite{Kaplunovsky:1994fg}.
Further, let us eliminate $C$ from the kinetic term of the matter fields by the redefinitions, $Q'' \equiv Q'C$ and $\bar{Q}''\equiv \bar{Q}'C$.
After the redefinitions, the integration of the matter fields does not generate the gaugino mass proportional to $F^C$
at one-loop level,
so that the gaugino mass is directly read off from the Wilsonian effective action.
By combining the counter terms of the anomalies to reach to 
$Q''_{\rm diff} = (2 {\cal E'})^{1/2} Q'C$ 
and 
$\bar{Q}''_{\rm diff} = (2 {\cal E'})^{1/2} \bar{Q}'C$, we eventually obtain
\begin{eqnarray}
\label{eq:counter SW}
\Delta S_C =   \frac{1}{16} \frac{1}{4\pi^2}\times \int {\rm d}^4 x\, {\rm d}^2\Theta\, 2{\cal E}'\,{\ln}\,C \,W'^\alpha W'_\alpha + {\rm h.c.}\ ,
\end{eqnarray}
where the corresponding path-integral measures are given by $[D Q''_{\rm diff}]$ and $[D \bar{Q}''_{\rm diff}]$.

In Ref.~\cite{deAlwis:2008aq}, 
it is claimed that there is no anomaly mediated gaugino mass derived in \cite{Randall:1998uk,Giudice:1998xp} 
by taking a gauge with $F^C = 0$. 
On the other hand, in Ref.~\cite{DEramo:2013mya}, taking another gauge with $M'=0$, the anomaly mediated gaugino mass is reproduced.
These arguments pose a puzzle, for the gaugino mass  should not depend on the gauge choice of $F^C$.

This puzzle is solved by remembering the discussion in Sec.~\ref{sec:measure}.
There, in order to read off the gaugino mass from the Wilsonian effective action, we have used the canonical measure $[D Q_{\rm SW}]\equiv [D  \left(2 {\cal E}\right)^{1/3} Q ]$.
Similarly, we should again use the measure,
\begin{eqnarray}
[DQ_c] \equiv [D  \left( 2 {\cal E'}\right)^{1/3} C Q' ] = [D  \left( 2 {\cal E'}\right)^{-1/6} Q''_{\rm diff}  ]\ ,
\end{eqnarray}
which is invariant under the ``approximate" super-Weyl symmetry.
The kinetic term of $Q_c$ is free from the mixings to both $M'$ and $F^C$, and hence, canonical.
Eventually, by translating the measure from $[DQ_{\rm diff}'']$ to $[DQ_{c}]$, the Wilsonian 
effective action obtains a correction ${\mit\Delta} S$, which adds up with ${\mit\Delta} S_C$,%
\footnote{One may obtain the following counter term directly from the relation,
\begin{eqnarray}
[DQ_{c}] = [D (2{\cal E})^{-1/6} C^{-1/2} Q_{\rm diff}]\ .
\end{eqnarray}
}
\begin{eqnarray}
\label{eq:SWinvariantForm}
{\mit\Delta} S + {\mit\Delta} S_C =   \frac{1}{16} \frac{1}{4\pi^2}\times \int {\rm d}^4 x\, {\rm d}^2\Theta \,2{\cal E}' 
\left({\ln}\left(2 {\cal E}'\right)^{1/3} + {\ln}\,C \right)W'^\alpha W'_\alpha + {\rm h.c.}\ .
\end{eqnarray}
This expression is manifestly invariant under the fictitious super-Weyl transformation.
Again the counter term is not invariant under the super-diffeomorphism, which is inevitable 
to cancel the anomaly of the super-diffeomorphism due to  $[DQ_c]$.
From this expression, we obtain the anomaly mediated gaugino mass 
\begin{eqnarray}
 M_\lambda/g^2 = + \frac{1}{2} \frac{1}{4\pi^2}
\left( {\ln}(2{\cal E}')^{1/3}+{\ln}\,C\right)|_{\Theta^2} = + \frac{1}{8\pi^2} \left( F^C - \frac{1}{3} M'^*\right) =+ \frac{1}{16\pi^2}\times 2m_{3/2}\ ,
\label{eq:SWinvGM}
\end{eqnarray}
which is independent of the gauge choice of $F^C$.

In our argument, the super-diffeomorphism variant counter term ${\mit \Delta }S$ is the key
to obtain the manifestly invariant expression of the anomaly mediated gaugino mass 
under the fictitious super-Weyl gauge symmetry.
It should be also stressed that the combination,
\begin{eqnarray}
 \int {\rm d}^4 x\, {\rm d}^2\Theta \,2{\cal E}' 
\left({\ln}\left(2 {\cal E}'\right)^{1/3} + {\ln}\,C \right)W'^\alpha W'_\alpha + {\rm h.c.}\ ,
\label{eq:invariant combination}
\end{eqnarray}
is invariant under the fictitious super-Weyl symmetry.
Thus, the mere knowledge of the anomaly of the fictitious super-Weyl gauge symmetry 
cannot determine the overall coefficient of Eq.~(\ref{eq:SWinvariantForm}).
It is crucial to start with the super-diffeomorphism invariant measure 
to obtain Eq.\,(\ref{eq:SWinvariantForm}).%
\footnote{Correspondingly, in the 1PI effective action, the fictitious super-Weyl gauge invariance alone cannot determine the gaugino mass term up to the contribution from Eq.~(\ref{eq:invariant combination}) with ${\ln}({2\cal E'})^{1/3}$ replaced by ${\rm ln} \,\Omega^{-1}$, where the chiral field $\Omega$ is defined in the following.
}

\subsubsection{Relation to the 1PI quantum effective action (I)}
\label{sec:1PI}
As is clear from Eq.\,(\ref{eq:SWinvGM}), the gaugino mass is simply read off from 
the counter term in the Wilsonian effective action, ${\mit\Delta} S_C$, 
by taking the gauge with $M'=0$ and $F^C = m_{3/2}$.
In the 1PI quantum effective action, on the other hand, it should be also possible 
to write down the gaugino mass term without using the compensator $C$.
To see how the gaugino mass appears in the 1PI action, let us 
consider a finite super-Weyl transformation of $R$,
\begin{eqnarray}
R' = - \frac{1}{8}e^{4 \Sigma} \left( {\cal D}^{\dag 2} - 8 R \right) e^{-2 \Sigma^\dag} + \cdots\ .
\end{eqnarray}
Here, ellipses denote terms which are irrelevant for the transformation of the lowest component of $R$.
Then, by taking $\Sigma$ such that 
\begin{eqnarray}
\label{eq:eliminate R}
\left( {\cal D}^2 - 8 R^\dag \right) e^{-2 \Sigma} =0\ ,
\end{eqnarray}
we can eliminate the lowest component of $R$.
The solution of Eq.\,(\ref{eq:eliminate R}) is given by\,\cite{Gates:1983nr,Butter:2013ura};
\begin{eqnarray}
e^{-2\Sigma} \equiv \Omega = 1 + \frac{1}{2\Box_+} \left({\cal D}^{\dag 2} - 8R\right) R^\dag\ , \nonumber \\
\Box _+ \equiv  \frac{1}{16} \left({\cal D}^{\dag 2} - 8R\right) \left({\cal D}^2 - 8R^\dag\right)\ .
\end{eqnarray}
Thus, by setting $C= \Omega^{-1}$, 
we can achieve the desirable gauge choice of the fictitious super-Weyl gauge symmetry where $M'=0$.
It should be noted that the apparent non-local expression of $\Omega$ does not cause problems because
the chiral field $\Omega$ is reduced to a local expression,
\begin{eqnarray}
\Omega \simeq 1 + \frac{1}{3} M^* \Theta^2\ ,
\label{eq:Omega}
\end{eqnarray}
in the flat limit. 
Thus, as long as we are interested in the flat limit, $\Omega$ can be treated as a local field.

In this gauge, $\Delta S_C$ is now expressed by,
\begin{eqnarray}
{\mit\Delta} S_{C=\Omega^{-1}} =  \frac{1}{16} \frac{1}{4\pi^2}\times \int {\rm d}^4x\, {\rm d}^2\Theta \,2{\cal E}' \,{\ln}\,\Omega^{-1}\, W'^\alpha W'_\alpha + {\rm h.c.}\ .
\end{eqnarray}
By expanding this expression around $\Omega =1$, we obtain 
\begin{eqnarray}
\label{eq:non-local}
{\mit\Delta} S_{C = \Omega^{-1}} \simeq -  \frac{1}{16} \frac{1}{8\pi^2}  \int {\rm d}^4x\, {\rm d}^2\Theta\, 2{\cal E} \,\frac{1}{\Box_+} \left({\cal D}^{\dag 2} - 8R\right) R^\dag \, W^\alpha W_\alpha + {\rm h.c.} \ ,
\end{eqnarray}
at the leading order.
Here, we have reverted ${\cal E}'$ and $W^{\prime\alpha}$ to 
 ${\cal E}$ and $W^\alpha$.
Since this term is expressed in terms of the gravity multiplet and independent of $C$, this provides 
an appropriate expression of the super-Weyl variance in the 1PI effective action.
In fact, the expression in Eq.~(\ref{eq:non-local}) reproduces the 1PI quantum effective action given in Ref.~\cite{Bagger:1999rd}.%
\footnote{Apparent difference by a factor of $4$ between our result and that in Ref.~\cite{Bagger:1999rd} is due to the difference of the normalization of the gauge multiplet.}
By substituting Eq.\,(\ref{eq:Omega}), 
we again obtain the anomaly mediated gaugino mass in Refs.~\cite{Randall:1998uk,Giudice:1998xp}.

\subsubsection{Relation with 1PI quantum effective action (II)}
The chiral field $\Omega$ is also useful to discuss the 1PI quantum effective action along the 
lines of Sec.\,\ref{sec:measure}, where we have not introduced the super-Weyl compensator $C$.
There, instead, we relied on the $F$-type super-Weyl invariant but super-diffeomorphism variant measure
to read off the gaugino mass from the Wilsonian effective action. 
The 1PI quantum effective action, however, must be invariant under the super-diffeomorphism by itself.
Thus, ${\mit \Delta} S$ should be replaced by a super-diffeomorphism invariant expression
in the 1PI quantum effective action.

To find an appropriate expression, let us remember that the chiral field $\Omega$ transforms as
\begin{eqnarray}
\delta _{\rm SW} \Omega =  -2\Sigma \Omega -  S^\alpha \frac{\partial}{\partial \Theta^\alpha}\Omega\ ,
\end{eqnarray}
under the super-Weyl transformation.
From this property, we can construct a measure
\begin{eqnarray}
[D Q_{\rm SW,diff}] \equiv [D \Omega^{1/2} \left(2 {\cal E}\right)^{1/2}Q ] = [D \Omega^{1/2} Q_{\rm diff} ]\ ,
\end{eqnarray}
which is invariant under both the $F$-type super-Weyl and the super-diffeomorphism transformations.%
\footnote{The component fields of $Q_{\rm SW,diff}$ defined by
$Q_{\rm SW,diff}=e^{1/2} [A_{Q_{\rm SW,diff}} + \sqrt{2} \Theta \chi_{Q_{\rm SW,diff}} + \Theta^2 F_{Q_{\rm SW,diff}}] $,
have the same canonical kinetic term with those of $Q_{SW}$ in Eq.\,(\ref{eq:SW component}).
}
Thus, in a similar way as Sec.\,\ref{sec:measure},
the Wilsonian effective action receives  a correction by 
changing the measure from $[DQ_{\rm diff}]$ to $[DQ_{\rm SW,diff}]$, 
\begin{eqnarray}
\label{eq:translation2}
[D Q_{\rm diff}] [D\bar{Q}_{\rm diff}] [D Q^\dag_{\rm diff}] [D\bar{Q}^\dag_{\rm diff}] &=&
[D Q_{\rm SW,diff}] [D\bar{Q}_{\rm SW,diff}] [D Q^\dag_{\rm SW,diff}] [D\bar{Q}^\dag_{\rm SW,diff}]
\times {\rm exp}
\left[
i {\mit\Delta} S_{\rm diff}
\right],\nonumber\\
{\mit\Delta} S_{\rm diff}&=&  \frac{1}{16} \frac{1}{4\pi^2}\times \int {\rm d}^4x\, {\rm d}^2\Theta\, 2{\cal E} \,{\ln}\,\Omega^{-1}\, W^\alpha W_\alpha + {\rm h.c.}\ .
\end{eqnarray}
Unlike ${\mit \Delta} S$, ${\mit\Delta} S_{\rm diff}$ is invariant under the super-diffeomorphism.
Thus, ${\mit\Delta} S_{\rm diff}$ is an appropriate expression of the super-Weyl breaking in the 
1PI quantum effective action.
Again, this expression reproduces the super-Weyl breaking term in the 1PI effective action in Ref.~\cite{Bagger:1999rd}.

\subsection{Non-abelian gauge theory}

Let us sketch the gaugino mass in a non-Abelian gauge theory.
In the non-Abelian gauge theory, the path-integral measure of the gauge multiplet should be taken into account.
The super-diffeomorphism invariant measure and the $F$-type super-Weyl invariant measure are given by
\begin{eqnarray}
[DV_{\rm diff}] = [D E^{1/2} V],~~
[DV_{\rm SW}] =  [D \left(2 {\cal E}\right)^{-1/6} \left(2 {\cal E}^\dag\right)^{-1/6} V_{\rm diff} ]\ ,
\end{eqnarray}
where $E$ is the super determinant of the super vielbein.
The super-Weyl transformation law of $E$ is given by (see Eq.~(\ref{eq:det super vielbein}))
\begin{eqnarray}
\label{eq:gauge measures}
\delta_{\rm SW} E = 2 \left(\Sigma + \Sigma^\dag \right) E.
\end{eqnarray}
Here, we collectively represents the gauge multiplet 
and the ghost multiplets by $V$, and so are $V_{\rm diff}$ and $V_{\rm SW}$ accordingly.

The translation from $[D V_{\rm diff}] $ to $[D V_{\rm SW}]$ is easily performed by the following trick.
Let us introduce a chiral compensator $C$ as in Sec.~\ref{sec:fictitious SW},
which defines $E'$ via,
\begin{eqnarray}
\label{eq:E SW}
E = C C^\dag E'\ .
\end{eqnarray}
By remembering that the super-Weyl transformation is anomalous, the gauge kinetic function receives 
a counter term depending on $C$ as \cite{Kaplunovsky:1994fg}
\begin{eqnarray}
\label{eq:gauge anomaly}
[D E^{1/2} V] &=& [D E'^{1/2} V ] \times e^{i {\mit\Delta} S^V_C}\ ,\nonumber\\
{\mit\Delta }S^V_C &=&  - \frac{1}{16} \frac{3T_G}{8\pi^2} \times\int {\rm d}^4 x\, {\rm d}^2\Theta \,2{\cal E}' \,{\ln}\,C\, W'^\alpha W'_\alpha + {\rm h.c.}\ ,
\end{eqnarray}
where $T_G$ is the Dynkin index of the adjoint representation.
It should be noted that ${\mit\Delta} S^V_C$ includes the rescaling anomaly form the ghost multiplets.
Then, by comparing Eqs.~(\ref{eq:gauge measures}), (\ref{eq:E SW}) and (\ref{eq:gauge anomaly}), 
we find that the counter term appearing along with the translation from $[D V_{\rm diff}] $ to $[D V_{\rm SW}]$ 
is given by replacing $C$ to $(2{\cal E})^{1/3}$,%
\footnote{The expression of the rescaling anomaly does not depend on whether 
the rescaling factor is a chiral superfield or a chiral density superfield.}
 which leads to
\begin{eqnarray}
\label{eq:}
\label{eq:translation2}
[D V_{\rm diff}] = [D V_{\rm SW} ] \times e^{i \Delta S^V_C}\ ,~~ C = \left(2 {\cal E}\right)^{1/3}\ .
\end{eqnarray}

By putting Eqs.~(\ref{eq:translation}) and (\ref{eq:translation2}) together, we obtain
\begin{eqnarray}
\prod_i[DQ^i_{\rm diff}] [DQ^{i\dag}_{\rm diff}] [DV_{\rm diff} ] = \prod_i[DQ^i_{\rm SW}] [DQ^{i\dag}_{\rm SW}] [DV_{\rm SW} ] \times e^{i{\mit \Delta} S}\ ,\nonumber \\
{\mit\Delta} S = - \frac{1}{16} \frac{3T_G -T_R}{8\pi^2}\times \int {\rm d}^4 x\, {\rm d}^2\Theta\, 2{\cal E} \,{\ln}\left(2 {\cal E} \right)^{1/3} W^\alpha W_\alpha + {\rm h.c.}\ ,
\end{eqnarray}
where $T_R$ is the total Dynkin index of matter fields $Q^i$. 
As a result, we find the gaugino mass,
\begin{eqnarray}
\label{eq:gaugino mass NA}
 M_\lambda/g^2 =- \frac{1}{2} \frac{3 T_G - T_R}{8\pi^2} {\ln}(2{\cal E})^{1/3}|_{\Theta^2} =  -\frac{3T_G - T_R}{16\pi^2}\times m_{3/2}\ ,
\end{eqnarray}
which reproduces the anomaly mediated gaugino mass found in Refs.~\cite{Randall:1998uk,Giudice:1998xp}.
We may also obtain the manifestly gauge independent expression in the formulation with the fictitious super-Weyl symmetry
for the non-abelian gauge theory by using Eq.\,(\ref{eq:translation2}) along the lines of Sec.\,\ref{sec:fictitious SW}.

\newpage
\section{Gaugino mass by additional matter fields}
\label{sec:correction}

\begin{screen}
This section is based on Ref.~\cite{Harigaya:2013asa};\\
{\it
 \underline{Keisuke Harigaya}, Masahiro Ibe and Tsutomu T. Yanagida,
  ``A Closer Look at Gaugino Masses in Pure Gravity Mediation Model/Minimal Split SUSY Model,''
  JHEP {\bf 1312}, 016 (2013),
  Copyright  (2013) by the authors.
}
\end{screen}

In this section, we derive the gluino, the wino and the bino mass in the presence of light vector-like matter fields and the QCD axion.

\subsection{Gaugino mass from light vector like matter}
\label{sec:correction vector}

If vector-like standard-model-gauge charged matter fields lighter than the gravitino mass exist, the gaugino mass receives corrections~\cite{Nelson:2002sa,Hsieh:2006ig,Gupta:2012gu}, as is the case with the higgsino threshold correction in Eq.~(\ref{eq: HT}).
We assume that there is a vecor-like matter fields 
$Q$ and $\bar{Q}$ which are charged 
under the standard model gauge symmetry.
We assume that $R$ charges of $Q$ and $\bar{Q}$ add up to $0$, 
so that the extra matter fields obtain 
a supersymmetric (Dirac) mass of order the gravitino mass from the $R$-breaking sector~\cite{Inoue:1991rk,Casas:1992mk}.
We also assume that the extra matter fields couple to the SUSY breaking sector only
through Planck suppressed interactions.%
\footnote{
If the extra matter fields couple to the SUSY breaking sector more strongly, soft masses of the MSSM fields are dominantly generated by the coupling, as is the case with the gauge mediation.}

\subsubsection{SUSY breaking mass spectrum of extra matter}

Generically, the threshold corrections from the extra matter field contribute to gaugino masses 
only when both a chirality flip of the fermion components of $Q\bar{Q}$, 
a Dirac mass, and a SUSY breaking
left-right mixing of the scalar components of $Q\bar{Q}$, a $b$ term, exist.
Therefore, we first demonstrate how the Dirac mass term and the $b$ term of
$Q\bar{Q}$ are obtained.

To illustrate how
they show up in the mass spectrum of extra matter, 
let us consider the simplest SUSY breaking sector with the following effective superpotential,
\begin{eqnarray}
W = \Lambda^2 Z + m_{3/2} \Mpl^2\ .
\end{eqnarray}
Here, $m_{3/2}$ denotes the gravitino mass representing the
breaking of the (discrete) $R$-symmetry
breaking.
The SUSY breaking field $Z$ obtains an $F$-term VEV
of $F_Z = -\Lambda^2$, and the flat universe condition gives $\Lambda^4 = 3 m_{3/2}^2\Mpl^2$.%
\footnote{
It is assumed that $|\vev{Z}|\ll \Mpl$.}
In the followings, we take $m_{3/2}$ and $\Lambda$ real and positive without loss of generality. 
The SUSY breaking field $Z$ is assumed to be charged under some symmetries
at the Planck scale or to be a composite field generated at some dynamical scale much 
lower than the Planck scale.

Due to the vanishing $R$-charge of $Q\bar{Q}$, 
the extra matter couples to the above SUSY breaking sector via
the super- and Kahler potentials;
\begin{eqnarray}
 W &=& \Lambda^2 Z\left(1 +y
  \frac{Q\bar{Q}}{\Mpl^2}\right) + m_{3/2}\Mpl^2\left(1 + y'\frac{Q\bar{Q}}{\Mpl^2}\right),\nonumber\\
 K &=& \lambda Q\bar{Q} + \lambda' Z^\dagger Z \frac{Q\bar{Q}}{\Mpl^2}
  +~{\rm h.c.} + \cdots,
\end{eqnarray}
where
$y$, $y'$, $\lambda$ and $\lambda'$ are dimensionless coupling constants. 
It should be noted that we can eliminate one of $y$, $y'$ and $\lambda$ 
through the Kahler-Weyl transformation 
when we are only interested in the masses of the extra matter fields
(see also Ref.\,\cite{DEramo:2013mya}
for a related discussion).
In fact, by  using the Kahler-Weyl transformation,%
\footnote{
Since the Kahler-Weyl transformation involves chiral rotations of
fermion fields in chiral multiplets, it induces gauge kinetic
functions which are proportional to $\lambda Q\bar{Q}/\Mpl^2$.
However, these terms do not contribute to gaugino masses. 
}
\begin{eqnarray}
 K\rightarrow K - \lambda Q\bar{Q} - \lambda^*Q^\dag \bar{Q}^\dag,~~~ W
  \rightarrow W {\rm exp}\left(\lambda Q\bar{Q}\right/\Mpl^2),
\end{eqnarray}
the super- and Kahler potential can be rewritten as,
\begin{eqnarray}
 W' &=& \left(y'+ \lambda\right)m_{3/2} Q\bar{Q} + \sqrt{3}\left(y  +
       \lambda \right)m_{3/2} Z \frac{Q \bar{Q}}{\Mpl} +
 \sqrt{3}m_{3/2}\Mpl Z +
 \cdots,\nonumber\\
K' &=& \lambda' Z Z^\dag \frac{Q\bar{Q}}{\Mpl^2} +~{\rm h.c.}\ .
\end{eqnarray}
Therefore, we obtain the supersymmetric Dirac mass, $\mu_Q$, and the SUSY breaking 
mixing mass parameter, $b$,
\begin{eqnarray}
\mu_Q &=& (y'+\lambda) m_{3/2}\ , \\
b &=& (3 y - y' + 2 \lambda - 3\lambda') m_{3/2}^2\ .
\end{eqnarray}
In deriving the expression of $b$, we have added up the contributions from the coupling to the SUSY breaking 
field and from the constant term in the superpotential through the supergravity interactions.%
\footnote{If $Q\bar{Q}$ couples to some flat directions, there also
exist contributions to the $b$ term by $F$ terms of the flat
directions~\cite{Pomarol:1999ie}. 
We assume, however, that $Q\bar{Q}$ do not couple to any flat directions.
}
As we will show, the phase of $b/\mu_Q$ is a very important parameter for the gaugino masses.

\subsubsection{Gaugino masses from threshold corrections}
In order to calculate the threshold corrections, 
let us take the mass diagonalized basis for the extra matters. 
Here, it should be noted that in addition to the above mentioned SUSY breaking $b$-term, the scalar components of the 
extra matter generically obtain soft squared masses 
of order the gravitino mass just as the MSSM matter fields do.
Thus, the mass terms of the scalar components $A,~{\bar A}$ and the
fermion components, $\psi,~\bar{\psi}$ are given by,
\begin{eqnarray}
 {\cal L}_{\rm mass-scalar} &=&
 - (|\mu_Q|^2+\tilde{m}_A^2) |A|^2  - (|\mu_Q^2| + \tilde{m}_{\bar{A}}^2) |\bar{A}|^2
  -\left( b A\bar{A} + {\rm h.c.}\right)
\nonumber\\
 &\equiv &- m_A^2 |A|^2  - m_{\bar{A}}^2 |\bar{A}|^2
  -\left( b A\bar{A} + {\rm h.c.}\right),\nonumber\\
 {\cal L}_{\rm mass-fermion} &=& - \mu_Q \psi \bar{\psi} + {\rm h.c.}, 
\end{eqnarray}
respectively.
Here, $\tilde{m}_A^2$ and $\tilde{m}_{\bar{A}}^2$ denote the soft squared masses.
The mass terms of the scalar components are diagonalized by rotating the
fields,
\begin{eqnarray}
 \left(
\begin{array}{c}
 A_+\\
 A_-
\end{array}
\right)&=&
\left(
\begin{array}{cc}
 {\cos} \beta_Q & -e^{-i (\delta+\delta')}{\sin} \beta_Q \\
 e^{i (\delta + \delta') }{\sin} \beta_Q & {\cos} \beta_Q
\end{array}\right)
 \left(
\begin{array}{c}
 A\\
 \bar{A}^\dag
\end{array}
\right),\nonumber\\
\tan\beta_Q &=& \frac{m_{\bar{A}}^2-m_A^2 +
 \sqrt{(m_{\bar{A}}^2-m_A^2)^2 + 4 |b|^2}}{2 |b|}>0,\nonumber\\
\delta &=& {\rm
 arg}(b/\mu_Q),~~
\delta' = {\rm
 arg}(\mu_Q),
\end{eqnarray}
which leads to the mass eigenvalues, 
\begin{eqnarray}
 m_{\pm}^2 &=& \frac{1}{2}\left(m_A^2 + m_{\bar{A}}^2 \pm
			   \sqrt{\left(m_{\bar{A}}^2-m_A^2\right)^2 + 4
			   |b|^2} \right).
\end{eqnarray}

The one-loop threshold correction from the extra matter with the above 
mass spectrum yields the gaugino masses~\cite{Poppitz:1996xw},
\begin{eqnarray}
 \Delta M_{\lambda}^{({\rm th})}
= \frac{g^2}{16\pi^2} T_Q 2e^{i \delta} {\rm
  sin}2\beta_Q |\mu_Q|
\left(
\frac{m_+^2}{m_+^2-|\mu_Q^2|}
{\rm ln}\frac{m_+^2}{|\mu_Q|^2}
-
\frac{m_-^2}{|\mu_Q|^2-m_-^2}
{\rm ln}\frac{|\mu_Q|^2}{m_-^2}
\right), 
\end{eqnarray}
at the renormalization scale just below their threshold. 
Here, $T_Q$ is a Dynkin index of $Q$, which is normalized to be $1/2$
for a fundamental representation, and $g$ is the gauge coupling constant 
evaluated at around the scale of the extra matter.
By adding the anomaly mediated effects of the extra matter, 
$\Delta M_{\lambda}^{({\rm AM})} = g^2/(16\pi^2) 2 T_Q m_{3/2}$,
we obtain the final result,%
\footnote{This formula can be applied to any cases, no matter the
origin of the Dirac mass, $b$ term, and soft squared mass terms.}
\begin{eqnarray}
\label{eq:deltam vector}
 \Delta M_{\lambda} 
&=& \frac{g^2}{16\pi^2} 2T_Q \left(e^{i \delta} {\rm
  sin}2\beta_Q |\mu_Q|
\left(
\frac{m_+^2}{m_+^2-|\mu_Q^2|}
{\rm ln}\frac{m_+^2}{|\mu_Q|^2}
-
\frac{m_-^2}{|\mu_Q|^2-m_-^2}
{\rm ln}\frac{|\mu_Q|^2}{m_-^2}
\right)
+m_{3/2}
\right).\nonumber\\
\end{eqnarray}

Several comments are in order.
First, it can be proven that
\begin{eqnarray}
\frac{m_+^2}{m_+^2-|\mu_Q^2|}
{\rm ln}\frac{m_+^2}{|\mu_Q|^2}
-
\frac{m_-^2}{|\mu_Q|^2-m_-^2}
{\rm ln}\frac{|\mu_Q|^2}{m_-^2}
>0.
\end{eqnarray}
Therefore, the phase of the gaugino mass contributed from the threshold
correction is always determined by the phase of $b/\mu_Q$.

Secondly, let us take the limit of small soft squared masses, i.e. $\tilde{m}_A^2,~\tilde{m}_{\bar{A}}^2\ll |\mu_Q|^2$.
In this limit, the diagonalized scalar masses and mixing angle are reduced to
\begin{eqnarray}
 m_{\pm}^2 = |\mu_Q|^2\pm |b|,~~
 {\rm tan}\beta_Q = 1.
\end{eqnarray}
With this mass spectrum, Eq.~(\ref{eq:deltam vector}) is also reduced to
\begin{eqnarray}
\label{eq:deltam nosquared}
 \Delta M_\lambda &=& \frac{g^2}{16\pi^2}2 T_Q\left[
\frac{b}{\mu_Q} F(|b/\mu_Q^2|)+m_{3/2}
\right],\nonumber\\
F(x)&\equiv& \frac{1+x}{x^2}{\rm ln}(1+x) +\frac{1-x}{x^2}{\rm ln}(1-x).
\end{eqnarray}
In order for the scalar components of $Q\bar{Q}$ not to be tachyonic, 
the $b$ term should satisfy $|b|<|\mu_Q|^2$, where 
the function $F$ takes values between $1$ to ${\ln}(4)\simeq1.4$.

Thirdly, let us consider the limit of $|y'|\gg 1$. 
In this case, the spectrum for the extra matter is similar to the case with a large Dirac mass term in the super-potential.
Therefore, we expect that $Q\bar{Q}$ decouples and $\Delta M_{\lambda} =0$ as expected from
the ultraviolet insensitivity properties of the anomaly mediation.
Actually, since the Dirac mass term and the $b$ term are given by $\mu_Q=y'm_{3/2}$ and $b/\mu_Q = - m_{3/2}$, 
and the soft squared mass terms are negligible, 
we obtain $\Delta M_\lambda=0$ from Eq.~(\ref{eq:deltam nosquared}).

Finally, let us take the limit of $|\lambda| \gg 1$, where the Dirac mass term and
the $b$ term are given by $\mu_Q = \lambda m_{3/2}$ and $b/\mu_Q = 2m_{3/2}$.
The soft squared mass terms are negligible and we obtain
\begin{eqnarray}
 \Delta M_\lambda = \frac{g^2}{16\pi^2} 6 T_Q m_{3/2}.
\end{eqnarray}
Here, the anomaly mediated effect and the threshold correction contribute to gaugino
masses additively.

\subsubsection{Bino, wino, gluino masses}

Let us assume that the vector-like matter fields belong to $SU(5)$ GUT
multiplets so that the coupling unification is preserved.
In this case, the contribution of the vector-like matter fields to the gaugino mass is given by,
\begin{eqnarray}
\label{eq:gaugino mass parametrization}
 \Delta M_i = \frac{g_i^2}{16\pi^2} e^{i\gamma} N_{\rm eff}m_{3/2},
\end{eqnarray}
and hence, satisfies the so-called GUT relation.
The definition of $N_{\rm eff}$ can be understood by comparing Eqs.\,(\ref{eq:deltam vector})
and (\ref{eq:gaugino mass parametrization}).%
\footnote{$N_{\rm eff}$ is not identical to the number of flavors,
$\sum_Q 2T_Q$.}
It should be noted that $N_{\rm eff}$
can be rather large
either from small $m_-^2$ or from many extra matter fields.
As we have discussed, the phase of $b/\mu_Q$ is a free parameter, 
and hence, we take $\gamma$ as a free parameter.

\begin{figure}[tb]
\begin{tabular}{cc}
\begin{minipage}{0.5\hsize}
\begin{center}
  \includegraphics[width=0.8\linewidth]{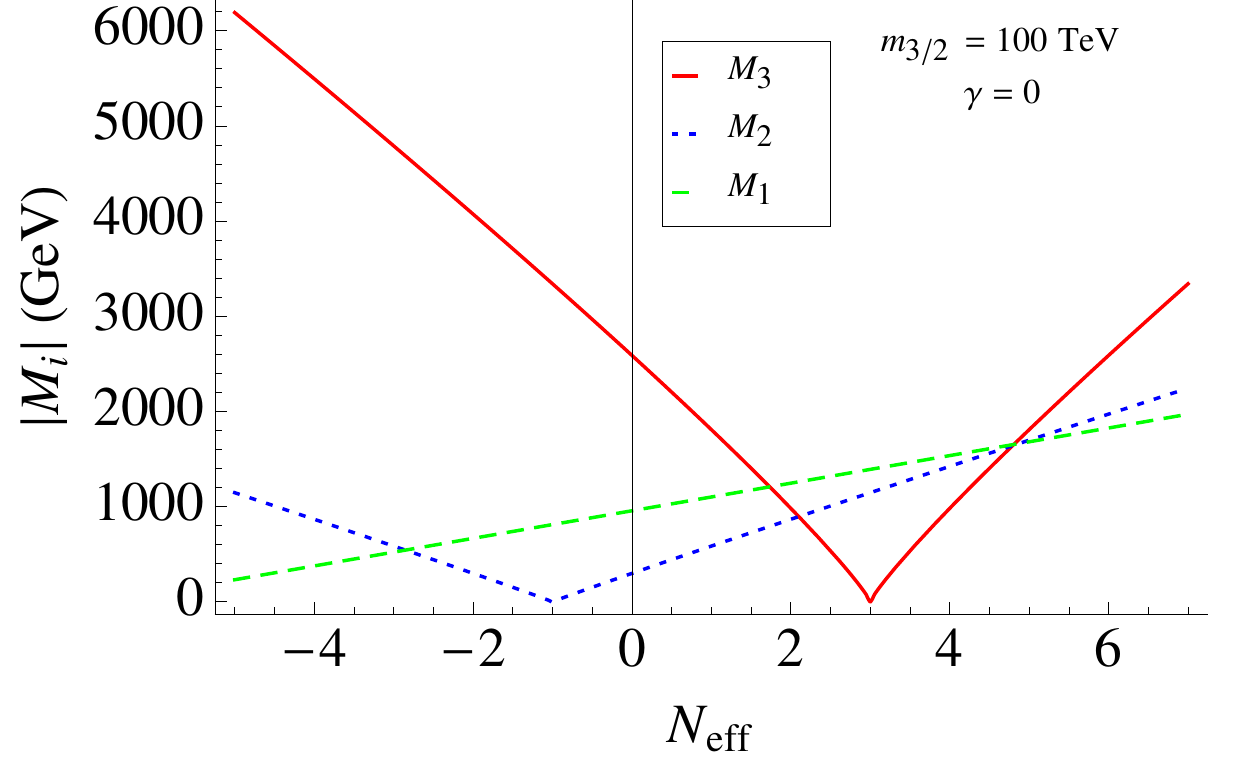}
\end{center}
\end{minipage}
 &
\begin{minipage}{0.5\hsize}
\begin{center}
  \includegraphics[width=0.8\linewidth]{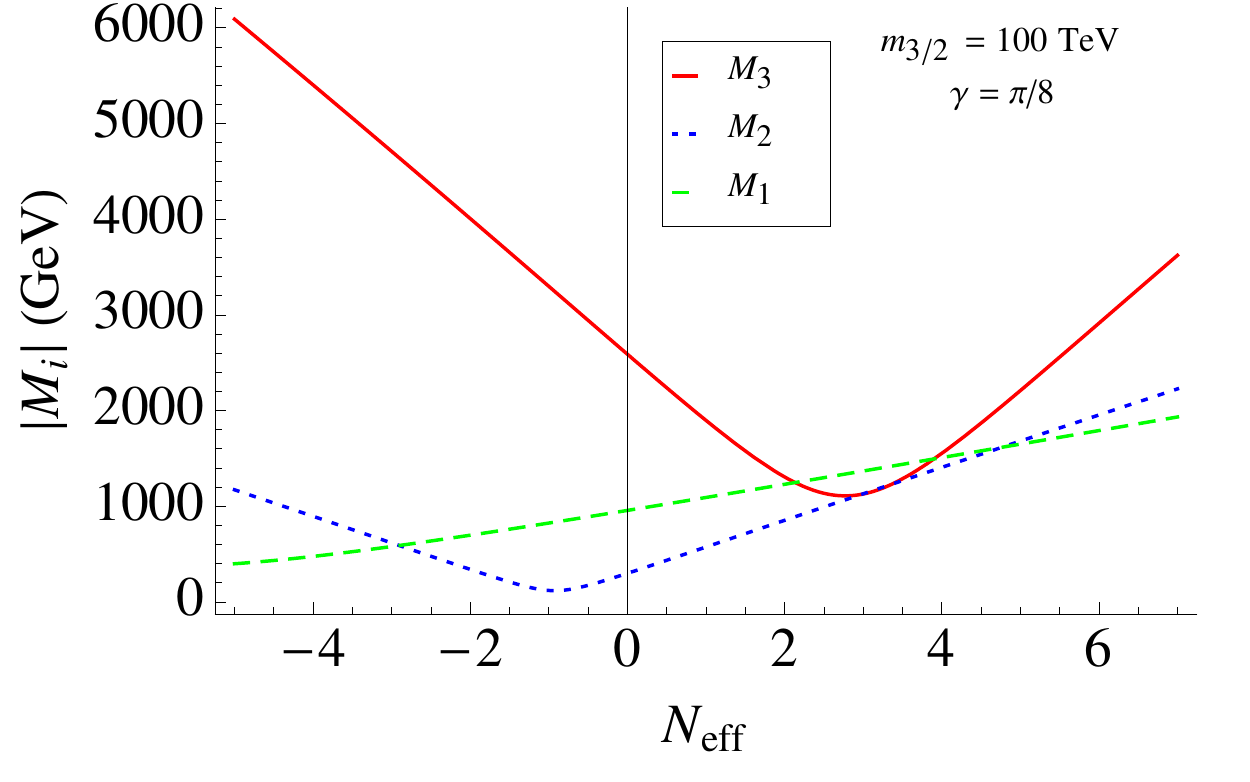}
\end{center}
\end{minipage}
\\
\begin{minipage}{0.5\hsize}
\begin{center}
  \includegraphics[width=0.8\linewidth]{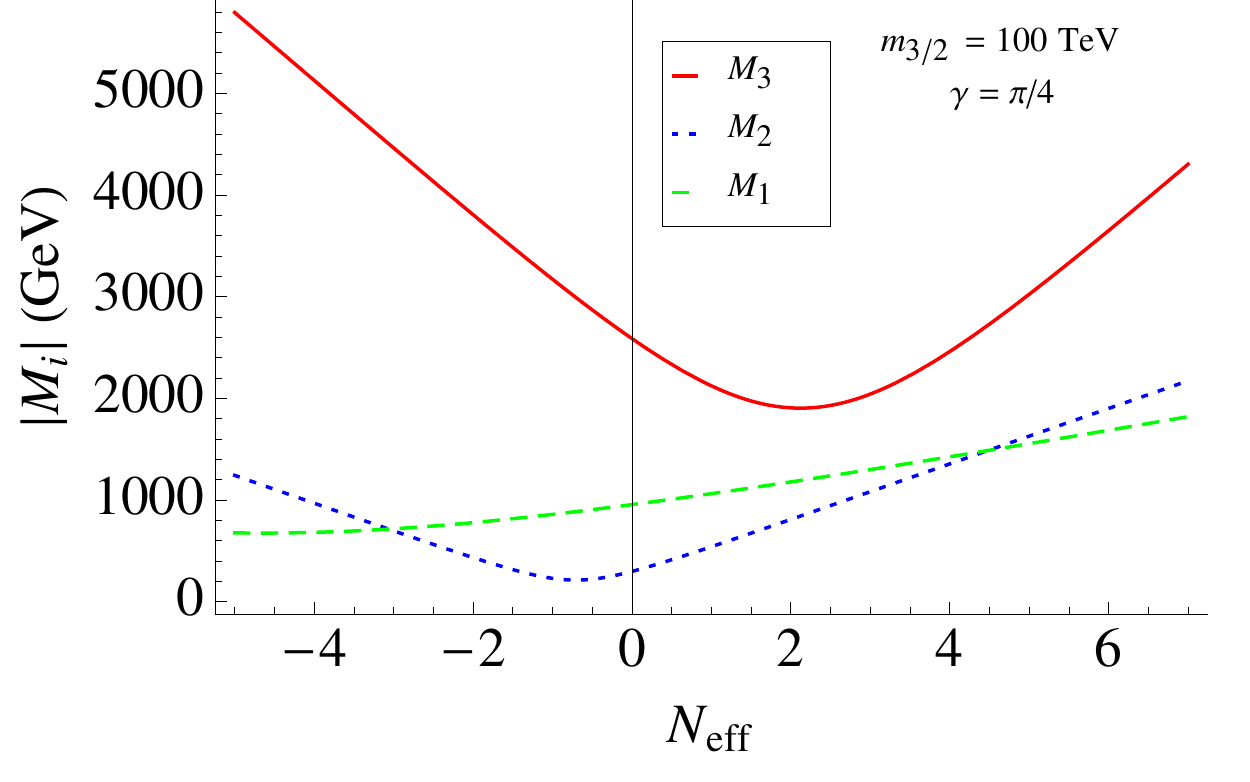}
\end{center}
\end{minipage}
 &
\begin{minipage}{0.5\hsize}
\begin{center}
  \includegraphics[width=0.8\linewidth]{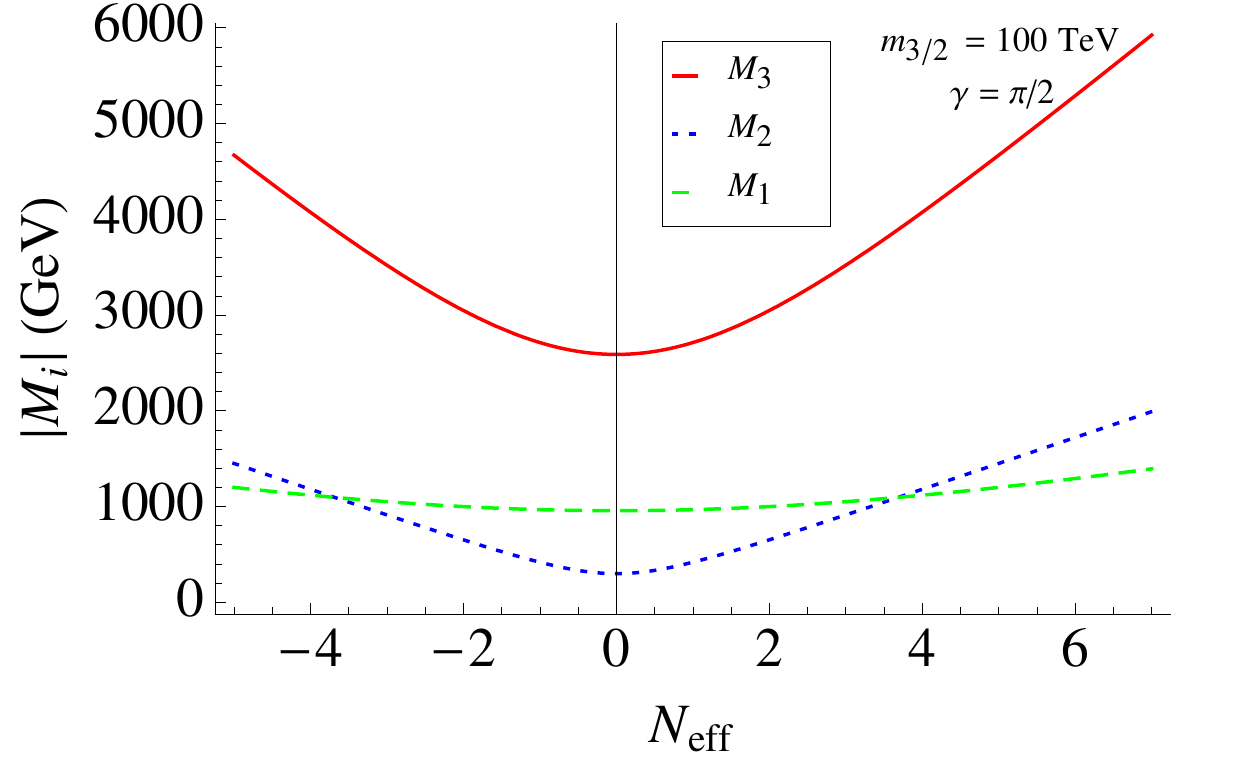}
\end{center}
\end{minipage}
\end{tabular}
\caption{\sl \small
The gluino, wino, and bino masses for $m_{3/2} = 100$ TeV with the threshold
 corrections from the extra vector-like matter in Eq.\,(\ref{eq:gaugino mass parametrization}).
We have neglected the higgsino threshold correction, for simplicity, i.e.~$L=0$.
}
\label{fig:vector-mass}
\end{figure}

In Figure~\ref{fig:vector-mass}, we show the physical gaugino masses 
in the presence of the extra matter fields for $m_{3/2}=100$ TeV
as a function of $N_{\rm eff}$ for given values of $\gamma$.
Here, we have neglected the higgsino threshold correction for
simplicity, i.e.~$L=0$.
It can be seen that the gluino mass can be much lighter than that
predicted in the purely anomaly mediated case, which enhances the detectability of the gluino at the LHC.

Note that the gaugino can be degenerated with each other.
In this case, the thermal abundance of the LSP is determined by coannihilations between gauginos.
We discuss this issue in the next section.

It should be noted that it is even possible for all three gauginos to be
degenerate for $\gamma \simeq 0$ and $N_{\rm eff} \simeq 4-5$.
This is bacause the MSSM contributions to the gluino mass is negative
while those to the wino and the bino masses are positive.
Thus, the addition of the extra matter contributions satisfying the GUT relation
can reduce the gluino mass while increasing the wino and bino masses.

\begin{figure}[t]
\begin{tabular}{cc}
\begin{minipage}{0.5\hsize}
\begin{center}
  \includegraphics[width=0.8\linewidth]{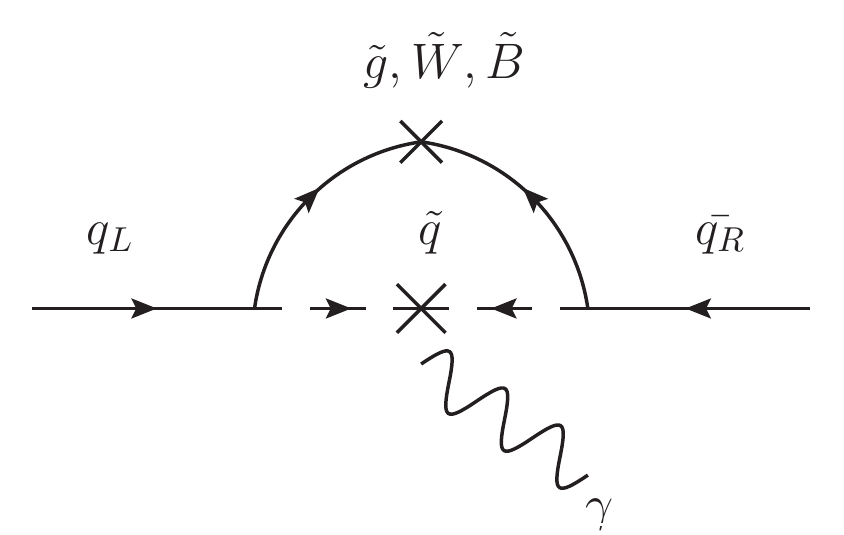}
\end{center}
\end{minipage}
 &
\begin{minipage}{0.5\hsize}
\begin{center}
  \includegraphics[width=0.6\linewidth]{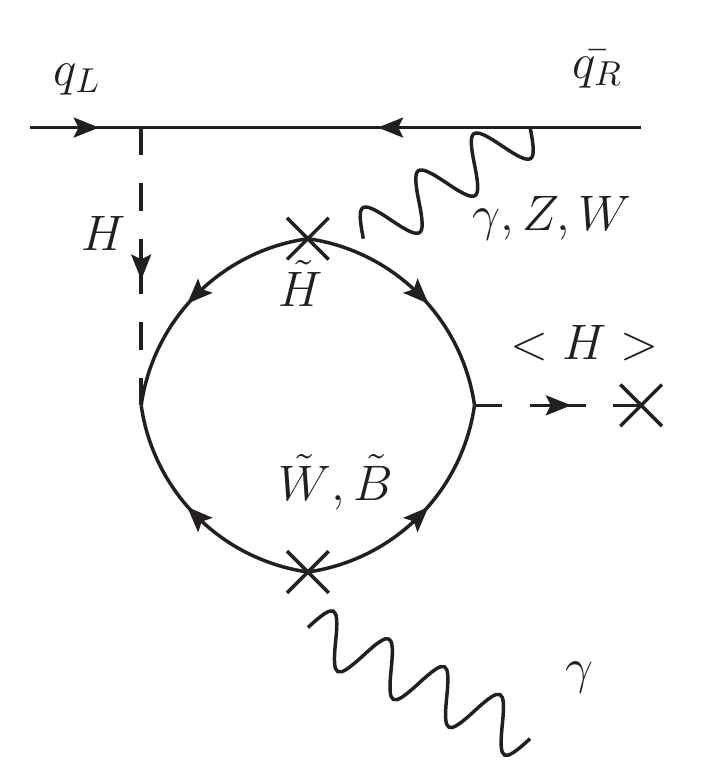}
\end{center}
\end{minipage}
\end{tabular}
\caption{\sl \small
Feynman diagram contributing to the neutron electric dipole moment.
}
\label{fig:edm}
\end{figure}

Let us comment on $CP$ violations from the phase of the gaugino masses.
First, we assume that some flavor symmetry controls the soft squared
mass terms so that they are nearly diagonal, since otherwise constraints
from the $K^0-\bar{K}^0$ mixing suggest that the soft squared mass
terms are larger than ${\cal O}(1000)$
TeV~\cite{Gabbiani:1996hi,Bhattacherjee:2012ed}, even if $\gamma=0$.
Under this assumption, a one-loop contribution to
the neutron electric dipole moment (left panel of Fig.~\ref{fig:edm}) is much smaller than
the experimental upper bound~\cite{Fuyuto:2013gla}.
A two loop Barr-Zee type contribution (right panel of Fig.~\ref{fig:edm}), which dominates over
the one loop contribution for large soft squared mass terms,
is also far smaller than the experimental upper bound for $\mu={\cal
O}(100)$ TeV~\cite{ArkaniHamed:2004yi}.

\subsection{Gaugino mass from QCD axion}
\label{sec:axion}

Here, we review the contribution of the QCD axion to
the gaugino masses, following Ref.~\cite{Nakayama:2013uta}.
In general SUSY QCD axion models, there is an
axion chiral multiplet which couples to
vector-like matter fields.
Since the axion multiplet is a flat direction and hence is not fixed, it generally obtain a non-zero $F$
term.
Thus, the gaugino mass receives threshold corrections from the
vector-like matters~\cite{Pomarol:1999ie}.
 
\subsubsection{KSVZ type models}
Let us consider the so-called KSVZ~\cite{Kim:1979if,Shifman:1979if} type
 axion model in which
the anomaly of the Peccei-Quinn (PQ) symmetry~\cite{Peccei:1977hh,Peccei:1977ur,Weinberg:1977ma,Wilczek:1977pj} of the QCD
is
mediated by additional standard model gauge charged matters $Q$ and $\bar{Q}$.
Here, we assume a superpotential,%
\footnote{The domain wall problem~\cite{Sikivie:1982qv} is absent for $n=1$.}
\begin{eqnarray}
W = \lambda X(\psi \bar{\psi}-v^2) + y \frac{\psi^n}{\Mpl^{n-1}}Q\bar{Q},
\end{eqnarray}
where $X$, $\psi$, $\bar{\psi}$ are chiral fields carrying (PQ,$R$) charges $(0,2)$,
$(1,r_\psi)$ and $(-1,-r_\psi)$, respectively.
Without loss of generality, we
take $\lambda$, $y$ and $v$ to be positive and real by field redefinitions.
We assume that the axion multiplet is the only flat direction,
$\lambda v \gg m_{3/2}$.
We also assume that $y \vev{\psi}^n/\Mpl^{n-1}\gg m_{3/2}$.

The scalar potential of the scalar components of $X$, $\psi$, and
$\bar{\psi}$ is given by
\begin{eqnarray}
 V &=& \lambda^2|\psi\bar{\psi}-v^2|^2 +
  \lambda^2|X|^2\left(|\psi|^2+|\bar{\psi}|^2\right)\nonumber\\
&& + m_{3/2}^2\left(a_X |X|^2+ a_\psi|\psi|^2 + a_{\bar{\psi}}|\bar{\psi}|^2\right)
+\left(2\lambda v^2m_{3/2}X + \tilde{b} m_{3/2}^2\psi\bar{\psi} + {\rm h.c.}\right).
\end{eqnarray}
Here, we assume that $X$, $\psi$ and $\bar{\psi}$ couple to the SUSY breaking
sector only through Planck suppressed interactions,
and hence, $a_X$, $a_{\psi}$ and $a_{\bar{\psi}}$ are at largest ${\cal O}(1)$. 
It should be noted that the $\tilde{b}$ term, $\tilde{b} m_{3/2}^2
\psi\bar{\psi}$ with $\tilde{b} ={\cal O}(1)$, can arise from the $R$ symmetry breaking
effect~\cite{Inoue:1991rk,Casas:1992mk} because the combination $\psi\bar{\psi}$ is
neutral under the PQ and $R$ symmetry.
As we will see, however, the $\tilde{b}$ term does not
affect gaugino masses.

The minimum of the potential is at
\begin{eqnarray}
\label{eq:vev axion}
 \vev{X} &=& -\frac{2 m_{3/2}v^2}{\lambda\left(|\vev{\psi}|^2 +
					|\vev{\bar{\psi}}|^2\right)}\left(1 +
 {\cal O}\left(\frac{m_{3/2}^2}{\lambda^2v^2}\right)\right),\nonumber\\
 \vev{\psi} &=& \left(
\frac{a_{\bar{\psi}}m_{3/2}^2 + \lambda^2 |\vev{X}|^2}{a_{\psi}m_{3/2}^2 + \lambda^2 |\vev{X}|^2}
\right)^{1/4}v \left(1 +
 {\cal O}\left(\frac{m_{3/2}^2}{\lambda^2v^2}\right)\right),\nonumber\\
 \vev{\bar{\psi}} &=& \left(
\frac{a_{\psi}m_{3/2}^2 + \lambda^2 |\vev{X}|^2}{a_{\bar{\psi}}m_{3/2}^2 + \lambda^2 |\vev{X}|^2}
\right)^{1/4}v\left(1 +
 {\cal O}\left(\frac{m_{3/2}^2}{\lambda^2v^2}\right)\right).
\end{eqnarray}
Here, we take $\vev{\psi}$ to be positive and real by field
redefinitions.
Note that at the leading order in $m_{3/2}/(\lambda v)$, the VEVs do
not depend on the $\tilde{b}$ term. This is because the direction
$\psi\bar{\psi}$ is fixed by the superpotential.

In order to calculate the gaugino masses, let us calculate the $b$ term of
$Q\bar{Q}$.
It is given by
\begin{eqnarray}
\label{eq:b term axion}
  {\cal L}_{b-{\rm term}} &=& y \frac{\vev{\psi}^n}{\Mpl^{n-1}} m_{3/2}A \bar{A}
   + n y\frac{\vev{\psi}^{n-1}}{\Mpl^{n-1}} \vev{F_\psi}
				    A \bar{A} + h.c.,\\
\label{eq:F term axion}
 F_\psi &=& -\left(W_{\psi^\dag}^\dag + m_{3/2} \psi\right) = - \lambda
  X^\dag \bar{\psi}^\dag - m_{3/2} \psi,
\end{eqnarray}
where $A$ and $\bar{A}$ are the scalar components of $Q$ and $\bar{Q}$,
respectively.

When we calculate the gaugino masses via a $Q\bar{Q}$ loop, the
contribution from the first term in Eq.~(\ref{eq:b term axion}) cancels
with the anomaly mediated contribution.%
\footnote{This cancellation happens only when $y
\vev{\psi}^n/{\Mpl^{n-1}}\gg m_{3/2}$. For gaugino masses with
$y\vev{\psi}^n/{\Mpl^{n-1}}\sim m_{3/2}$, see the previous subsection}
This is nothing but the decoupling of heavy vector-like matter~\cite{Giudice:1998xp}.
The contribution from the second term, on the other hand, does not
cancel, which yields the correction to the gaugino masses given by
\begin{eqnarray}
\label{eq:gaugino mass axion pre}
 \Delta M_\lambda &=& -\frac{ny \vev{\psi}^{n-1}
  \vev{F_\psi}/\Mpl^{n-1}}{y \vev{\psi}^n m_{3/2}/\Mpl^{n-1}}\times
  \frac{g^2}{16\pi^2} 2 T_Q m_{3/2} = \frac{g^2}{16\pi^2}2 T_Q
  \times\frac{-n \vev{F_\psi}}{\vev{\psi}}.
\end{eqnarray}

From Eqs.~(\ref{eq:vev axion}) and (\ref{eq:F term axion}), the $F$ term
of $\psi$ is given by
\begin{eqnarray}
\label{eq:F term axion result}
 F_{\psi} = - m_{3/2}\vev{\psi} \frac{a_{\bar{\psi}}-a_{\psi}}{a_{\bar{\psi}}+a_{\psi} + 2
\lambda^2 |\vev{X}|^2/m_{3/2}^2} \left(1 +
 {\cal O}\left(\frac{m_{3/2}^2}{\lambda^2v^2}\right)\right)\equiv - m_{3/2} \vev{\psi} \epsilon,
\end{eqnarray}
where $\epsilon$ is of order one, unless the soft squared mass terms of
$\psi$ and $\bar{\psi}$ accidentally coincide with each other.

By substituting Eq.~(\ref{eq:F term axion result}) into Eq.~(\ref{eq:gaugino mass axion pre}), we obtain the contribution from the axion model to the gaugino masses,
\begin{eqnarray}
\label{eq:gaugino mass axion}
 \Delta M_\lambda = \frac{g^2}{16\pi^2}2T_Q n\epsilon m_{3/2}
\end{eqnarray}
Note that the phase is aligned with the anomaly mediated contribution. 
This is because the phases of $\vev{\psi}$ and $\vev{F_\psi}$ are aligned
with each other.

Let us comment on the case with several flavors of vector-like matters,
as is the case with axion models presented in
Ref.~\cite{Harigaya:2013vja,Harigaya:2015soa}.
Even if there are several flavors of vector-like matters, we can always
diagonalize their mass matrices.
Each mass eigenstates contributes to the gaugino masses as given in Eq.~(\ref{eq:gaugino mass axion}).
The correction to the gaugino masses is simply multiplied by the number of the flavors.

\subsubsection{Bino, wino, gluino masses}

We assume that matter fields $Q\bar{Q}$ belong to $SU(5)$ GUT multiplets.
In the presence of the axion model described above, the gaugino masses
receive threshold corrections at the scale of the mass of $Q\bar{Q}$.
However, $M_\lambda/g^2$ is a renormalization invariant in
SUSY theory at an one-loop level.
Hence, it is not necessary to solve the renormalization equations from the
mass scale of $Q\bar{Q}$ to the gravitino mass scale for an one-loop analysis.
We can treat the correction given by Eq.~(\ref{eq:gaugino mass axion}) as if it is
generated at the gravitino mass scale, and solve the renormalization
equations (\ref{eq:renormalization eq}).
Therefore, in this axion model, gaugino masses are parameterized by Eq.~(\ref{eq:gaugino mass parametrization}) with $\gamma =0$.
Physical gaugino masses are given by the upper
left panel of Fig.~\ref{fig:vector-mass}.
In axion models with a large number of additional
matter~\cite{Harigaya:2013vja,Harigaya:2015soa}, $N_{\rm eff}$
would be considerable.

\newpage
\section{Gaugino coannihilation}
\label{sec:pheno}

\begin{screen}
This section is based on Ref.~\cite{Harigaya:2014dwa}; \\
{\it  \underline{Keisuke Harigaya}, Kunio Kaneta and Shigeki Matsumoto,
  ``Gaugino coannihilations,''
  Phys.\ Rev.\ D {\bf 89}, 115021 (2014),
  Copyright (2014) by the American Physical Society.
}
\end{screen}

As we have seen in the previous section, the gaugino mass receives corrections from light vector-like matter fields or a QCD axion. Then gauginos may be degenerated with each other.
In that case, the thermal relic abundance of the LSP is determined by the coannihilation of gauginos.
In this section, we calculate the LSP abundance in the coannihilation region of gauginos, and discuss the phenomenology of gauginos at the LHC and cosmic-ray experiments.

We treat gaugino masses
as free parameters and present model-independent results.
We thus consider the following three coannihilations below: bino-gluino coannihilation, wino-gluino coannihilation, and bino-wino coannihilation
In calculating the annihilation cross section, we take the Sommerfeld effect~\cite{Hisano:2003ec,Hisano:2004ds} into account. As we will see, the Sommerfeld effect change the thermal LSP abundance considerably.%
\footnote{
Let us list differences from previous works.
In Refs.~\cite{Profumo:2004wk,Feldman:2009zc}, neutralino-gluino coannihilation is considered without including the Sommerfeld effect.
In Ref.~\cite{Ibe:2013pua}, bino-wino coannihilation is considered with including the Sommerfeld effect, while ignoring masses of standard model particles.
}

\subsection{Lagrangian of gauginos}

Before discussing the thermal relic abundance of the gaugino dark matter, we write down the low-energy effective lagrangian of the heavy sfermion scenario at the scale around the gaugino masses. As already mentioned in introduction, the higgsino is assumed to be much heavier than the gauginos, 
and thus the mixing between bino and wino is approximately given by $m_Z^2/(\mu |\Delta M|) \simeq 10^{-2}(\mu / 100\,{\rm TeV})^{-1}(|\Delta M|/10\,{\rm GeV})^{-1}$, where $\Delta M$ is the mass difference between bino and wino. Even if bino and wino (whose masses are ${\cal O}(10^{2-3})$ GeV) are nearly degenerate, the mixing is less than ${\cal O}(1)\,\%$ in the parameter region of  interest. Therefore, their mass eigenstates are well approximated by their weak eigenstates%
\footnote{Note that the mixing is significant if higgsino is light, which is discussed in  Ref.~\cite{Baer:2005zc}. If the sign of $M_1$ and $M_2$ is opposite, bino and wino do not mix each other \cite{Baer:2005jq}}.
The lightest and the second lightest neutralinos are then pure neutral gauginos, while the lightest chargino is the pure charged wino. In following discussion, we denote bino, neutral wino, charged wino, and gluino fields as $\widetilde{B}$, $\widetilde{W}^0$, $\widetilde{W}^-$, and $\widetilde{G}^a$ with $M_1$, $M_2$, $M_2^c$, and $M_3$ being their physical masses, respectively. The mass difference between charged and neutral winos is generated by a quantum correction of the standard model (SM)~\cite{Cheng:1998hc,Feng:1999fu,Gherghetta:1999sw}, and has been calculated at two-loop level~\cite{Yamada:2009ve,Ibe:2012sx}. When the wino mass $|M_2|$ is much larger than the electroweak scale, the difference is about 170\,MeV without depending on $M_2$.

The effective lagrangian involves SM interactions, renormalizable interactions of the gauginos which play important roles to calculate their annihilation cross sections, and higher-dimensional interactions obtained by integrating out heavy fields with masses of ${\cal O}(m_{3/2})$ (sfermions, higgsino, heavy higgs bosons):
\begin{eqnarray}
{\cal L}_{\rm eff} &=& {\cal L}_{\rm SM} + {\cal L}_{\rm bino}
+ {\cal L}_{\rm wino} + {\cal L}_{\rm gluino} + {\cal L}_{\rm H.O.},
\label{eq: effective lagrangian} \\
{\cal L}_{\rm bino} &=& (1/2) \,
\overline{\widetilde{B}} (i\slashed{\partial} - M_1) \widetilde{B}, \\
{\cal L}_{\rm wino} &=& (1/2) \,
\overline{\widetilde{W}^0} (i\slashed{\partial} - M_2) \widetilde{W}^0
+ \overline{\widetilde{W}^-} (i\slashed{\partial} - M_2^c) \widetilde{W}^-
\nonumber\\
&& - g_2 \, \overline{\widetilde{W}^-}
\left(s_W \slashed{A} - c_W \slashed{Z} \right)
\widetilde{W}^-
- g_2 \, (\overline{\widetilde{W}^-} \slashed{W}^- \widetilde{W}^0 + h.c. ), \\
{\cal L}_{\rm gluino} &=& (1/2) \,
\overline{\widetilde{G}^a} (i\slashed{\partial} - M_3) \widetilde{G}^a
+ i \, (g_3/2) \, f^{abc}
\overline{\widetilde{G}^a} \slashed{G}^b \widetilde{G}^c.
\end{eqnarray}
$A$, $W^-$, and $Z$ are photon, $W$, and $Z$ boson fields, while
$s_W = \sin \theta_W$ ($c_W = \cos \theta_W$) is
the sine (cosine) of the Weinberg angle. The SM lagrangian is denoted by ${\cal L}_{\rm SM}$. The last term ${\cal L}_{H.O.}$ involves higher-dimensional interactions: e.g. four Fermi interactions including two gauginos and two SM fermions. The operators play important roles to maintain chemical equilibrium between the lightest and next lightest SUSY particles during the coannihilation period via decay, inverse decay, and conversion processes. Since detailed forms of the higher-dimensional interactions are not important for our discussion,
we do not explicitly write them down.

\subsection{Bino-gluino coannihilation}
\label{subsec: bino-gluino}

It is known that the thermal relic abundance of dark matter with coannihilation processes is obtained by solving the following Boltzmann equation~\cite{Griest:1990kh}:
\begin{eqnarray}
\frac{dY}{dx} = -\frac{\langle \sigma_{\rm eff}\,v \rangle}{H \, x}
\left( 1 - \frac{x}{3g_{*s}}\frac{dg_{*s}}{dx} \right)
s \, (Y^2-Y^2_{eq}).
\label{eq: Boltzmann}
\end{eqnarray}
$Y$ is the dark matter yield defined by the ratio between the number density of the dark matter particle and the entropy density of the universe, $s = g_{*s}\,(2\pi^2/45)(m^3/x^3)$, with $x$ being the inverse temperature of the universe in unit of the dark matter mass, $x=m/T$. The Hubble parameter $H$ and the equilibrium yield $Y_{eq}$ are given by $H = (g_*/90)^{1/2}(\pi/\Mpl)(m^2/x^2)$ and $Y_{eq} = (g_{\rm eff}\,m^3/s)\,x^{-3/2}\,e^{-x}/(2\pi)^{3/2}$, respectively.
The massless degrees of freedom for energy and entropy are denoted by $g_*$ and $g_{*s}$, respectively.
We evaluate them according to Refs.~\cite{Hindmarsh:2005ix,Arbey:2009gu} using lattice data of the QCD phase transition~\cite{Karsch:2000ps}. The effective annihilation cross section $\sigma_{\rm eff}$ is given by
\begin{eqnarray}
\sigma_{\rm eff}\,v = \sum_{i,\,j} (\sigma_{ij}\,v) \,
\frac{g_i\,g_j}{g^2_{\rm eff}}
(1 + \Delta_i)^{3/2}(1 + \Delta_j)^{3/2}
\exp[-x\,(\Delta_i + \Delta_j)],
\label{eq: effective annihilation}
\end{eqnarray}
where $\sigma_{ij}$ is the annihilation cross section between particles `$i$' and `$j$' with $g_i$ and $g_j$ being their spin (color) degrees of freedom,%
\footnote{Not to be confused with the gauge coupling constant.}
$v$ is the relative velocity between the particles, and $g_{\rm eff}$ is the effective degree of freedom for `dark matter particles', $g_{\rm eff} = \sum_{i} g_i\,(1+\Delta_i)^{3/2}\exp[-x\Delta_i]$ with $\Delta_i = (m_i - m)/m$. The mass of the particle `$i$' is denoted by $m_i$, and $m_1 = m$ corresponds to the dark matter mass. The cross section with the bracket, $\langle \sigma_{\rm eff}\,v \rangle$, in Eq.~(\ref{eq: Boltzmann}) represents the one which is averaged by the dark matter velocity distribution at the temperature $T$.

For the case of bino-gluino coannihilation in the heavy sfermion scenario, the annihilations of $\widetilde{B} \, \widetilde {B} \to$ SMs and $\widetilde{B} \, \widetilde{G} \to$ SMs are suppressed due to heavy sfermions and higgsinos. Only the annihilation $\widetilde{G} \, \widetilde{G} \to$ SMs contributes to the effective annihilation cross section. It is worth noting here that the chemical equilibrium between coannihilating particles during the freeze-out epoch is maintained thanks to higher-dimensional operators in the lagrangian (\ref{eq: effective lagrangian}): the (inverse) decay rate of the gluino and the conversion rate between bino and gluino are enough larger than the expansion rate of the universe $H$, so that the ratio of number densities between the coannihilating particles is determined only by the temperature $T$.%
\footnote{In the limit of infinite sfermion masses, the chemical quilibrium is not maintained.
By requiring that the conversion rate of a bino into a gluino is large enough, we obtain the upper bound on sfermion masses;
\begin{eqnarray}
m_{\rm sfermion} <  1000~{\rm TeV} \left(\frac{M_1}{{\rm TeV}}\right)^{3/4} \left(\frac{T}{10^{-2} M_1 }\right)^{1/4},
\end{eqnarray}
where $T$ is the temperature of the universe.
}

In the gluino annihilation, the Sommerfeld effect may enhance or suppress its cross section~\cite{Hisano:2003ec,Hisano:2004ds}. The effect can be interpreted as the one distorting wave-functions of incident particles due to long-range force acting between them, and it is incorporated through the following formula at leading order%
\footnote{Coannihilation between gluino and neutralino (corresponding to bino and wino in our case) has  been already considered in Refs.~\cite{Profumo:2004wk,Feldman:2009zc} without including the Sommerfeld effect,
while the case of gluino being LSP is studied with including the Sommerfeld effect \cite{Baer:1998pg}.
The effect has been included in the coannihilation between bino ($SU(2)_L$-singlet) and gluino ($SU(3)_c$-octet) in Ref.~\cite{deSimone:2014pda}.}:
\begin{eqnarray}
\sigma v = (\sigma_0 v) \times \lim_{r \to \infty}|\psi(r)|^2,
\label{eq: gluino annihilation}
\end{eqnarray}
where $\sigma_0$ is the self-annihilation cross section of the gluino calculated in a usual perturbative way, while $|\psi(r)|^2$ is so-called the Sommerfeld factor. The factor is calculated by solving the following Shr${\rm \ddot{o}}$dinger equation,
\begin{eqnarray}
\left[-\frac{1}{M_3} \frac{d^2}{dr^2} + V(r) \right] \psi(r) = E \, \psi(r),
\end{eqnarray}
with the boundary condition: the wave-function $\psi(r)$ has only an out-going wave at $r \to \infty$ with its normalization fixed to be $\psi(0) = 1$.

Potential $V(r)$ in the above Shr${\rm \ddot{o}}$dinger equation depends on which color representation the incident gluino pair has. The product of two color adjoint representations is decomposed into $1 \oplus 8_A \oplus 8_S \oplus 10 \oplus \overline{10} \oplus 27$. With the fact that the s-wave process dominates the annihilation and the gluino is a Majorana fermion, the representations $1$, $8_S$, and $27$ must form spin-0 states, while other representation $8_A$, $10$, and $\overline{10}$ must form spin-1 states. The potential is then given by
\begin{eqnarray}
V_R(r) \simeq c_R \, \alpha_3/r,
\end{eqnarray}
with the coefficient $c_R = -3$, $-3/2$, $-3/2$, $0$, $0$, and $1$ for representations $1$, $8_S$, $8_A$, $10$, $\overline{10}$, and $27$, respectively (see Appendix~\ref{sec:gluino ani}). It then turns out that the potential gives repulsive force for the representation $27$, and its annihilation cross section is highly suppressed. For the representations $10$ and $\overline{10}$, the potential vanishes, and their initial wave-functions are not distorted. For the representations $1$, $8_S$, and $8_A$, the potential gives attractive force, and their annihilation cross sections are expected to be enhanced. In fact, the Shr${\rm \ddot{o}}$dinger equation can be solved analytically when $V(r)$ is approximated by the Coulomb potential, and the Sommerfeld factor becomes \cite{Sommerdeld:1931} (see also Appendix~\ref{sec:gluino ani})
\begin{eqnarray}
\lim_{r \to \infty}|\psi(r)|^2 = 
\frac{2\pi c_R \alpha_3/v}{\exp [2\pi c_R \alpha_3/v] - 1},
\end{eqnarray} 
with $v$ being the relative velocity between the incident gluino pair. The factor is actually enhanced by $1/v$ for a negative $c_R$, while it is suppressed for a positive $c_R$. Here, we should mention which energy scale we should use to evaluate $\alpha_3$ in the factor, because higher order QCD corrections to $V(r)$ significantly depends on the scale. According to the prescription in Ref.~\cite{Nagano:1999nw}, we take the scale $\mu$ obtained by solving the following self-consistency equation:
\begin{eqnarray}
\mu &=& (M_3/2) \, |c_R| \, \alpha_3(\mu).
\end{eqnarray} 
In order to evaluate the factor more accurately, we should calculate the potential including higher order QCD corrections as well as finite temperature corrections, because the freeze-out phenomena occurs before QCD phase transition (say, in the symmetric phase), which is postponed to future work.

Since the Sommerfeld effect depends on the representation of the incident gluino pair, the annihilation cross section, $\sigma_0\,v$, in Eq.~(\ref{eq: gluino annihilation}) must be calculated in each representation. The cross section is given by
\begin{eqnarray}
\sigma_0\,v|_{R = 1~\,} &=& 4 \pi \alpha_3^2 c_R^2/M_3^2, \\
\sigma_0\,v|_{R = 8_S} &=& 4 \pi \alpha_3^2 c_R^2/M_3^2, \\
\sigma_0\,v|_{R = 8_A} &=&
 (\pi \alpha_3^2/M_3^2) {\textstyle \sum_f}
(2 + m_f^2/M_3^2) (1 - m_f^2/M_3^2)^{1/2}, \\
\sigma_0\,v|_{R = 27 \,} &=& 4 \pi \alpha_3^2 c_R^2/M_3^2,
\end{eqnarray}
while the cross sections for the representations $10$ and $\overline{10}$ vanish. The cross section for $R = 8_A$ comes from annihilations to various quark pairs, while those for other representations ($R = 1$, $R = 8_S$, and $R = 27$) are from annihilation to a gluon pair. As a result, the contribution to the effective annihilation cross section in Eq.~(\ref{eq: effective annihilation}) from the gluino self-annihilation is given by
\begin{eqnarray}
\sigma_{\tilde{G}\tilde{G}} \, v = (1/256)
(\sigma\,v|_{R = 1} + 8\,\sigma\,v|_{R = 8_S}
+ 3 \times 8\,\sigma\,v|_{R = 8_A} + 27\,\sigma\,v|_{R=27}),
\end{eqnarray}
which is consistent with Ref. \cite{Baer:1998pg}.

\begin{figure}[t]
\begin{center}
\includegraphics[scale=.57]{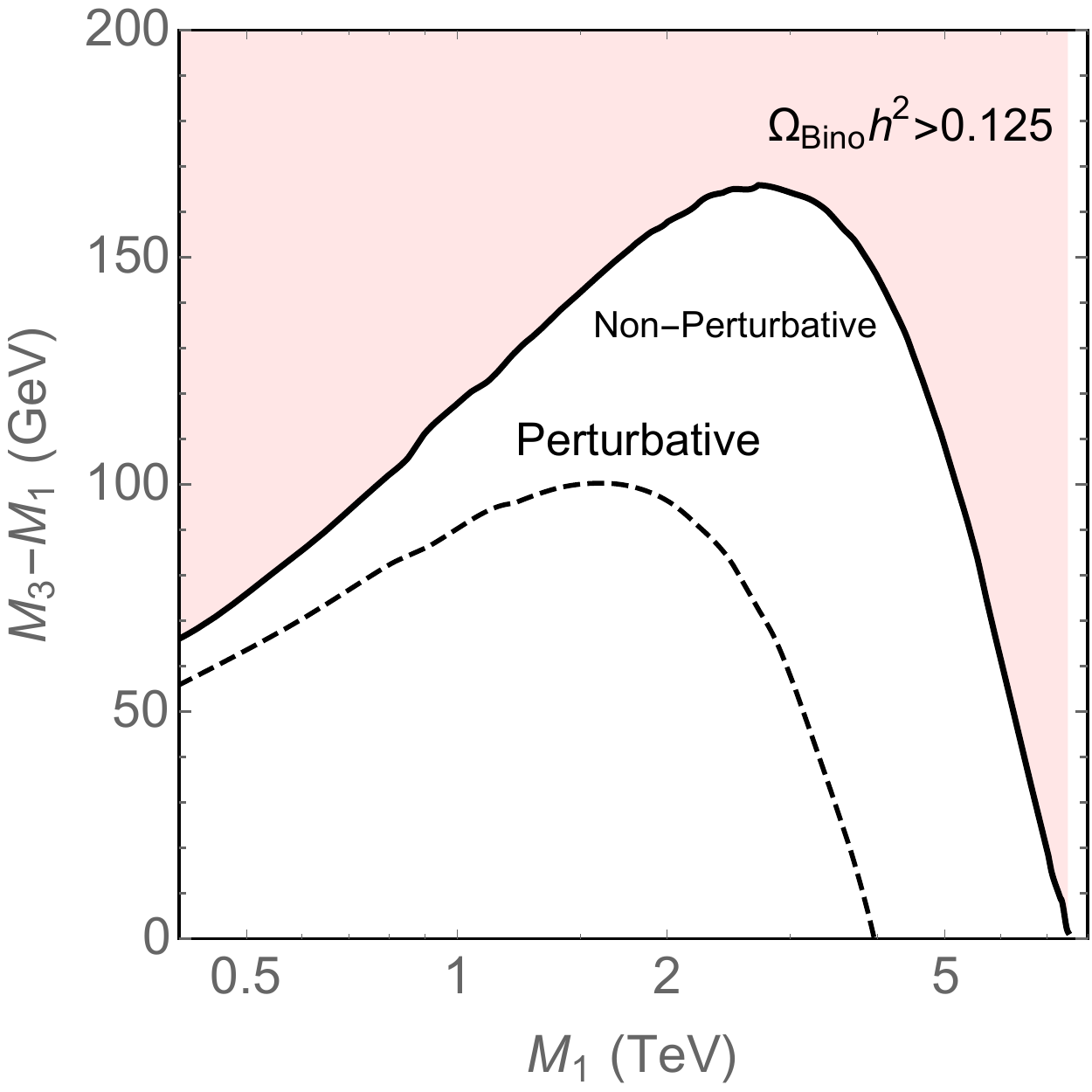}
~~
\includegraphics[scale=.55]{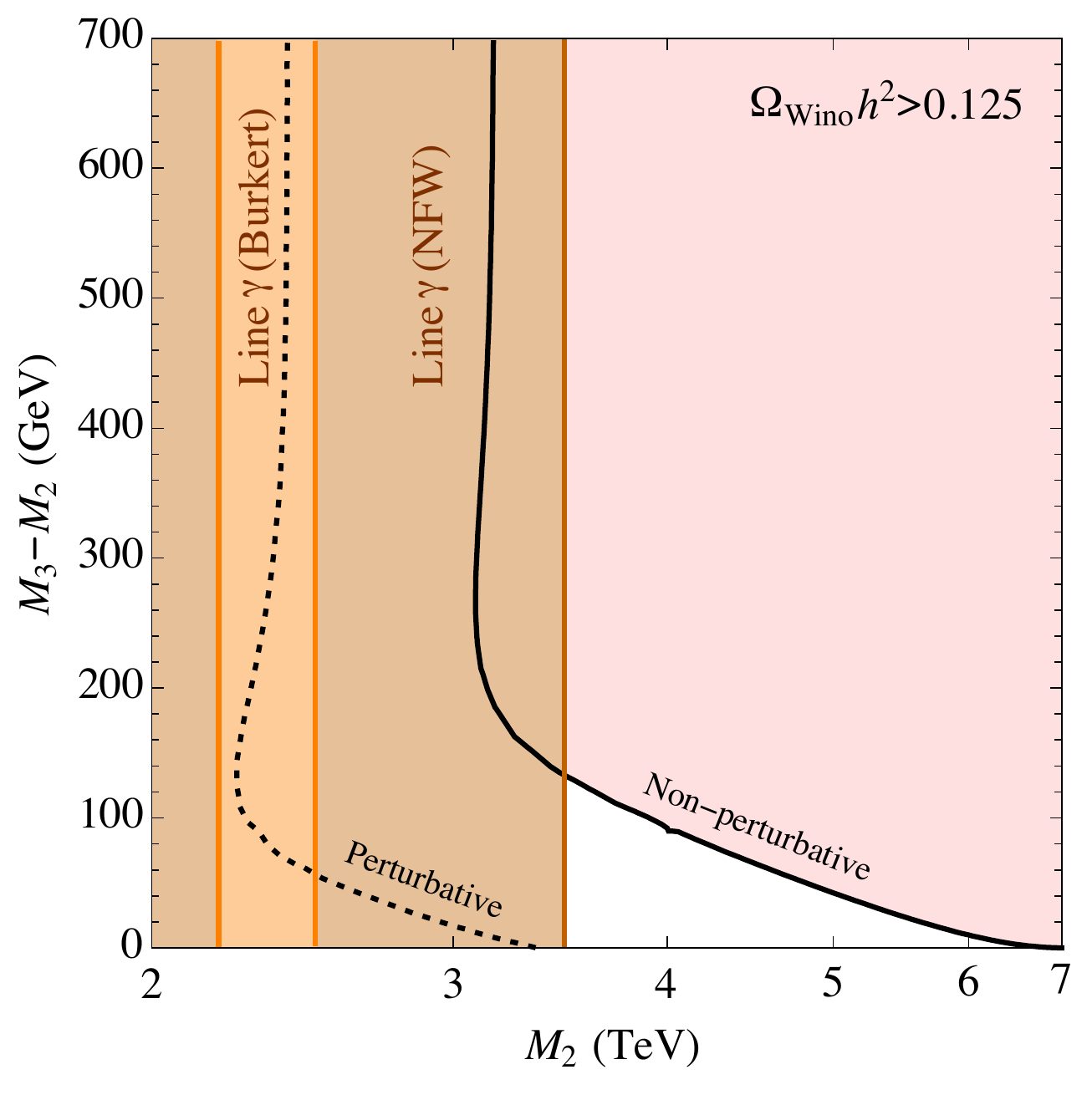}
\caption{\sl \small {\bf Left panel:} Coannihilation region between bino and gluino. The bino dark matter is over-produced in the region above the black line. For comparison, the result without the Sommerfeld effect is shown as the black dotted line.
{\bf Right panel:} Coannihilation region between wino and gluino. The black solid and dotted lines have the same meanings as those of the left panel. A limit on the wino dark matter obtained from the monochromatic line-gamma ray search (by observing the galactic center) at the H.E.S.S. experiment is also shown. See text for more details.}
\label{fig: gluino coannihilations}
\end{center}
\end{figure}

With the annihilation cross section discussed above and solving the Boltzmann equation (\ref{eq: Boltzmann}), we obtain the final yield of the dark matter particle, $Y(\infty)$. The thermal relic abundance of the dark matter is then given by $\Omega h^2 = m\,s_0\,Y(\infty)/(\rho_c\,h^{-2})$ with $s_0=2889$\,cm$^{-3}$ and $\rho_c\,h^{-2} = 1.054 \times 10^{-5}$\,GeV\,cm$^{-3}$. In the left panel of Fig.~\ref{fig: gluino coannihilations}, the coannihilation region of bino and gluino is shown. Along the black solid line, the resultant bino abundance coincides with the observed upper limit, $\Omega^{\rm (obs.)} h^2 = 0.125$. In the region below (above) the line, the abundance is smaller (larger) than the value. As a reference, we have shown the result neglecting the Sommerfeld effect~\cite{Profumo:2004wk,Feldman:2009zc}, which is denoted by the black dotted line. It can be seen that the bino dark matter can be as heavy as 7--8\,TeV due to the coannihilation.

Note that the mass difference between the bino and the gluino is required to be smaller than $O(100)$ GeV.
With such small mass difference, the gluino becomes rather long-lived due to phase space suppression in the decay of the gluino~\cite{Nagata:2015hha}.
The decay length of the gluino is given by~\cite{Gambino:2005eh,Sato:2012xf,Sato:2013bta}
\begin{eqnarray}
c \tau \simeq 3\times \left( \frac{M_3-M_1}{100~{\rm GeV}}\right)^{-5}  \left( \frac{m_{\rm sfermion}}{100~{\rm TeV}}\right)^{4}~{\rm cm}.
\end{eqnarray}
A gluino with such a large decay length yields a visible displaced vertex in detectors of hadron colliders.
In Ref.~\cite{Nagata:2015hha}, curent and future-expected lower bound on the gluino mass from the LHC experiment is estimated as
\begin{eqnarray}
M_3 \gsim
\left\{
\begin{array}{ll}
1.1~{\rm TeV}& (\text{current, 8 TeV, 20 fb}^{-1})\\
1.5~{\rm TeV}& (\text{future, 14 TeV, 300 fb}^{-1})~,
\end{array}
\right.
\end{eqnarray}
for $m_{\rm sfermion}= 100$ TeV and the mass difference which explains the thermal abundance of the bino.


\subsection{Wino-gluino coannihilation}
\label{subsec: wino-gluino}

Calculation of the dark matter abundance in wino-gluino coannihilation region is essentially the same as that in the previous subsection. Only the difference is that annihilations of wino dark matter and its $SU(2)_L$ partners also contribute to the effective annihilation cross section in Eq.~(\ref{eq: effective annihilation}). The coannihilation between wino and gluino is again suppressed because of heavy sfermions and higgsinos. In the wino annihilations, there are six annihilation modes: $\widetilde{W}^0 \widetilde{W}^0$, $\widetilde{W}^+ \widetilde{W}^-$, $\widetilde{W}^0 \widetilde{W}^\pm$, and $\widetilde{W}^\pm \widetilde{W}^\pm$. Remembering the fact that the neutral wino is a Majorana fermion, initial states of $\widetilde{W}^0 \widetilde{W}^0$ and also $\widetilde{W}^\pm \widetilde{W}^\pm$ form only spin-0 states. Initial states of other modes, on the other hand, form both spin-0 and spin-1 states. See appendix~\ref{app: wino annihilations} for concrete expressions of their annihilation cross sections. As in the gluino annihilation, the wino annihilations also receive the Sommerfeld effect. In the annihilations, the potentials $V(r)$ in their Schr$\ddot{\rm o}$dinger equations are generated by exchanging photons (Coulomb potential) and $W/Z$ bosons (Yukawa potential) between the incident particles.
Since the Sommerfeld effect on the annihilations have already been discussed in the literature~\cite{Hisano:2006nn}, we omit to write down those explicitly.

The coannihilation region between wino and gluino is shown in the right panel of Fig.~\ref{fig: gluino coannihilations}. 
The relic abundance of neutral wino is below the observed upper limit on the left side of the black solid line, when the Sommerfeld effect is included. 
For comparison, the result without the Sommerfeld effect is shown by the black dotted line.
At the right ends of the lines, gluino and wino are almost degenerated with each other.
In this case, due to the large annihilation cross section of gluino in comparison with that of wino, the dark matter abundance is essentially determined by the annihilation cross section of gluino~\cite{Profumo:2004wk}.
It can be seen that wino can be as heavy as 7\,TeV because of the coannihilation. When the mass difference between wino and gluino is large enough, the solid line asymptotically approaches $M_2 \simeq 3.1$\,TeV, which is the mass predicted by the usual wino dark matter. A bumpy structure can be seen on the black solid (dotted) line at $M_3 - M_2 \sim 200$\,GeV ($M_3 - M_2 \sim 100$\,GeV), which originates from the gluino contribution; it is somewhat suppressed by the Boltzmann factor in this region and its annihilation cross section becomes comparable to wino's, leading to the suppression of the effective annihilation cross section due to the increase of $g_{\rm eff}$.

Another limit on the wino dark matter is also shown in the plot, which is obtained from the monochromatic gamma-ray search (by observing the galactic center) at the H.E.S.S. experiment~\cite{Abramowski:2013ax}. The limit depends strongly on the dark matter profile at the center. The orange band is the limit adopting the NFW (cuspy) profile~\cite{Navarro:1995iw}, while the brown band is the one adopting the Burkert (cored) profile~\cite{Burkert:1995yz}. The limits are estimated with allowing 2$\sigma$-deviation from circular velocity data of our galaxy~\cite{Nesti:2013uwa}. It is interesting to see that, even if we take the limit adopting the NFW (cuspy) profile, we can find the parameter region consistent with the thermal relic abundance of dark matter.

\subsection{Bino-wino coannihilation}
\label{subsec: bino-wino}

Calculation of the dark matter abundance in bino-wino coannihilation region is also the same as those in previous subsections. In this region, only the wino annihilations contribute to the effective annihilation cross section in Eq.~(\ref{eq: effective annihilation}). Other annihilation processes between binos and between bino and wino are suppressed again because of heavy sfermions and higgsinos. Since both bino and wino can be dark matter in this coannihilation, we discuss the two cases separately.

\begin{figure}[t]
\begin{center}
\includegraphics[scale=0.56]{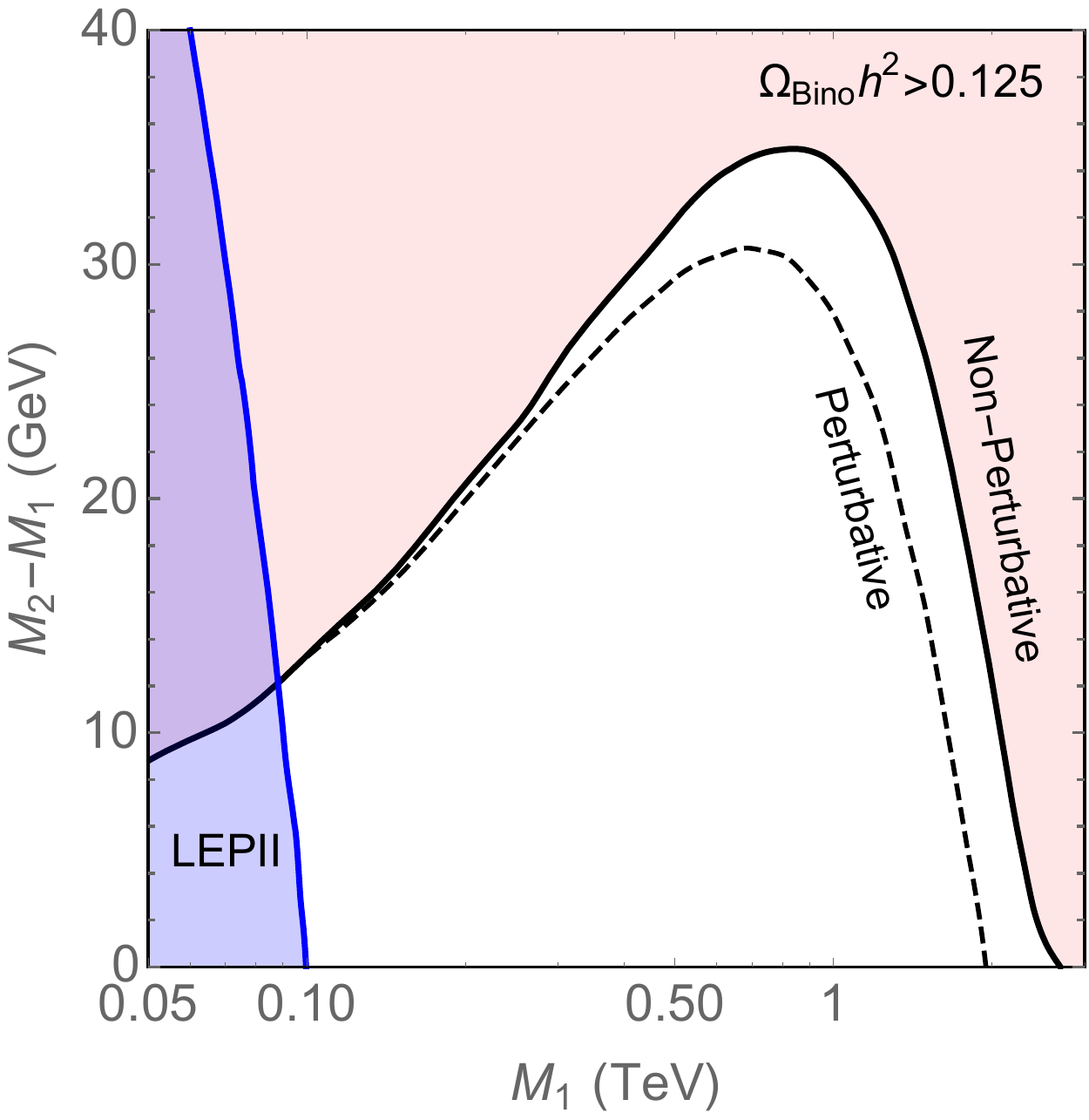}
~~
\includegraphics[scale=0.785]{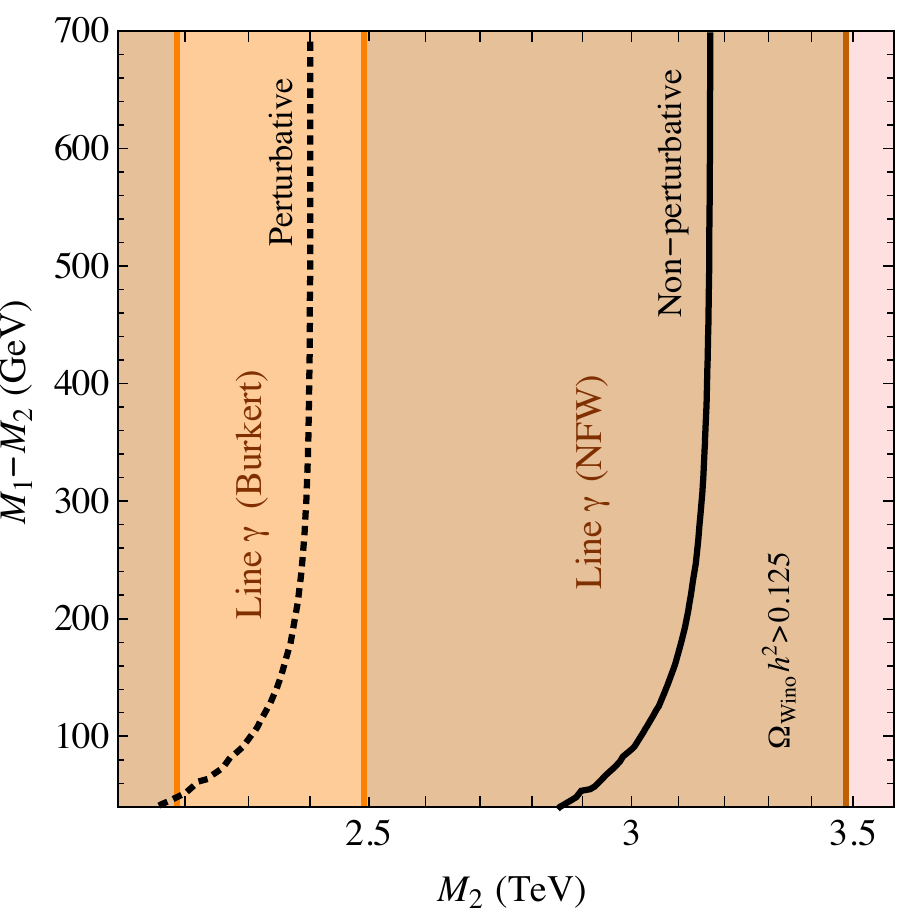}
\caption{\sl \small {\bf Left panel:} Coannihilation region between bino and wino with the bino being dark matter. Black solid and dotted lines have the same meanings as those of previous figures. Limit from the LEP\,II experiment is also shown as blue line. {\bf Right panel:} Coannihilation region between bino and wino with the wino being dark matter. The black solid and dotted lines have the same meaning as those of the left panel. A limit on the wino dark matter obtained by the monochromatic line-gamma ray search (observing the galactic center) at the H.E.S.S. experiment is also shown.}
\label{fig: bino-wino coannihilations}
\end{center}
\end{figure}

Bino-wino coannihilation with the bino being dark matter is similar to bino-gluino coannihilation, as seen in the left panel of Fig.~\ref{fig: bino-wino coannihilations}. Black solid and dotted lines have the same meanings as those of previous figures. The bino dark matter can be as heavy as 3\,TeV due to the coannihilation.
We have also shown other limits obtained by collider physics. The blue region has been excluded by the LEP\,II experiment, in which the wino pair production was searched for via
an initial state radiation of a photon~\cite{Heister:2002mn}.
In the viable parameter region, there is no constraint from the LHC experiment at present.
This is because the mass difference between the bino and the wino is very small.
However, for the mass difference between the bino and the wino of $O(10)$ GeV,
the neutral wino becomes rather long-lived due to the heavy higgsino mass as well as phase space suppression~\cite{Nagata:2015pra}.
The long-lived neutral wino can be searched at the LHC with displaced vertices.
Ref.~\cite{Nagata:2015pra} suggests that the wino with a mass of $600\mathchar`-800$ GeV can be probed at the 14 TeV running of the LHC.

The coannihilation between bino and wino with the wino being dark matter is, on the other hand, similar to wino-gluino coannihilation, which is shown in the right panel of Fig.~\ref{fig: bino-wino coannihilations}. Black solid and dotted lines have the same meanings as before. The difference from the wino-gluino coannihilation can be seen at the region that coannihilating particles are highly degenerated in mass. In wino-gluino coannihilation, the effective annihilation cross section is enhanced by the gluino annihilation at this region, while it is suppressed by very small (almost zero) annihilation of bino in bino-wino coannihilation. As a result, the wino mass coinciding with the observed upper limit is decreased to 2.8\,TeV, which is smaller than the mass predicted by the usual wino dark matter, $M_2 \simeq 3.1$\,TeV. When the mass difference between bino and wino is large enough, the black solid line approaches this value.
In the plot, a limit from the H.E.S.S. experiment is also shown as in the case of wino-gluino coannihilation. It then turns out that, if we take the limit adopting the NFW profile, all region is excluded, though the use of the NFW profile seems too aggressive to conclude that the coannihilation region has completely been ruled out.

\newpage
\section{Summary and discussion}

In this thesis, we have discussed the gaugino mass in the heavy sfermion scenario.
As we have reviewed in Sec.~\ref{sec:PGM}, the heavy sfermion scenario is
not only consistent with the observed higgs mass, but also is free from cosmological problems such as the Polonyi problem and the gravitino problem.
In the heavy sfermion scenario, gauginos are as heavy as $O(1)$ TeV and are primary targets of experimental searches.

In Sec.~\ref{sec:AMSB}, we have derived the anomaly mediated gaugino mass,
which is the essential ingredient of the heavy sfermion scenario.
We have derived it in the superspace formalism of supergravity with a Wilsonian effective action.
We have shown that in the heavy sfermion scenario, the gaugino mass is protected by the approximate super-Weyl symmetry.
The gaugino mass is induced by the quantum anomaly of the approximate super-Weyl symmetry.
We have treated the anomaly with the path-integral formulation and reproduced the gaugino mass derived with other formalisms in the literature.
In our derivation, it is essential to construct a super-diffeomorphism invariant path-integral measure.

In Sec.~\ref{sec:correction}, we have derived the gluino, the wino and the bino mass in the presence of light vector-like matter fields and the QCD axion.
We have shown that gaugino masses receive corrections as large as the anomaly mediation.
The gluino mass can be smaller than the purely anomaly mediated case, which enhances the detectability of the gluino at the LHC.

Including these corrections, is it possible that gauginos are degenerated in their masses with each other.
In this case, the thermal abundance of the LSP is determined by coannihilations between gauginos.
By calculating the thermal abundance of the LSP, we can predict mass differences between gauginos.
The information on the mass differences is important for the experimental search of gauginos in the gaugino coannihilation region.

In Sec.~\ref{sec:pheno}, we have calculated the thermal LSP abundance in the gaugino coannihilation region and discussed the phenomenology of gaugino searches at the LHC and cosmic ray experiments.
In the calculation, we have taken the Sommerfeld effect into account.
Here we summarize the phenomenology:
\begin{itemize}
\item
In the bino-gluino coannihilation region, the mass difference between the bino LSP and the gluino is typically $O(100)$ GeV.
Search for the gluino with a rather long decay length plays an important role.
\item
In the wino-gluino coannihilation region, the wino mass can be as large as $7$ TeV.
The monochromatic gamma-ray search is important.
\item
In the bino-wino coannihilation region with the bino LSP, the mass difference between the bino LSP and the wino is typically $O(10)$ GeV. Long-lived neutral wino search at the LHC is important.
\item
Phenomenology of the bino-wino coannihilation region with the wino LSP is the same as that of the purely anomaly mediated case with the wino LSP.
\end{itemize}

Finally, we stress that the observation of gauginos is important not only for testing the heavy sfermion scenario, but also for investigating physics beyond the MSSM.
Within the MSSM, the relation between the masses of gauginos are restricted;
for example, it is difficult for the gluino to be light (see Fig.~\ref{fig: gaugino masses}).
If gaugino masses deviating from the prediction of the MSSM are observed,
it indicates the existence of physics beyond the MSSM.

\section*{Acknowledgments}
The author thanks  Masahiro Ibe, Kunio Kaneta, Shigeki Matsumoto and Tsutomu T.~Yanagida for
the collaborations on which Sec.~\ref{sec:AMSB}, \ref{sec:correction} and \ref{sec:pheno} of this thesis are based,
and Masahiro Kawasaki and Mihoko M. Nojiri for useful discussions during his doctoral course.
He also thanks Masahiro Ibe, Masahiro Kawasaki and Shigeki Matsumoto for their reading and commenting on the manuscript, and
Koichi Hamaguchi, Mitsuhiro Kato, Yoichi Kazama, Toshinori Mori and Satoru Yamashita for
judging his doctoral thesis examination.
He is especially grateful to Tsutomu T.~Yanagida for advising and educating him during his doctoral course.
This work is supported by World Premier International Research Center Initiative (WPI Initiative), MEXT, Japan,
and in part by a JSPS Research Fellowships for Young Scientists.

\appendix

\section{MSSM Higgs}
\label{sec:higgs}

In this section, we calculate the minimum of the MSSM higgs potential and the mass spectrum of MSSM higgses, following Ref.~\cite{Martin:1997ns}.

\subsection{Minimum of the potential and ${\rm tan}\beta$}
Let us first find the minimum of the MSSM higgs potential.
At the tree level , the potential of the up-type higgs $H_u = (h_u^+, h_u^0)^T$ and the down-type higgs $H_d = (h_d^0, h_d^-)^T$ is given by
\begin{eqnarray}
\label{eq:higgs potential app}
V (H_u,H_d) &=&
   \left(|\mu|^2 + m_{H_u}^2\right) \left(|h_u^0|^2 + |h_u^+|^2\right)
+ \left(|\mu|^2 + m_{H_d}^2\right) \left(|h_d^0|^2 + |h_d^-|^2\right)\nonumber\\
&& + \left[ b_H \left( h_u^+ h_d^- - h_u^0 h_d^0 \right) + {\rm h.c.}\right] \nonumber \\
&& + \frac{1}{8} (g^2 + g'^2)\left(|h_u^0|^2 + |h_u^+|^2 - |h_d^0|^2 -|h_d^-|^2  \right)^2
 + \frac{1}{2}g^2 | h_u^+ h_d^{0*} + h_u^0 h_d^{-*}  |^2,
\end{eqnarray}
where $\mu$, $m_{H_u}^2$, $m_{H_d}^2$, $b_H$, $g$ and $g'$ are the higgsino mass, the soft scalar squared masses of the up-type and the down-type higgs, the holomorphic soft mass, the $SU(2)_L$ and $U(1)_Y$ gauge coupling constant, respectively.  

By an $U(1)_{\rm PQ}$ rotation, we take $b_H>0$ without loss of genericity.
By the $SU(2)_L$ rotation, we take $\vev{h_u^+} =0$. Then $\vev{h_d^-} =0$ obviously. For $h_u^+ = h_d^- =0$, the potential of $h_0^0$ and $h_d^0$ is given by
\begin{eqnarray}
\label{eq:higgs potential2}
V (h_u^0,h_d^0) &=&
   \left(|\mu|^2 + m_{H_u}^2\right)|h_u^0|^2 
+ \left(|\mu|^2 + m_{H_d}^2\right) |h_d^0|^2 + \left( - b_H h_u^0 h_d^0  + {\rm h.c.}\right) \nonumber \\
&& + \frac{1}{8} (g^2 + g'^2)\left(|h_u^0|^2  - |h_d^0|^2   \right)^2.
\end{eqnarray}

For the potential to be bounded from below along the $D$-flat direction $|h_u^0|= |h_d^0|$, $b_H$ should not be too large;
\begin{eqnarray}
2b_H  < 2|\mu|^2 + m_{H_u}^2 + m_{H_d}^2.
\end{eqnarray}
Also, for the quadratic term to be tachyonic at the origin, it is required that
\begin{eqnarray}
b_H^2 > \left( |\mu|^2 + m^2_{H_u}\right)  \left( |\mu|^2 + m^2_{H_d}\right).
\end{eqnarray}

We denote the VEVs of $h_u^0$ and $h_d^0$ as  $\vev{h_u^0} = v_u = v {\rm sin}\beta / \sqrt{2}$, $\vev{h_d^0} = v_d = v {\rm sin}\beta / \sqrt{2}$. Since $b_H >0$, $v_u v_d >0$. By a $U(1)_Y$ rotation, we take $v_u,v_d>0$. We use a convention where $0 < \beta < \pi/2$.

The minimization condition $\partial V / \partial h_u^0 = \partial V / \partial h_d^0 =0 $ requires that 
\begin{eqnarray}
\label{eq:higgs min}
m^2_{H_u} + |\mu|^2 - b_H {\rm cot}\beta - \frac{m_Z^2}{2} {\rm cos}(2\beta) =0 ,\nonumber\\
m^2_{H_d} + |\mu|^2 - b_H {\rm tan}\beta + \frac{m_Z^2}{2} {\rm cos}(2\beta) =0.
\end{eqnarray}
Here, we have used the relation $v_u^2 + v_d^2 =v^2 /2 = 2 m_Z^2 / (g^2 + g^{'2}) $.
From Eq.~(\ref{eq:higgs min}), ${\rm tan}\beta$ and $m_Z^2$ are given by
\begin{eqnarray}
\label{eq:tanb}
{\rm sin}(2\beta) = \frac{2 b_H}{m^2_{H_u} + m^2_{H_d} + 2 |\mu|^2}, \\
m_Z^2 = \frac{|m^2_{H_u} - m^2_{H_d}|}{\sqrt{1- {\rm sin}^2 (2\beta)}} - m^2_{H_u}- m^2_{H_d}.
\end{eqnarray}
Note that ${\rm tan}\beta$ is large if $b_H \ll m^2_{H_u} + m^2_{H_d} + 2 |\mu|^2 $ while 
${\rm tan}\beta=O(1)$ if $b_H \sim m^2_{H_u} + m^2_{H_d} + 2 |\mu|^2$.

\subsection{MSSM higgs mass}
The mass eigenstate of MSSM higgses, that is, the CP even neutral scalars $h^0$ and $H^0$, the CP odd neutral scalar $A^0$, the neutral would-be Nambu-Goldstone boson $G^0$, the charged scalar$H^+$ and the charged would-be Nambu-Goldstone boson $G^+$, are given by
\begin{eqnarray}
\begin{pmatrix}
h_u^0 \\ h_d^0
\end{pmatrix} &=&
\begin{pmatrix}
v_u \\ v_d
\end{pmatrix}
+ \frac{1}{\sqrt{2}}
R_\alpha
\begin{pmatrix}
h^0 \\ H^0
\end{pmatrix}
+ \frac{i}{\sqrt{2}}
R_{\beta_0}
\begin{pmatrix}
G^0 \\ A^0
\end{pmatrix},~
\begin{pmatrix}
h_u^+ \\ h_d^-
\end{pmatrix} =
R_{\beta_+}
\begin{pmatrix}
G^+ \\ H^+
\end{pmatrix},\nonumber \\
R_\gamma &\equiv&
\begin{pmatrix} 
{\rm cos}\gamma & {\rm sin}\gamma\\
-{\rm sin} \gamma & {\rm cos}\gamma.
\end{pmatrix} 
\end{eqnarray}
By the minimization condition, one can show that $\beta_0 = \beta_+ = \beta$.
The masses of $h^0$, $H^0$, $A^0$ and $H^+$ are given by
\begin{eqnarray}
m^2_{A^0} &=& \frac{2b_H}{{\rm sin} (2\beta)} = 2|\mu|^2 + m^2_{H_u} + m^2_{H_d},\\
m^2_{h^0} &=& \frac{1}{2}\left(
m^2_{A^0} + m_Z^2 - \sqrt{(m^2_{A^0} - m_Z^2)^2 + 4 m_Z^2 m^2_{A^0} {\rm sin}^2 (2\beta)}
\right),\\
m^2_{H^0} &=& \frac{1}{2}\left(
m^2_{A^0} + m_Z^2 + \sqrt{(m^2_{A^0} - m_Z^2)^2 + 4 m_Z^2 m^2_{A^0} {\rm sin}^2 (2\beta)}
\right),\\
m^2 _{H^+} &=& m_{A^0}^2 + m_W^2.
\end{eqnarray}
The mixing angle $\alpha$ is given by
\begin{eqnarray}
\frac{{\rm sin} (2\alpha)}{{\rm sin} (2\beta)} = - \frac{m^2_{H^0} + m^2_{h^0}}{m^2_{H^0} - m^2_{h^0}},~~
\frac{{\rm tan} (2\alpha)}{{\rm tan} (2\beta)} =  \frac{m^2_{A^0} + m_Z^2}{m^2_{H^0} - m^2_Z}.
\end{eqnarray}
We take a convention  where $- \pi < \alpha <0$.

Note that the mass of $h^0$ is bounded from above,
\begin{eqnarray}
m_{h^0}^2 < m_Z^2 {\rm cos}^2 (2\beta),
\end{eqnarray}
where the bound is saturated for $m_{A^0}^2 \gg m_Z^2 $ or $m_{A^0}^2 = m_Z^2 $.
As we discuss in Sec.~\ref{sec:higgs mass}, this bound is evaded if the quantum correction by the SUSY breaking effect is large.

Let us consider the decoupling limit, $m^2_{A^0} \gg m_Z^2$.
Then $m^2_{A^0} \simeq m^2_{H^0} \simeq m^2_{H^+} \gg m_Z^2$.
Below the energy scale $m_{A^0}$, the higgs sector is expected to be described by the standard model higgs doublet.
Indeed, $\alpha \simeq \beta- \pi/2$ in the decoupling limit and hence $h^0$ behaves as the standard model higgs $h$.
The mass of $h$ is given by
\begin{eqnarray}
m_h^2 = m_Z^2 {\rm cos}^2 (2\beta).
\end{eqnarray}

\section{Review on supergravity}
\label{sec:SUGRA review}

In this section, we sketch the superspace formulation of supergravity.
For detailed calculations and discussions,  see Refs.~\cite{Gates:1983nr,Wess:1992cp}.
We follow the notation in Ref.~\cite{Wess:1992cp}, 
except for the notation of complex conjugate (we use $\dagger$).

\subsection{Gravity theory from local Lorentz symmetry}
Before constructing supergravity, let us construct the gravity theory from a theory with a local Lorentz symmetry.
This construction enables us to easily include spinor representations of the Lorentz symmetry in the theory.
The gravity theory is obtained from the theory with the diffeomorphism invariance and the local Lorentz symmetry, imposing torsion constraints and gauge-fixing the local Lorentz symmetry.

We denote the local Lorentz vector index by $a,b,\cdots$ and the coordinate (Einstein) vector index by $m,n,\cdots$.
The infinitesimal diffeomorphism and the infinitesimal local Lorentz transformation  of a Lorentz vector $V^a$ and a coordinate vector $U_m$ are given by
\begin{eqnarray}
\delta V^a (x) = - \xi^n (x) \partial_n V^a(x) + V^b(x) {L_b}^a(x),\\
\delta U_m(x) = -\xi^n (x) \partial_n U_m(x)  - \left(\partial_m \xi^n\left(x\right)\right) U_n(x),
\end{eqnarray}
where $\xi^m$ and $L_{ab} = - L_{ba}$  parameterize the the diffeomorphism and the local Lorentz transformation, respectively.
The local Lorentz transformations of a undotted Weyl spinor $\psi ^\alpha$ and a dotted Weyl spinor $\psi^\dag_{\dot{\alpha}}$ are given by
\begin{eqnarray}
\delta \psi^\alpha(x) = \psi^\beta(x) {L_\beta}^\alpha(x),~~{L_{\beta}}^{\alpha} = - \frac{1}{2} {{\sigma^{ab}}_\beta}^\alpha L_{ab},\nonumber \\
\delta {\psi^\dag}_{\dot{\alpha}}(x) = \delta {\psi^\dag}_{\dot{\beta}}(x) {L^{\dot{\beta}}}_{\dot{\alpha}}(x),~~
 {L^{\dot{\beta}}}_{\dot{\alpha}}(x) = -\frac{1}{2} {{{\sigma^{ab}}}^{\dot{\beta}}}_{\dot{\alpha}} L_{ab}.
\end{eqnarray}
Note that $L_{\alpha\beta} = L_{\beta\alpha}$ and $L_{\dot{\alpha}\dot{\beta}} = L_{\dot{\beta}\dot{\alpha}}$

We introduce the vielbein field, which is a vector under the local Lorentz symmetry and the coordinate transformation, ${e_m}^a(x)$.
Then we {\it define} the metric by
\begin{eqnarray}
g_{mn} = {e_m}^a e_{na}.
\end{eqnarray}
We use the vielbein to go back and forth between the local Lorentz index and the Einstein index.

The connection associated with the local Lorentz symmetry ${\omega_m}_{ab}= -{\omega_m}_{ba}$
is called the spin connection;
\begin{eqnarray}
\label{eq:lorentz covariant}
D_m V^a = \partial_m V^a + {{\omega_m}^a}_b V^b = \partial_m V^a -  V^b {{\omega_m}_b}^a ,\nonumber\\
D_m \psi^\alpha = \partial_m \psi^\alpha - \psi^\beta \frac{-1}{2} {{\sigma^{ab}}_\beta}^\alpha \omega_{mab} .
\end{eqnarray}
where $\psi$ is a spinor field.
The local Lorentz transformation of the spin connection is given by
\begin{eqnarray}
\delta {{\omega_m}_a}^b = \partial_m {L_a}^b + {{\omega_m}_a}^c {L_c}^b - {L_a}^c {{\omega_m}_c}^b.
\end{eqnarray}
One can check that the derivative in Eq.~(\ref{eq:lorentz covariant}) is covariant under the local Lorentz transformation.

The affine connection is obtained from the consistency of the covariant derivative of $V^a$ and $V^m = e^m_a V^a$ as
\begin{eqnarray}
{\Gamma^p}_{mn} = e_a^p (\partial_m e_n^a +{{ \omega_m}^a}_b e^b_n ).
\end{eqnarray}
The curvature tensor is defined as the field strength of the local Lorentz symmetry,
\begin{eqnarray}
R_{mnab} = \partial _m \omega_{nab} -  \partial _n \omega_{mab} + \omega_{mac}{{\omega_n}^c}_b -  \omega_{nac}{{\omega_m}^c}_b.
\end{eqnarray}
The Ricci tensor and the Ricci scalar are defined by
\begin{eqnarray}
R_{ma} =  R_{mnab} e^{bn}, ~~ R = R_{ma}e^{am}.
\end{eqnarray}

In the gravity theory, the spin connection is an unnecessary degree of freedom.
We eliminate it by {\it imposing} a covariant torsion constraint.%
\footnote{Instead of imposing a torsion constraint, one can construct the spin connection from the vielbein so that the constructed field transforms as a gauge field of the local Lorentz symmetry.}
The torsion ${T_{mn}}^a$ is defined by the covariant exterior derivative of the vielbein,
\begin{eqnarray}
d e^a + {\omega^a}_b \wedge e^b \equiv T^a,
\end{eqnarray}
where we have used the differential form to simplify the expression.
The conventional constraint is that the torsion vanishes identically.
Then the spin connection is given by the vielbein as
\begin{eqnarray}
\label{eq:omega e}
{\omega_m}^{ab} = 2 e^{n[a} \partial_{[m}{e_{n]}}^{b]}-e^{n[a}e^{b]p}e_{m c}\partial_n e_p^c.
\end{eqnarray}

A theory with the diffeomorphism invariance
is obtained by
constructing an action invariant under the diffeomorphism and the local Lorentz transformation, and imposing Eq.~(\ref{eq:omega e}).
At this point, the local Lorentz symmetry is maintained.
A part of the vielbein is gauged-away by 
the remaining local Lorentz symmetry.
The vielbein ${e_m}^a$ has 16 combonents, while the local Lorentz transformation is parameterize by 6 parameters, $L_{ab} = -L_{ba}$. Thus, 10 components of the vielbein remain after gauge-fixing the local Lorentz transformation.
Note that the number of components is the same as that of the metric $g_{mn}$.

For example, let us consider a perturbation around a flat space;
\begin{eqnarray}
e_{ma} = \eta_{ma} + h_{ma}.
\end{eqnarray}
Then the transformation of the perturbation $h_{ma}$ is given by
\begin{eqnarray}
\delta h_{ma} = L_{ma}.
\end{eqnarray}
By taking $L_{ma} = (h_{ma} - h_{am})/2$, one can eliminate the anti-symmetric part of $h_{ma}$.

\subsection{Global supersymmetry}

\subsubsection{SUSY algebra and superspace}
The SUSY algebra is composed of
the translation $P_m$
and the global SUSY transformation $Q_\alpha$ and $Q^\dag_{\dot{\beta}}$.
Their commutation relation is given by
\begin{eqnarray}
\{ Q_\alpha,Q^\dag_{\dot{\beta}} \} =2 \sigma_{\alpha\dot{\beta}}^m P_m.
\end{eqnarray}
The algebra is conveniently represented in the superspace labelled by $z = (x^m, \theta^\alpha, \theta^\dag_{\dot{\alpha}} )$, where $\theta^\alpha$ and $\theta^\dag_{\dot{\alpha}}$ are Grassmann variables.
In the superspace, the representation of $P_m$, $Q$ and $Q^\dag$ is given by
\begin{eqnarray}
\label{eq:SUSY rep}
P_m = i \partial_m,~~
Q_\alpha = \frac{\partial}{\partial \theta^\alpha} - i \sigma_{\alpha\dot{\alpha}}^m \theta^{\dag \dot{\alpha}} \partial_m,~~
Q^{\dag \dot{\alpha}} = \frac{\partial}{\partial \theta_{\dot{\alpha}}^\dag} - i \theta^\alpha \sigma_{\alpha\dot{\beta}}^m \epsilon^{\dot{\beta}\dot{\alpha}} \partial_m.
\end{eqnarray}

A superfield is a function of the superspace $z$.
Component fields, which are functions only of $x$, are defined by the expansion coefficients of the superfield with respect to $\theta$ and $\theta^\dag$.
For a superfield $G(x,\theta,\theta^\dag) = A(x) + \theta\chi(x) + \cdots $, the SUSY transformation law of $G$ and its component is defined by
\begin{eqnarray}
\label{eq:SUSY trans}
\delta_\zeta G (x, \theta,\theta^\dag) \equiv (\zeta Q + \zeta^\dag Q^\dag) G \equiv \delta_\zeta A (x) + \theta \delta_\zeta \chi(x) + \cdots,
\end{eqnarray}
where $\zeta$ is the parameter of the SUSY transformation.

\subsubsection{Chiral and vector multiplet}
The superfield is in general a reducible representation of the SUSY.
In order to construct a generic SUSY invariant action, it is necessary to construct irreducible representations.
Here, we introduce two kinds of the irreducible representation of the SUSY.

For that purpose, we first introduce the following differential operators,
\begin{eqnarray}
D_\alpha = \frac{\partial}{\partial \theta^\alpha} + i \sigma_{\alpha\dot{\alpha}}^m \theta^{\dag \dot{\alpha}} \partial_m,~~
D^{\dag \dot{\alpha}} = - \frac{\partial}{\partial \theta_{\dot{\alpha}}^\dag} - i \theta^\alpha \sigma_{\alpha\dot{\beta}}^m \epsilon^{\dot{\beta}\dot{\alpha}} \partial_m,
\end{eqnarray}
which satisfy
\begin{eqnarray}
\label{eq:D operator}
\{ D^{(\dag)} , Q^{(\dag)}\} = \{D,D\} = \{D^\dag, D^\dag\}=0,~~
\{ D_\alpha,D^\dag_{\dot{\alpha}} \} = - 2 i {\sigma_{\alpha\dot{\alpha}}}^m \partial_m.
\end{eqnarray}
Chiral (or scalar) multiplets $\Phi$ are defined by a constraint,
\begin{eqnarray}
\label{eq:chiral condition}
D^\dag_{\dot{\alpha}}\Phi = 0.
\end{eqnarray}
Due to the anti-commutation relation in Eq.~(\ref{eq:D operator}), a chiral multiplet is transformed into a chiral multiplet by the SUSY transformation.
Due to the linearity of the operator $D^\dag$, a product of chiral fields is again a chiral field.
For a general superfield $X$, the following superfield is chiral,
\begin{eqnarray}
D_{\dot{\alpha}}^\dag D^{\dot{\alpha}^\dag} X,
\end{eqnarray}
due to the anti-commutation relation in Eq.~(\ref{eq:D operator}).

The real (or vector) multiplet $V$ is defined by the constraint
\begin{eqnarray}
V = V^\dag.
\end{eqnarray}
From the definition of the SUSY transformation in Eq.~(\ref{eq:SUSY trans}), a real multiplet is transformed into a real multiplet by the SUSY transformation.

There is a convenient coordinate system called the chiral coordinate, which is convenient for the calculation involving chiral superfields.
Consider a coordinate system $(y,\theta, \theta^\dag)$, where
\begin{eqnarray}
y^m = x^m + i \theta \sigma^m \theta^\dag.
\end{eqnarray}
In this coordinate system, $D$, $D^\dag$, $Q$, $Q^\dag$ are given by
\begin{eqnarray}
D_\alpha = \frac{\partial}{\partial \theta^\alpha} + 2 i \sigma_{\alpha\dot{\alpha}}^m \theta^{\dag \dot{\alpha}} \frac{\partial}{\partial y^m},~~
D^{\dot{\alpha}\dag} = - \frac{\partial}{\partial \theta_{\dot{\alpha}}^\dag},\\
\label{eq:SUSY trans chiral}
Q_\alpha = \frac{\partial}{\partial \theta^\alpha},~~
Q^{\dag \dot{\alpha}} = \frac{\partial}{\partial \theta_{\dot{\alpha}}^\dag} - 2 i \theta^\alpha \sigma_{\alpha\dot{\beta}}^m \epsilon^{\dot{\beta}\dot{\alpha}}  \frac{\partial}{\partial y^m}.
\end{eqnarray}
Then the definition of chiral fields in Eq.~(\ref{eq:chiral condition}) is simply stated as an independence from $\theta^\dag$ in the chiral coordinate.
Chiral fields are in general expressed as
\begin{eqnarray}
\label{eq:chiral field}
\Phi (y, \theta) = A(y) + \sqrt{2} \theta \chi(y) + \theta^2 F(y).
\end{eqnarray}

\subsubsection{SUSY invariant action}

For a given superfield, the SUSY transformation law of its highest component is a total derivative,
as can be seen from Eqs.~(\ref{eq:SUSY rep}) and (\ref{eq:SUSY trans}).
Then, for a given real superfield $V(x,\theta,\theta^\dag)$, the following action is SUSY invariant and hermitian,
\begin{eqnarray}
\int {\rm d}^4 x {\rm d}^2 \theta {\rm d}^2 \theta^\dag V(x,\theta,\theta^\dag)
\equiv
\int {\rm d}^8 z V(x,\theta,\theta^\dag).
\end{eqnarray}
We refer to this type of Lagrangian term as ``$D$ terms".
The kinetic term of a chiral multiplet $\Phi$ is, for example, given by
\begin{eqnarray}
{\cal L}_{\rm kin} = \int {\rm d}^8 z  \Phi^\dag \Phi.
\end{eqnarray}

Also, for a given chiral field, its $\theta^2$ component transforms into a total derivative by the SUSY transformation,
as can be checked using Eqs.~(\ref{eq:SUSY trans}), (\ref{eq:SUSY trans chiral}) and (\ref{eq:chiral field}).
Then for a given chiral superfield $\Xi(y,\theta)$, the following action is SUSY invariant and hermitian,
\begin{eqnarray}
\int {\rm d}^4 x {\rm d}^2 \theta \Xi W + {\rm h.c.}
\equiv
\int {\rm d}^6 z  \Xi + {\rm h.c.}
\end{eqnarray}
We refer to this type of Lagrangian term as ``$F$ terms".
A mass term of a chiral multiplet $\Phi$ is, for example, given by
\begin{eqnarray}
\int {\rm d}^6 z \frac{1}{2}m \Phi^2 + {\rm h.c.}.
\end{eqnarray}

\subsubsection{Non-renormalization theorem}

In the supersymmetric theory, so-called the non-renormalization theorem holds.
It states that parameters of $F$ terms are not renormalized by perturbative quantum loop corrections.%
\footnote{Precisely speaking, there is a renormalization scheme such that parameters of $F$ terms are not renormalized.}
The non-renormalization theorem can be proven by a diagrammatic method~\cite{Grisaru:1979wc}.
Here, we show an intuitive proof of the theorem by the holomorphy~\cite{Seiberg:1993vc}.

To be concrete, consider the following tree-level Lagrangian,
\begin{eqnarray}
{\cal L} = \int {\rm d}^8 z Z_B \Phi_B^\dag \Phi_B +\left[ \int {\rm d}^6z \left( \frac{1}{2}m_B \Phi_B^2 + \frac{1}{3}\lambda_B \Phi_B^3 \right) + {\rm h.c.} \right],
\end{eqnarray}
where $\Phi_B$ is a bare chiral multiplet and $Z_B$, $m_B$, $\lambda_B$ are bare constants.
We formally promote the constants $m$ and $y$ to background (i.e.~non-dynamical) chiral multiplets, $m_B(y,\theta)$ and $\lambda_B(y,\theta)$. In the end, we turn-off their $y$ and $\theta$ dependence.
Similarly, we promote $Z_B$ to a real multiplet.
Then the action has a formal $U(1)_R$ symmetry and a $U(1)_A$ symmetry shown in Table~\ref{tab:formal charge}.

\begin{table}[tb]
\begin{center}
\begin{tabular}{|c||c|c|c|}
& $\Phi$ & $m$ & $y$ \\ \hline
$U(1)_{R}$ & $1$ & $0$ & $-1$ \\
$U(1)_{A}$ & $1$ & $-2$ & $-3$
\end{tabular}
\end{center}
\caption{Formal charge assignment of $\Phi$, $m$ and $y.$}
\label{tab:formal charge}
\end{table}%

After quantum corrections are taken into account, the Wilsonian effective action becomes
\begin{eqnarray}
{\cal L} = \int {\rm d}^8 z Z_r \Phi_B^\dag \Phi_B + \left[\int {\rm d}^6z W (m_B, \lambda_B, \Phi_B ) + {\rm h.c.}\right],
\end{eqnarray}
where $Z_r$ is now a renormalized one.
Here, $W$ is a function of $m_B$, $\lambda_B$ and $\Phi_B$.
$W$  cannot depend on their conjugate and $Z_B$ due to SUSY.

Let us determine the form of $W$ from the $U(1)_R$ and $U(1)_A$ symmetry.
A function of $m_B$, $\lambda_B$ and $\Phi_B$ consistent with the symmetries are in general given by
\begin{eqnarray}
c_k \times m_B^{1-k} y_B^k \Phi_B^{2+k},
\end{eqnarray}
where $c$ and $k$ are constants.
The theory should be regular in the limit $\lambda_B\rightarrow0$.
Thus, $k$ is a non-negative integer.
$k = 2,3,\cdots$ leads to negative power of $m$, which corresponds to tree level exchanges of $\Phi$.
Such terms are absent as long as the cut of the Wilsonian effective action is larger than the mass of $\Phi$.
Thus, $W$ is in general given by
\begin{eqnarray}
W =  c _0 m_B \Phi_B^2 + c_1\lambda_B \Phi_B^3.
\end{eqnarray}

In the limit $\lambda_B\rightarrow 0$, the action asymptotically approach to the bare action.
Thus, it is required that $c_0$ and $c_1$ coincide with the constants in the tree-level action, $c_0 = 1/2$ and $c_1 =1/3$.
This shows that $m_B$ and $\lambda_B$ are not renormalized in a perturbation theory.
With similar techniques, one can show the non-renormalization theorem for other theories.

\subsection{Supergravity}

\subsubsection{super-diffeomorphism and super local Lorentz symmetry}

The construction of supergravity is parallel to that of gravity, but done in the superspace.
We denote the coordinate of the superspace by $z^M = (x^m, \theta^\mu, \theta^\dag_{\dot{\mu}})$.
The reparameterization invariance about the super coordinate is called the super-diffeomorphism invariance.

The vector and the spinor index of the Lorentz symmetry is denoted by $a,b,\cdots$ and $\alpha, \beta,\cdots$, respectively.
The Lorentz indices are collectively denoted by $A,B,\cdots$.
The local Lorentz symmetry is now extended to the superspace and is parametrized by functions of the super coordinate.
We refer to the extend local Lorentz symmetry as the ``super local Lorentz symmetry".%
\footnote{This terminology is not common, but we use it to separate the extended local Lorentz symmetry from an ordinary local Lorentz symmetry.}

The infinitesimal super-diffeomorphism and the infinitesimal super local Lorentz transformation of a Lorentz vector $V^A$ and a coordinate vector $U_M$ are given by
\begin{eqnarray}
\label{eq:sdiff and slocal lorentz}
\delta V^A (z) &=& - \xi^N(z) \partial_N V^A(z) + V^B(z){L_B}^A(z), \nonumber \\ 
\delta U_M (z) &=& - \xi^N(z) \partial_N U_M(z) - \left(\partial_M \xi^N\left(z\right)\right) U_N(z),
\end{eqnarray}
where $\xi^M$ and $L_{AB} = - (-)^A L_{BA}$  parameterize the super-diffeomorphism and the super local Lorentz transformation, respectively.

Supergravity is obtained from the theory with the super-diffeomorphism invariance and the super local Lorentz symmetry, imposing torsion constraints and fixing gauges.%
\footnote{
The super-diffeomorphism and the super local Lorentz symmetry are linear in fields.
On the other hand, after the gauge fixing, the remaining symmetries are non-linear in order to preserve the gauge conditions (see Eq.~(\ref{eq:sugra trans})).
It would be possible to directly construct the resultant  theory with symmetries non-linear in fields, but it would be
much simpler to start from a theory with all symmetries linear in fields.
}

\subsubsection{Vielbein, connection and torsion constraint}

The super vielbein and the super spin connection are the basic ingredients of supergravity.
The super vielbein field is a superfield with a superspace Einstein index $M$ and a local Lorentz index $A$, ${E_M}^A(z)$.
The super vielbein is used to
go back and forth between the local Lorentz index and the Einstein index.
The super spin connection ${\phi_{MA}}^B$ is the connection associated with the super local Lorentz symmetry,
\begin{eqnarray}
{\cal D}_M V_A = \partial_M V_A  - {\phi_{MA}} ^B V_B,
\end{eqnarray}
where ${\cal D}$ is  the superspace covariant derivative and $V_A$ is a superfield with a local Lorentz index $A$.
The field strength, or the curvature, is defined by
\begin{eqnarray}
{R_A}^B = d {\phi_A}^B + {\phi_A}^C {\phi_C}^B.
\end{eqnarray}

Since the super vielbein and the super spin connection include too much degree of freedom, we impose torsion constraints to obtain the supergravity with minimal contents, namely the graviton and the gravitino.
The torsion is defined by the covariant exterior derivative of the super vielbein,
\begin{eqnarray}
T^A = d E^A + E^B {\phi_B}^A.
\end{eqnarray}
It is known that the following constraints are appropriate ones~\cite{Wess:1992cp},
\begin{eqnarray}
\label{eq:torsion constraint}
T_{\alpha\dot{\beta}}^c &=& T_{\dot{\beta}\alpha}^c =  2 i \sigma_{\alpha \dot{\beta}}^c,\nonumber\\
T_{\underline{\alpha\beta}}^{\underline{\gamma}}& =& T_{\alpha\beta}^c = T_{\dot{\alpha}\dot{\beta}}^c= T_{\underline{\alpha}b}^c = T_{a \underline{\beta}}^c = T_{ab}^c =0,
\end{eqnarray}
where $\underline{\alpha}$ denotes either $\alpha$ or $\dot{\alpha}$.
Note that the constraint preserves the super-diffeomorphism and the super local Lorentz symmetry.

With the torsion constraint, covariant superfields, namely the torsion and the curvature are expressed by the superfields  $R$, $G_{\alpha\dot{\alpha}}$ and $W_{\alpha \beta\gamma}$ satisfying the following conditions~\cite{Wess:1992cp},
\begin{eqnarray}
{\cal D}_{\dot{\alpha}}^\dag R = 0,~~ \left(G_{\alpha\dot{\alpha}}\right)^\dag = G_{\alpha\dot{\alpha}},~~ {\cal D}_{\dot{\alpha}}^\dag W_{\beta\gamma\delta} = 0,~\nonumber\\
{\cal D}^\alpha G_{\alpha \dot{\beta}} = {\cal D}^\dag_{\dot{\beta}}R^\dag,~~
{\cal D}^\alpha W_{\alpha\beta\delta} + \frac{i}{2}\left( {\cal D}_{\beta\dot{\beta}}{G_\delta}^{\dot{\beta}} + {\cal D}_{\delta \dot{\beta}} {G_\beta}^{\dot{\beta}} \right)=0,
\end{eqnarray}
with $W_{\alpha\beta\gamma}$ being symmetric in its indices.

\subsubsection{Supergauge transformation}

The supergauge transformation is a combination of the super-diffeomorphism and the super local Lorentz symmetry which is covariant under the super-diffeomorphism and the super local Lorentz symmetry.
It is just a convenient rearrengement of the symmetries.

A superfield with a Lorentz index transforms under the super-diffeomorphism and the super local Lorentz transformation as
\begin{eqnarray}
\label{eq:sdiff lorentz vector}
\delta V^A (z) &=& - \xi^M(z) \partial_M V^A(z) + V^B(z){L_B}^A(z)\nonumber \\
&=& - \xi^M {\cal D}_M V^A + V^B \xi^C {\phi_{CB}}^A + V^B {L_B}^A.
\end{eqnarray}
If one takes field-dependent ${L_B}^A$ such that
\begin{eqnarray}
\label{eq:supergauge1}
{L_B}^A = - \xi^C {\phi_{CB}}^A,
\end{eqnarray}
the transformation of $V^A$ is given by
\begin{eqnarray}
\label{eq:supergauge2}
\delta_\xi V^A = - \xi^B {\cal D}_BV^A.
\end{eqnarray}
We refer to this transformation as the supergauge transformation.

\subsubsection{Gauge fixing and supergravity transformation}

With the supergauge symmetry and the super local Lorentz symmetry, many components of the super vielbein and the super spin connection are gauged away.
We keep the lowest component ($\theta$ and $\theta^\dag$ independent part) of the supergauge symmetry and that of the super local Lorentz symmetry unused in order  to manifestly preserve the diffeomorphism, the supergravity (explained below) and the local Lorentz symmetry.
Higher components are used to eliminate gauge degree of freedoms.
Especially, the lowest component of the super vielbein and the super spin connection are cast into the form,
\begin{eqnarray}
\label{eq:vielbein connection low}
{E_M}^A| =
\begin{pmatrix}
{e_m}^a & \frac{1}{2}{\psi_m}^\alpha & \frac{1}{2}{\psi_{m\dot{\alpha}}}^\dag \\
0 & \delta_\mu^\alpha & 0 \\
0 & 0 & \delta^{\dot{\mu}}_{\dot{\alpha}}
\end{pmatrix},\nonumber \\
{\phi_{mA}}^B| = {\omega_{mA}}^B,~~
{\phi_{\mu A}}^B = {{\phi^{\dot{\mu}}}_A}^B| =0,
\end{eqnarray}
where $X|$ denotes the lowest component of the superfield $X$.
${\omega_{mA}}^B$ is the spin connection which is expressed by the vielbein ${e_m}^a$ and the gravitino ${\psi_m}^\alpha$ due to the torsion constraint.
The torsion, curvature and covariant derivative, out of which an action is constructed, are expressed by the following fields,
\begin{eqnarray}
{e_m}^a:~\text{vielbein},~~
{\psi_m}^\alpha:~\text{gravitino},~~
M:~\text{auxiliary scalar} ,~~
b_a:~\text{auxiliary vector},\nonumber
\end{eqnarray}
which we collectively refer to as the supergravity multiplet.
Here, $M$ and $b_a$ are defined by
\begin{eqnarray}
R| = -\frac{1}{6}M,~~ G_a| = -\frac{1}{3}b_a.
\end{eqnarray}

A supergravity transformation is a combination of the supergauge transformation and the super local Lorentz symmetry with parameters,
\begin{eqnarray}
\xi^a| =0,~~ \xi^\alpha| = \zeta^\alpha,~~
L_{AB}|=0,
\end{eqnarray}
and higher components are chosen so that the gauge condition in Eq.~(\ref{eq:vielbein connection low}) is preserved.
By a small calculation, the supergravity transformation is determined as
\begin{eqnarray}
\label{eq:sugra trans}
\xi^\alpha(z) = \zeta^\alpha(x),~~
\xi^a(z) = 2i \left[ \theta\sigma^a \zeta^\dag(x) - \zeta(x) \sigma^a \theta^\dag  \right],\nonumber\\
L_{\alpha\beta} (z) = \frac{1}{3}\left\{
\theta_\alpha \left[
2 \zeta_\beta(x) M^*(x) - b_{\beta\dot{\gamma}}(x) \zeta^{\dot{\gamma}\dag}(x)
\right]
+ \left(\alpha \leftrightarrow \beta \right)
\right\}.
\end{eqnarray}
It is at this point where non-linearily appears in the transformation law.

Let us comment on the diffeomorphism and the local Lorentz symmetry.
The diffeomorphism preserves the gauge condition in Eq.~(\ref{eq:vielbein connection low}), and hence remains unchanged after the gauge fixing procedure.
The local Lorentz transformation, on the other hand, does not preserve the gauge condition because it transforms ${E_\mu}^\alpha|$ as ${\delta_\mu}^\alpha \rightarrow {\delta_\mu}^\beta {L_\beta}^\alpha(x)$.
Thus, the local Lorentz transformation must involve a compensating supergauge transformation to preserve the gauge condition.
The transformation of the super vielbein under the supergauge and the super local Lorentz transformation is given by
\begin{eqnarray}
\delta {E_M}^A = - {\cal D}_M \xi^A - \xi^B {T_{BM}}^A + {E_M}^B {L_B}^A.
\end{eqnarray}
Especially,
\begin{eqnarray}
\delta {E_\mu}^\alpha| = - \partial_\mu \xi^\alpha| -  \delta_\mu^\gamma\xi^m| {T_{m\gamma}}^\alpha| + \delta_\mu^\beta {L_\beta}^\alpha|.
\end{eqnarray}
Putting $\xi^m|=0$ to avoid a mixing with the diffeomorphism,
the compensating transformation is given by the supergauge transformation with
\begin{eqnarray}
\label{eq:lorentz compensate}
\xi^\alpha(z) = \delta_\mu^\beta {L_\beta}^\alpha(x) \theta^\mu.
\end{eqnarray}

So far, we have encountered various transformations.
For convenience, we summarize the property of transformations in Table~\ref{tab:summary trans}.

\begin{table}[tb]
\begin{center}
\begin{tabular}{|cl|c|c|c|}
\hline
& & parameter & see Eq(s). & descriptions \\  
 \hline
1)&super-diffeomorphism & $\xi^M(x,\theta,\theta^\dag)$ & (\ref{eq:sdiff and slocal lorentz})&\\
2)&super local Lorentz & $L_{AB}(x,\theta,\theta^\dag)$ & (\ref{eq:sdiff and slocal lorentz}) & \\
3)&supergauge & $\xi^M(x,\theta,\theta^\dag)$  & (\ref{eq:supergauge2}) & 1) and 2) with $L_{AB} (\xi)$  \\
\hline
4)&supergravity &$\zeta^\alpha(x)$ & (\ref{eq:sugra trans}) & part of 2) and 3) \\
5)&diffeomorphism & $\xi^m(x)$ & (\ref{eq:sdiff and slocal lorentz}) & part of 1) \\
6)&local Lorentz & $L_{\alpha\beta}(x)$ & (\ref{eq:sdiff and slocal lorentz}), (\ref{eq:lorentz compensate}) & part of 2) and 3) \\
\hline
\end{tabular}
\end{center}
\caption{Summary of transformations}
\label{tab:summary trans}
\end{table}%

\subsubsection{Chiral and real multiplet}

Chiral superfields are defined by the constraint,
\begin{eqnarray}
{\cal D}^\dag_{\dot{\alpha}}\Phi =0.
\end{eqnarray}
Their components are defined by
\begin{eqnarray}
A = \Phi|,~~ \chi_\alpha = \frac{1}{\sqrt{2}}{\cal D}_\alpha\Phi|,~~ F = - \frac{1}{4}{\cal D}^\alpha {\cal D}_\alpha\Phi|.
\end{eqnarray}
For a given superfield $X$ without local Lorentz indices, the following superfield is a chiral superfield,
\begin{eqnarray}
\left({\cal D}_{\dot{\alpha}}^\dag D^{\dot{\alpha}\dag} - 8R \right) X.
\end{eqnarray}
Real superfields are defined by the constraint,
\begin{eqnarray}
V^\dag = V.
\end{eqnarray}

\subsubsection{Super-diffeomorphism and super local Lorentz invariant action}

Let us construct a super-diffeomorphism and super local Lorentz invariant action
in the super coordinate $z^M = (x^m, \theta^\mu, \theta^\dag_{\dot{\mu}})$.
Under the super-diffeomorphism and the super local Lorentz transformation, an Einstein and Lorentz scalar $V(z)$ transforms as
\begin{eqnarray}
\delta V(z) = - \xi (z)^M \partial_M V(z),
\end{eqnarray}
where $\xi$ parameterizes the super-diffeomorphism.
It can be seen that a mere integration of $V(z)$ by $\int{\rm d}^8z = \int{\rm d}^4x {\rm d}^2 \theta {\rm d}^2 \theta^\dag$ does not yield an invariant action.

In order to obtain the supergravity action, we consider the super determinant of the super vielbein,%
\footnote{
Consider a matrix ${\cal M}$ with bosonic and fermionic components,
\begin{eqnarray}
{\cal M} =
\begin{pmatrix}
A_{ab} & B_{a\beta} \\
C_{\alpha b} & D_{\alpha \beta}
\end{pmatrix},
\end{eqnarray}
where Latin indices and Greek indices denote bosonic and fermionic indices.
The super determinant is defined by
\begin{eqnarray}
{\rm sdet}{\cal M} = \frac{ {\rm det}_{ab} A_{ab}}{{\rm det}_{\alpha\beta}(D_{\alpha\beta} - C_{\alpha b}A^{-1 ba}B_{a\beta} )}
\end{eqnarray}
}
\begin{eqnarray}
\label{eq:det super vielbein}
E= {\rm sdet}\left({E_M}^A\right).
\end{eqnarray}
The transformation law of the super vielbein is given by
\begin{eqnarray}
\delta {E_M}^A = - \xi^N\partial_N {E_M}^A - \left(\partial_M \xi^N\right){E_N}^A + {E_M}^B {L_B}^A.
\end{eqnarray}
Then the transformation of $E$ is given by
\begin{eqnarray}
\delta E &=& E \times {\rm str}  \left({E_A}^M \delta {E_M}^B\right) = - \partial_M \left(\left(-\right)^M\xi^M E \right).
\end{eqnarray}

The product of an Einstein and Lorentz scalar $V(z)$ and $E$ transforms as
\begin{eqnarray}
\delta \left(E V \right) = - \partial_M \left(\left(-\right)^M\xi^M EV\right),
\end{eqnarray}
which is a total derivative.
Hence, the super-diffeomorphism and the super local Lorentz invariant action is given by
\begin{eqnarray}
\label{eq:action real}
\int  {\rm d}^4x {\rm d}^2 \theta {\rm d}^2 \theta^\dag E V + {\rm h.c.},
\end{eqnarray}
where $V$ is an Einstein and Lorentz scalar.

\subsubsection{Chiral representation}

The expression of the action in Eq.~(\ref{eq:action real}) is not convenient for practical calculations.
One needs tedious calculations to expand given superfields by $\theta$ and $\theta^\dag$.
Calculations are simplified by constructing an action in a chiral representation.
A draw back is that a part of a manifest super-diffeomorphism and super local Lorentz invariance are lost because we construct an action in a gauge-fixed form.%
\footnote{It might be possible to preserve the manifest super-diffeomorphism and super local Lorentz invariance by introducing compensator fields.}
A manifest diffeomorphism, supergauge transformation, and local Lorentz transformation invariance are maintained.%
\footnote{Otherwise, one cannot eliminate the gauge degree of freedoms in the vielbein and the gravitino.}

The chiral coordinate $(x^m,\Theta^\alpha)$  is a coordinate system such that the expansion of a chiral scalar field $\Phi$ is given by
\begin{eqnarray}
\Phi = A(x) + \sqrt{2} \Theta^\alpha \chi_\alpha(x) + \Theta^2 F(x).
\end{eqnarray}
Then the supergravity transformation of chiral scalar fields is given by~\cite{Wess:1992cp}
\begin{eqnarray}
\label{eq:scalar sugra}
\delta \Phi &=& - \eta^M(x,\Theta)\partial_M \Phi,\nonumber \\
\eta^m &=& 2i \Theta \sigma^m \zeta^\dag + \cdots,
\nonumber \\
\eta^\alpha &=& \zeta^\alpha +\cdots,
\end{eqnarray}
where the ellipses denote terms given by $\zeta$ and the supergravity multiplet.
Here and hereafter, indices $M,N,\cdots$ denote the chiral coordinate $(x^m,\Theta^\alpha)$.
In the flat limit, a global supergravity transformation ($\zeta^\alpha(x) = {\rm const.} $) in the chiral representation reduces to a global SUSY transformation in the chiral representation (see Eq.~(\ref{eq:SUSY trans chiral})).

In order to construct an action invariant under the supergravity transformation, we define chiral densities.
Chiral densities are functions of $(x, \Theta)$ whose supergravity transformation is given by
\begin{eqnarray}
\label{eq:density sugra}
\delta \Delta = - \partial_M \left[ \eta^M \Delta \left(-\right)^{M} \right].
\end{eqnarray}
A product of a chiral density $\Delta$ and a chiral scalar $\Phi$ is also a chiral density;
\begin{eqnarray}
\delta (\Delta \Phi) =  \partial_M \left[ \eta^M \Delta \Phi \left(-\right)^{M} \right].
\end{eqnarray}
Then, a supergravity invariant action is given by
\begin{eqnarray}
\int {\rm d}^4 x {\rm d}^2 \Theta \Delta \Phi.
\end{eqnarray}

In order to achieve the diffeomorphism invariance, we consider a chiral density whose lowest component is a determinant of the vielbein, $e = {\rm det}\left({e_m}^a\right)$.
Such chiral density ${\cal E}$ is  constructed by comparing the supergravity transformation law of the supergravity multiplet with Eq.~(\ref{eq:density sugra}),
\begin{eqnarray}
2 {\cal E} = e\left[
1 + i \Theta \sigma \psi_a^\dag - \Theta^2 \left\{
M^\dag + \psi^\dag_a \bar{\sigma}^{ab}\psi^\dag_b
\right\} 
\right].
\end{eqnarray}
Note that ${\cal E}$ is scalar under the local Lorentz transformation.
Apparently, the diffeomorphism of ${\cal E}$ and $\Phi$ is given by
\begin{eqnarray}
\delta {\cal E } &=&  -\partial_M \left[ \eta^M {\cal E} \left(-\right)^{M} \right],~~ \delta \Phi = - \eta^M\partial_M \Phi,\nonumber \\
\eta^m (z) &=& \eta^m (x),~~ \eta^\alpha(z)=0.
\end{eqnarray}
Then, an action invariant under the diffeomorphism, the local Lorentz transformation, and the supergravity transformation is given by
\begin{eqnarray}
\int {\rm d}^4 x {\rm d}^2 \Theta 2 {\cal E} \Phi + {\rm h.c.}.
\end{eqnarray}

For example, the Einstein gravity and the kinetic term of the gravitino are given by
\begin{eqnarray}
S_{\rm EG} = - 3\Mpl^2 \int {\rm d}^4 x {\rm d}^2 \Theta 2 {\cal E} R + {\rm h.c.}.
\end{eqnarray}
The kinetic term of a chiral scalar multiplet $\Phi$ is given by
\begin{eqnarray}
S_{\rm kin} = - \frac{1}{8} \int {\rm d}^4 x {\rm d}^2 \Theta 2 {\cal E}
\left({\cal D}_{\dot{\alpha}}^\dag D^{\dot{\alpha}\dag} - 8R \right)
\Phi^\dag \Phi + {\rm h.c.}.
\end{eqnarray}

\section{Super-Weyl transformation}
\label{sec:SW}
In this section, we summarize the super-Weyl transformation law of various superfields and their components.

\subsection{Vielbein, connection and covariant derivative}
The super-Weyl transformation of the vielbein and the connection is the transformation of them such that
\begin{eqnarray}
\delta {E_M}^a \propto {E_M}^a
\end{eqnarray}
and the torsion constraint in Eq.~(\ref{eq:torsion constraint}) is preserved.
The super-Weyl transformation of the vielbein is given by
\begin{eqnarray}
\label{eq:SW vielbein}
\delta {E_M}^a &=& \left(\Sigma + \Sigma^\dag \right){E_M}^a,\nonumber\\
\delta {E_M}^\alpha &=&
\left(2\Sigma^\dag - \Sigma \right){E_M}^\alpha +
\frac{i}{2} {E_M}^a {\left(\epsilon \sigma_a \right)^\alpha}_{\dot{\alpha}} {\cal D}^{\dag \dot{\alpha}}\Sigma^\dag \nonumber \\
\delta {E_M}^{\dot{\alpha}} &=&
\left(2\Sigma - \Sigma^\dag \right){E_M}^{\dot{\alpha}} +
\frac{i}{2} {E_M}^a \sigma_a^{\dot{\alpha} \beta} {\cal D}_\beta\Sigma,
\end{eqnarray}
where $\Sigma$ is an arbitrary chiral multiplet.
The super-Weyl transformation of the connection is given by
\begin{eqnarray}
\label{eq:SW connection}
\delta {\phi_{MB}}^A &=& {E_M}^C {\Omega_{CB}}^A,\nonumber \\
\Omega_{\gamma \alpha \beta} &=& - \left( \epsilon_{\gamma \alpha} {\cal D}_\beta + \epsilon_{\gamma \beta} {\cal D}_\alpha  \right) \Sigma,~~
\Omega_{\dot{\gamma} \dot{\alpha}  \dot{\beta}} = \left( \epsilon_{\dot{\gamma} \dot{\alpha}} {\cal D}^\dag_{\dot{\beta}} + \epsilon_{\dot{\gamma} \dot{\beta}} {\cal D}^\dag_{\dot{\alpha}}  \right) \Sigma^\dag,
\nonumber\\
\Omega_{\gamma \dot{\alpha} \dot{\beta}} &=& \Omega_{\dot{\gamma} \alpha \beta} = 0,\nonumber \\
\Omega_{\gamma a b} &=& 2{\left( \sigma_{ab}  \right)_\gamma}^\beta {\cal D}_\beta \Sigma,~~
\Omega_{\dot{\gamma} ab} = -2 {\left( \bar{\sigma}_{ab}\right)^{\dot{\beta}}}_{\dot{\gamma}} {\cal D}^\dag_{\dot{\beta}} \Sigma^\dag    ,\nonumber\\
\Omega_{c\alpha \beta} &=& - \left( \sigma_{cb} \epsilon \right)_{\alpha\beta} {\cal D}^b \left( \Sigma + \Sigma^\dag \right),~~
\Omega_{c \dot{\alpha}\dot{\beta}} = \left( \epsilon \bar{\sigma}_{cb} \right)_{\dot{\alpha} \dot{\beta}} {\cal D}^b  \left( \Sigma + \Sigma^\dag \right),\nonumber\\
\Omega_{cab} &=& \left(\eta_{bc}{\cal D}_a - \eta_{ac}{\cal D}_b \right) \left( \Sigma + \Sigma^\dag \right).
\end{eqnarray}

\subsection{Supergravity multiplet}
The super-Weyl transformation of the gravity multiplet is obtained by calculating the transformation of torsions from Eqs.~(\ref{eq:SW vielbein}) and (\ref{eq:SW connection}).
The transformations of the superspace curvature $R$ and the superspace vector $G_{\alpha \dot{\alpha}}$ are given by
\begin{eqnarray}
\label{eq:SW RG}
\delta R &=& -  4 \Sigma R - \frac{1}{4} \left({\cal D}^{\dag 2} - 8R  \right)\Sigma^\dag,\nonumber \\
\delta G_{\alpha \dot{\alpha}} &=& - \left( \Sigma + \Sigma^\dag \right) G_{\alpha \dot{\alpha}} - i {\cal D}_{\alpha \dot{\alpha}}\left( \Sigma -\Sigma^\dag  \right).
\end{eqnarray}

By taking the lowest component of Eqs.~(\ref{eq:SW vielbein}) and (\ref{eq:SW RG}), we obtain the transformation law of component fields, 
\begin{eqnarray}
\label{eq:SW sugra}
\delta {e_m}^a &=& \left(\Sigma + \Sigma^\dag\right)| {e_m}^a,\nonumber\\
\delta {\psi_a}^\alpha &=& \left( \Sigma^\dag - 2 \Sigma \right)| {\psi_a}^\alpha - i {\bar{\sigma}_a}^{\dot{\alpha} \alpha} {\cal D}^\dag_{\dot{\alpha}} \Sigma^\dag|,\nonumber\\
\delta M &=& -2 \left( 2\Sigma - \Sigma^\dag \right)| M  + \frac{3}{2} {\cal D}^{\dag 2} \Sigma^\dag| , \nonumber\\
\delta b_{\alpha \dot{\alpha}} &=& -  \left(\Sigma + \Sigma^\dag\right)|b_{\alpha \dot{\alpha}} + 3 i {\cal D}_{\alpha \dot{\alpha}}\left( \Sigma -\Sigma^\dag  \right)|,
\end{eqnarray}
where ${e_m}^a$, ${\psi_a}^\alpha$, $M$ and $b_{\alpha \dot{\alpha}}$ are the vielbein, the gravitino, the auxiliary scalar, and the auxiliary vector, respectively. $X|$ denotes the lowest component of a superfield $X$.

\subsection{Chiral multiplet}

The super-Weyl transformation of a chiral multiplet $Q$ is defined by
\begin{eqnarray}
\delta Q = w Q,
\end{eqnarray}
where $w$ is the Weyl weight of $Q$.
By remembering the definition of component fields,
\begin{eqnarray}
A = Q|,~~ \chi_\alpha = \frac{1}{\sqrt{2}} {\cal D}_\alpha Q|,~~ F=-\frac{1}{4} {\cal D}^2 Q|,
\end{eqnarray}
and the transformation law of the connection in Eq.~(\ref{eq:SW connection}),
we obtain
\begin{eqnarray}
\label{eq:SW chiral}
\delta A &=& w \Sigma| A,\nonumber \\
\delta \chi_\alpha &=& \left(  \Sigma - 2 \Sigma^\dag \right)| \chi_\alpha + w \frac{1}{\sqrt{2}} {\cal D}_\alpha \left(\Sigma Q \right)|,\nonumber\\
\delta F &=&  2\left( \Sigma - 2 \Sigma^\dag \right)| F - \sqrt{2} {\cal D}^\alpha \Sigma| \chi_\alpha + w \frac{-{\cal D}^2}{4}\left(\Sigma Q\right)|
\end{eqnarray}

\subsection{Vector multiplet}

The Super-Weyl transformation of a vector multiplet $U$ is defined by
\begin{eqnarray}
\delta U = w' (\Sigma + \Sigma^\dag) U,
\end{eqnarray}
where $w'$ is the Weyl weight of $U$. Then the transformation of the chiral projection of $U$ is given by
\begin{eqnarray}
\delta \left[ \left({\cal D}^{\dag 2} - 8R \right) U \right] = \left( {\cal D}^{\dag 2} - 8R  \right) \left[ (w'-4) + (w'+2) \Sigma^\dag \right] U
\end{eqnarray}

\subsection{Gauge multiplet}

The super-Weyl transformation of a gauge multiplet $V$ is defined by
\begin{eqnarray}
\delta V = 0.
\end{eqnarray}
Then the superspace fields strength defined by
\begin{eqnarray}
W_\alpha \equiv -\frac{1}{4} \left({\cal D}^{\dag 2}- 8R\right) \left(
e^{-2V} {\cal D}_\alpha e^{2V}
\right),
\end{eqnarray}
transforms as 
\begin{eqnarray}
\delta W^\alpha = - 3 \Sigma W^\alpha.
\end{eqnarray}
From the definition of component fields,
\begin{eqnarray}
\lambda^\alpha =  \frac{i}{2}W^\alpha|,~~ D = -\frac{1}{4}{\cal D}^\alpha W_\alpha|,~~v_{\alpha\dot{\alpha}} = -\frac{1}{4}\left[ {\cal D}_\alpha, {\cal D}^\dag_{\dot{\alpha}} \right],
\end{eqnarray}
and Eq.~(\ref{eq:SW connection}), we obtain
\begin{eqnarray}
\label{eq:SW gauge} 
\delta \lambda_\alpha &=& - 3 \Sigma| \lambda_\alpha,\nonumber\\
\delta D &=& - 2 \left(\Sigma + \Sigma^\dag\right)| D , \nonumber\\
\delta v_{\alpha \dot{\alpha}} &=& -  \left(\Sigma + \Sigma^\dag\right )|v_{\alpha \dot{\alpha}}
\end{eqnarray}

\subsection{Transformation with a fixed chiral coordinate}

We may also consider the super-Weyl transformation of chiral superfields with fixing the definition of the chiral coordinate $(x,\Theta)$
while transforming components fields according to Eqs.~(\ref{eq:SW sugra}),~(\ref{eq:SW chiral}) and (\ref{eq:SW gauge}).
This transformation is convenient when the supergravity action is given in a chiral coordinate.

The expansion of the chiral density ${\cal E}$ is given by
\begin{eqnarray}
{\cal E} = \frac{1}{2}e \left(
1 + i \Theta \sigma^a \psi^\dag - \Theta^2 \left(M^\dag + \psi^\dag_a \sigma^{ab} \psi^\dag_b\right) 
\right).
\end{eqnarray}
We define the transformed chiral density ${\cal E}'$ by
\begin{eqnarray}
{\cal E}' = \frac{1}{2}e' \left(
1 + i \Theta \sigma^a \psi'^\dag - \Theta^2 \left(M'^\dag + \psi'^\dag_a \sigma^{ab} \psi'^\dag_b\right) 
\right),
\end{eqnarray}
where $X = X' + \delta X$ for a component field $X$.
Then the transformation of ${\cal E}$ is written as
\begin{eqnarray}
\label{eq:chiral SW1}
\delta {\cal E} &=& {\cal E} - {\cal E}' = 
6 \Sigma {\cal E} + \frac{\partial}{\partial \Theta^\alpha}\left(S^\alpha {\cal E}\right), \nonumber\\
S^\alpha &\equiv& \Theta^\alpha \left( 2 \Sigma^\dag - \Sigma \right)| +\Theta^2 {\cal D}^\alpha \Sigma|.
\end{eqnarray}
In the similar way, we obtain transformation laws of  $R$, $Q$ and the chiral projection of $U$ as
\begin{eqnarray}
\label{eq:chiral SW2}
\delta R &=&  -4 \Sigma R  - \frac{1}{4}\left( {\cal D}^{\dag 2} - 8 R\right) \Sigma^\dag - S^\alpha \frac{\partial}{\partial \Theta^\alpha}R\ ,\\
\delta  Q &=& w\Sigma Q - S^\alpha \frac{\partial}{\partial \Theta^\alpha}Q\ , \nonumber\\
\delta \left[ \left({\cal D}^{\dag 2} - 8R \right) U \right] &=& \left( {\cal D}^{\dag 2} - 8R  \right) \left[ (w'-4) + (w'+2) \Sigma^\dag \right] U - 
S^\alpha \frac{\partial}{\partial \Theta^\alpha} \left[ \left({\cal D}^{\dag 2} - 8R \right) U \right]. \nonumber
\end{eqnarray}
The transformation given in Eqs. (\ref{eq:chiral SW1}) and (\ref{eq:chiral SW2}) is the one we use in Sec.~\ref{sec:AMSB}.

\section{Super-diffeomorphism invariant measure}
\label{sec:Sdiff measure}
In this appendix, we show that the measure given in Eq.~(\ref{eq:SD measure}) 
is invariant under the super-diffeomorphism.%
\footnote{
As we have mentioned in Sec.~\ref{sec:measure}, the ``super-diffeomorphism" given in Eqs.~(\ref{eq:Sdiff}) and (\ref{eq:Sdiff2})
is not the full super-diffeomorphism, but a part of it.
}

\subsection{Diffeomorphism}

Before discussing supergravity, we construct the diffeomorphism invariant measure of
a real scalar field $\phi(x)$ and a Weyl fermion $\psi(x)$~\cite{Fujikawa:1984qk}.
Under an infinitesimal diffeomorphism, $\phi$, $\psi$ and a vielbein ${e_m}^a$ transforms as
\begin{eqnarray}
\phi (x) &\rightarrow& \phi'(x) = \phi(x) - \xi^n(x) \partial_n \phi(x),\nonumber\\
\psi (x)& \rightarrow& \psi'(x) = \psi(x)  - \xi^n(x) \partial_n \psi(x),\nonumber\\
{e_m}^a(x) &\rightarrow& {e_m}^{a'}(x) =  {e_m}^a (x)  - \xi^n(x) \partial_n {e_m}^a(x) - \left(\partial_m \xi^n\left(x\right)\right) {e_n}^a(x),
\end{eqnarray}
where $\xi^m$ parameterize the diffeomorphism. 
Then the determinant of the vielbein $e\equiv {\rm det}({e_m}^a)$ transforms as
\begin{eqnarray}
e(x) \rightarrow e'(x) = e(x)  - \partial_m \left( \xi^m e (x) \right).
\end{eqnarray}

Let us first discuss the measure of the scalar $\phi(x)$.
In order to define the path-integral measure, we consider a complete set $\{\phi_n\}$ normalized as
\begin{eqnarray}
\int {\rm d}^4 x \phi_m(x) \phi_n(x) = \delta_{nm}.
\end{eqnarray}
The following field,
\begin{eqnarray}
\tilde{\phi} = e^{1/2} \phi,
\end{eqnarray}
transforms under the diffeomorphism as
\begin{eqnarray}
\tilde{\phi} \rightarrow \tilde{\phi}' =  \tilde{\phi}- \xi^m \partial_m \tilde{\phi} - \frac{1}{2} \left(\partial_m \xi^m\right) \tilde{\phi}.
\end{eqnarray}
We expand $\tilde{\phi}$ by $\{\phi_n\}$,
\begin{eqnarray}
\tilde{\phi}(x) = \sum_n a_n  \phi_n(x),
\end{eqnarray}
and define the path-integral measure by
\begin{eqnarray}
\label{eq:measure scalar}
[D \tilde{\phi}] \equiv \prod_n {\rm d}a_n.
\end{eqnarray}

Let us calculate the Jacobian in the transformation of the measure.
$a_n$ and $a_n'$ are related by
\begin{eqnarray}
a_n' &=& \int {\rm d}^4 x \tilde{\phi}' (x) \phi_n(x)\nonumber\\
 &=& a_n - \int {\rm d}^4 x \phi_n(x) \left[
\xi^m(x) \partial_m \tilde{\phi} (x) + \frac{1}{2}\left(\partial_m \xi^m\right) \tilde{\phi}(x)
\right]\nonumber\\
&=& \left[
\delta_{nk} -
\int {\rm d}^4 x  \phi_n(x) 
\left\{
\xi^m(x) \partial_m + \frac{1}{2} \left( \partial_m \xi^m \left(x\right) \right)
\right\}\phi_k(x)
\right]a_k \nonumber \\
&\equiv & (\delta_{nk} + M_{nk} )a_k.
\end{eqnarray}
Then the Jacobian is given by
\begin{eqnarray}
\prod_m {\rm d}a_m' &=& \prod_m {\rm d}a_m \times J,\\
\label{eq:expansion J}
{\rm ln} J &=& {\rm ln~det}(\delta_{nk}  +M_{nk}) =  {\rm tr}(M_{nk}) \nonumber \\
&=& - \sum_n\int {\rm d}^4 x  \phi_n(x) 
\left\{
\xi^m(x) \partial_m + \frac{1}{2} \left( \partial_m \xi^m \left(x\right) \right)
\right\}\phi_n(x) \nonumber \\
&=& - \frac{1}{2}\sum_n \int {\rm d}^4 x \partial_m \left\{
\xi^m \phi_n \phi_n (x)
\right\}\nonumber \\
&=& 0.
\end{eqnarray}
In  the last line, we have assumed that $\xi^m(x)$ vanishes sufficiently fast at infinity.
We have shown that the measure given in Eq.~(\ref{eq:measure scalar}) is invariant under the diffeomorphism.

For later convenience, we show the invariance of the measure in a slightly differrent way.
The transformation of the measure in Eq.~(\ref{eq:measure scalar}) is given by
\begin{eqnarray}
\label{eq:functional J}
[D \tilde{\phi}'] &=& \prod_x [d \tilde{\phi}'(x)] = \prod_{x} [d \tilde{\phi}(x)] {\rm Det}_{x' x''} \frac{\delta  \tilde{\phi}'(x'') }{\delta \tilde{\phi} (x') } \nonumber \\
&=& [D \tilde{\phi}] {\rm Det}_{x' x} \left[
1 - \xi^m(x) \partial_m  - \frac{1}{2} \partial_m \xi^m(x)
\right] \delta^4 \left(x'-x\right)\nonumber \\
&=&  [D \tilde{\phi}] J, \nonumber \\
{\rm ln}J &=& -{\rm Tr}_{x'x} \left[  \xi^m(x) \partial_m  + \frac{1}{2} (\partial_m \xi^m(x)) \right]\delta^4 \left(x'-x\right)\nonumber \\
&=&- \int {\rm d}^4 x {\rm d}^4 x' \delta^4 \left(x'-x\right)  \left[  \xi^m(x) \partial_m  + \frac{1}{2} (\partial_m \xi^m(x)) \right]\delta^4 \left(x'-x\right).
\end{eqnarray}
The product of delta functions is handled by the completeness relation,
\begin{eqnarray}
\sum_n \phi_n(x) \phi_n(x') = \delta^4 (x'-x).
\end{eqnarray}
By replacing one of the delta function in Eq.~(\ref{eq:functional J}), we obtain ${\ln J} =0$.

Similarly, the diffeomorphism invariant measure of the Weyl fermion $\psi$ is given by
\begin{eqnarray}
[D\tilde{\psi}][D\tilde{\psi}^\dag],\\
\tilde\psi (x) \equiv e^{1/2}(x) \psi(x).
\end{eqnarray}
The diffeomorphism invariance of the measure can be proven in the same way as the case of the scalar.
A crucial difference is that the cancellation like that in Eq.~(\ref{eq:expansion J})
occurs between the contribution from $[D\tilde{\psi}]$ and $[D\tilde{\psi}^\dag]$.

Finally, we show the diffeomorphism invariance of the measure $[D\tilde{\phi}]$ from a very definition of the path-integral measure.
We define a path-integral measure of scalar fields by
\begin{eqnarray}
\int [D C]  {\rm exp}\left[ \frac{1}{2} i A \int {\rm d}^4x C(x)^2 \right] = N,
\end{eqnarray}
where $A$ and $N$ are constants.
Then the measure $[D \tilde{\phi}]$ satisfies
\begin{eqnarray}
\int [D \tilde{\phi}]  {\rm exp}\left[ \frac{1}{2} i A \int {\rm d}^4x \tilde{\phi}(x)^2 \right] 
= \int [D \tilde{\phi}]  {\rm exp}\left[ \frac{1}{2} i A \int {\rm d}^4x e \phi(x)^2 \right] 
= N.
\end{eqnarray}
Since the exponent as well as $N$ are diffeomorphism invariant, the measure $[D \tilde{\phi}]$ is also diffeomorphism invariant.

\subsection{Super-diffeomorphism}
Next, let us discuss the super-diffeomorphism invariance of the measure in Eq.~(\ref{eq:SD measure}),
\begin{eqnarray}
[D Q_{\rm diff}] = 
[D\left(2{\cal E}\right)^{1/2}Q].
\end{eqnarray}
Under the transformation given in Eq.~(\ref{eq:Sdiff}),
\begin{eqnarray}
\label{eq:Sdiff2}
Q \rightarrow Q'  &=& Q - \eta^M(x,\Theta)\partial _M Q\ ,\nonumber \\
{\cal E} \rightarrow {\cal E}' &=& {\cal E}- \eta^M(x,\Theta)\partial _M {\cal E} - (-)^M \left(\partial_M \eta^M\left(x, \Theta\right)\right){\cal E}\ ,
\end{eqnarray}
the variable $Q_{\rm diff}$ transforms as
\begin{eqnarray}
Q_{\rm diff} \rightarrow Q_{\rm diff}' &=& Q_{\rm diff}- \eta^M(x,\Theta)\partial _M Q_{\rm diff} - \frac{1}{2}(-)^M \left(\partial_M \eta^M\left(x, \Theta\right)\right)Q_{\rm diff}\ .
\label{eq:QdiffSD}
\end{eqnarray}
Then, the path-integral measure $[DQ_{\rm diff}]$ transforms as (see Eq.~(\ref{eq:functional J})),
\begin{eqnarray}
[DQ_{\rm diff}'] &=& [DQ_{\rm diff}]\times {\rm exp}\left[\,{\rm sTr}_{z',z}\,{\cal O}(z',z)\,\right]\ ,\\
\label{eq:Ozz}
{\cal O}(z',z) &\equiv& - \left[ \eta^M\partial_M + \frac{1}{2} (-1)^M \left(\partial_M\eta^M\right) \right]\delta^6(z'-z)\ ,\\
\label{eq:sTr}
{\rm sTr}\,{\cal O}(z',z) &=& \int {\rm d}^6z {\rm d}^6z' \delta^6 (z'-z) {\cal O}(z',z)\ .
\end{eqnarray}
where we have collectively represented $x$ and $\Theta$ by $z$.
A naive conclusion is that the super-trace vanishes due to the saturation of Grassmann 
variables $\Theta$ and $\Theta'$ from the delta functions in Eqs.\,(\ref{eq:Ozz}) and (\ref{eq:sTr}).
However, since there is also a factor of $\delta^4(x'-x)$, which is well-defined only after integrating over $x$ or $x'$, 
one should carefully investigate the integration.

To examine the the integration, let us expand the delta function by plane waves,
\begin{eqnarray}
\delta^6 (z'-z) &=& \int \frac{{\rm d}^4 k}{(2\pi)^4} d^2 \tau \Psi_{-k,-\tau}(z') \Psi_{k,\tau}(z)\ ,\\
\Psi_{k,\tau}(z) &\equiv& {\rm exp}(ikx+ 2i \tau \Theta)\ .
\end{eqnarray}
By substituting this expression into Eq.\,(\ref{eq:sTr}), the above super-trace is expressed by,
\begin{eqnarray}
{\rm sTr}\,{\cal O}(z',z) = - \int {\rm d}^6z \int \frac{{\rm d}^4 k}{(2\pi)^4} d^2 \tau \Psi_{-k,-\tau}(z)
\left[
\eta^M \partial_M + \frac{1}{2} (-)^M \left(\partial_M\eta^M\right)
\right]
\Psi_{k,\tau}(z)\ .
\end{eqnarray}
Now, let us notice an identity,
\begin{eqnarray}
\label{eq:identity}
\int {\rm d}^6z \Psi_{k,\eta}(z)
\left[
\eta^M \partial_M + \frac{1}{2} (-)^M \left(\partial_M\eta^M\right)
\right]
\Psi_{k,\eta}(z)&=& \frac{1}{2}(-)^M \int {\rm d}^6 z \partial_M 
\left[\Psi_{k,\eta}\left(z\right)\eta^M\Psi_{k,\eta}\left(z\right)\right]\nonumber\\
&=&0\ ,
\end{eqnarray}
where we have used the property that an integration of a total derivative vanishes.
By using this identity several times, the super-trace can be rearranged as
\begin{eqnarray}
{\rm sTr}\,{\cal O}(z',z)  &=& -\frac{1}{2}\int {\rm d}^6z \int \frac{{\rm d}^4 k}{(2\pi)^4} d^2 \tau \left(\Psi_{k,\tau}\left(z\right)+\Psi_{-k,-\tau}\left(z\right)\right)\nonumber\\
&&
\left[
\eta^M \partial_M + \frac{1}{2} (-)^M \left(\partial_M\eta^M\right)
\right]
\left(\Psi_{k,\tau}\left(z\right)+\Psi_{-k,-\tau}\left(z\right)\right)\nonumber\\
&=&
-\frac{1}{4}(-)^M\int {\rm d}^6z \int \frac{{\rm d}^4 k}{(2\pi)^4} d^2 \tau \partial_M
\left[
\left(\Psi_{k,\tau}\left(z\right)+\Psi_{-k,-\tau}\left(z\right)\right)
\eta^M
\left(\Psi_{k,\tau}\left(z\right)+\Psi_{-k,-\tau}\left(z\right)\right)
\right]\nonumber\\
&=&0\ .
\end{eqnarray}
This shows that the measure given in Eq.~(\ref{eq:SD measure}) is actually invariant under the super-diffeomorphism.
It should be noted that the transformation law in Eq.\,(\ref{eq:QdiffSD}) is crucial to use Eq.\,(\ref{eq:identity}),
and hence, the super-diffeomorphism invariance does not hold for measures 
with different weights $[D(2{\cal E})^{n}Q]$ $(n \neq 1/2)$.
In fact, the super-diffeomorphism transformation of  $[D(2{\cal E})^{n}Q]$ $(n \neq 1/2)$
is accompanied by Konishi-Sizuya anomaly~\cite{Konishi:1985tu}.
This argument provides an superfield expression of the arguments in Ref.~\cite{Fujikawa:1984qk} reviewed in the provious subsection.

As is the case with the diffeomorphism invariance,
there is a quicker route to show the super-diffeomorphisim invariance of $[DQ_{\rm diff}]$.
The path-integral measure of superfields in the chiral superspace is defined by
\begin{eqnarray}
\label{eq:path integral}
\int [D C ] {\rm exp} \left[ \frac{i}{2} A\int {\rm d}^6 z  C^2 \right] =N\ ,
\end{eqnarray}
where $A$ and $N$ are normalization constants.
Then the measure $[D Q_{\rm diff}]$ satisfies
\begin{eqnarray}
\int [D Q_{\rm diff} ] {\rm exp} \left[ \frac{i}{2} \int {\rm d}^6 z Q_{\rm diff}^2 \right] =
 \int [D Q_{\rm diff} ] {\rm exp} \left[ \frac{i}{2} \int {\rm d}^6 z\, 2{\cal E}\, Q Q  \right] = 
N\ .
\end{eqnarray}
Now, since $\int {\rm d}^6z 2 {\cal E}QQ$ is invariant under 
the super-diffeomorphism as $Q$ is a chiral scalar multiplet,
so is the path-integral measure $[D Q_{\rm diff}]=[D(2{\cal E})^{1/2}Q]$.

In fact, under the super-diffeomorphism, 
\begin{eqnarray}
{\cal E'} = {\cal E} - \delta_{\rm SD} {\cal E}\ ,~~Q' = Q - \delta_{\rm SD} Q\ ,
\end{eqnarray}
we have the following identities
\begin{eqnarray}
N &=& \int [D \left(2 {\cal E}\right)^{1/2} Q ] {\rm exp} \left[ \frac{i}{2} \int {\rm d}^6 z\, 2{\cal E} \,Q Q  \right]\ , \nonumber\\
&=&\int [D \left(2 {\cal E'}\right)^{1/2} Q' ] {\rm exp} \left[ \frac{i}{2} \int {\rm d}^6 z \,2{\cal E'}\, Q' Q'  \right] \ , \nonumber\\
&=&\int [D \left(2 {\cal E'}\right)^{1/2} Q' ] {\rm exp} \left[ \frac{i}{2} \int {\rm d}^6z  \,2{\cal E}\, Q Q  \right]\ .
\end{eqnarray}
Here, the second equality is just a change of variable.
We have used a super-diffeomorphism invariance of the exponent in the third equality. 
Thus, from these identities, we find that 
\begin{eqnarray}
[D \left(2 {\cal E'}\right)^{1/2} Q' ] = D [\left(2 {\cal E}\right)^{1/2} Q ]\ ,
\end{eqnarray}
which again shows the super-diffeomorphism invariance of the measure $[DQ_{\rm diff}]$.
In the same token, we can derive the super-diffeomorphism invariance of the measure 
of a scalar multiplet $V$ in a real superspace,
\begin{eqnarray}
[D V_{\rm diff}] = [DE^{1/2} V]\ .
\end{eqnarray}

\section{Anomaly mediated gaugino mass in SUSY breaking with singlet}
\label{sec:AMSB singlet}

In this section, we discuss the anomaly mediated gaugino mass when SUSY is broken by $F$ terms of singlet chiral fields.

\subsection{Classical action}
We start from the supergravity action with the super-Weyl compensator $C$,
\begin{eqnarray}
\label{eq:SW classical action2}
{\cal L} &=& \int {\rm d}^2 \Theta\,2 {\cal E}  \frac{3}{8} \left({\cal D}^{\dag2} - 8R\right)C C^\dag{\exp}\left[-K/
3\right] 
 + 
\frac{1}{16g^2}\int {\rm d}^2 \Theta\, 2{\cal E}\, W^\alpha W_\alpha + {\rm h.c.}.
\end{eqnarray}
Here, we take the unit with $\Mpl=1$.
We parametrize the components of $C$ as
\begin{eqnarray}
 C = \phi + \sqrt{2}\Theta \chi + \Theta\Theta F, 
\end{eqnarray}
in the chiral coordinate.
Note that these components are gauge degree of freedoms, and
we can freely choose them.
$\phi$
can be used to choose the frame of gravity theory.
The most convenient choice is,
\begin{eqnarray}
 \phi = {\rm exp}[K/6],
\end{eqnarray}
which yields the Einstein frame gravity.
$\chi$
can be chosen in order to remove
mixings between the gravitino and chiral matter fields.

In order to find a convenient choice for $F$,
let us solve
the equation of motion of the auxiliary component of the gravity multiplet
and the chiral matter multiplets, $M$ and $F^i$.
From Eqs. (\ref{eq:expansion}), (\ref{eq:chiral projection}) and (\ref{eq:SW classical action2}), Lagrangian terms which depend on
$M$, $F^i$ and $F$ are given by
\begin{eqnarray}
 e^{-1}{\cal L} &=& e^{-K/3}\left[
|\phi|^2 K_{i\bar{i}}F^iF^{\bar{i}\dag} - 3 |F-\frac{1}{3}\phi K_iF^i|^2
 -\frac{1}{3}|\phi|^2|M|^2
 \right.\nonumber\\
&&\left. 
+\phi^\dag (F-\frac{1}{3}\phi K_iF^i) M +\phi (F-\frac{1}{3}\phi K_iF^i)^\dag M^\dag
\right]\nonumber\\
&& + \left[3\phi^2 F W + \phi^3 W_i F^i -M^\dag \phi^3 W + {\rm h.c.}\right]
\end{eqnarray}
By solving the equation of motion of $M$ and $F^i$, we obtain
\begin{eqnarray}
 M &=& 3(\phi^\dag)^{-1}\left(F^\dag-e^{K/3} \phi^2W\right) -K_{\bar{i}}F^{\bar{i}\dag}\\
F^i &=& -K^{ i{\bar{j}}} \phi^{-1}\phi^{\dag 2}e^{K/3} \left[
W_{\bar{j}}^\dag+K_{\bar{j}}W^\dag
\right]
\end{eqnarray}
In this formulation, it is NOT necessary to solve the equation of motion of $F$,
because it is a gauge degree of freedom.

We take a gauge in which $M=0$,
 that is,
\begin{eqnarray}
 F= e^{K/3}\phi^{\dag 2}W^\dag + \frac{1}{3}\phi K_iF^i.
\end{eqnarray}
This choice simplifies calculation and
hence is frequently adopted in the literature.
In the gauge where $\phi = {\rm exp}(K/6)$ and $M=0$, $F^i$ and $F$ are given by
\begin{eqnarray}
F^i &=&  - e^{K/2}K^{ i{\bar{j}}}  \left[
W_{\bar{j}}^\dag+K_{\bar{j}}W^\dag
\right],\\
\label{eq:compensator F}
F &=& e^{K/6} \left[
e^{K/2} W^\dag + \frac{1}{3}K_iF^i
\right] = e^{K/6} \left[
m_{3/2} + \frac{1}{3}K_iF^i
\right] .
\end{eqnarray}
We adopt this gauge in the following.

\subsection{Anomaly mediation from the compensator}

Note that the Lagrangian in Eq.~(\ref{eq:SW classical action2}) possesses the following formal super-Weyl symmetry,
\begin{eqnarray}
C' &=& e^{-2\Sigma} C,\nonumber\\
{\cal E}' &=& e^{6\Sigma}{\cal E} + \cdots,\nonumber\\
R' &=& - \frac{1}{8}e^{-4 \Sigma} \left( {\cal D}^{\dag 2} - 8 R \right) e^{2 \Sigma^\dag},\nonumber\\
W^{'\alpha} &=& e^{-3\Sigma} W^\alpha \nonumber \\
Q^{'i} &=& Q^i, \nonumber \\
{\cal D}'_\alpha &=& e^{\Sigma - 2\Sigma^\dag} {\cal }{\cal D}_\alpha + \cdots.
\end{eqnarray}
This symmetry must be anomaly free. Otherwise, one cannot gauge away the super-Weyl compensator $C$ to come back to the original theory without $C$.

To see the anomaly of the formal super-Weyl symmetry, let us consider the $U(1)_R$ part, $\Sigma = i \alpha$ where $\alpha$ is a real constant. Fermion parts of the chiral multiplet, $\chi^i$, and the gaugino $\lambda$ transforms as 
\begin{eqnarray}
\chi^{i'}_\alpha = \frac{1}{\sqrt{2}}({\cal D}_\alpha Q^i)'| = e^{3 i \alpha} \chi_\alpha^i, \nonumber \\
\lambda'_\alpha = \frac{i}{2} W_\alpha'| = e^{-3i \alpha } \lambda_\alpha.
\end{eqnarray}
This transformation is anomalous;
\begin{eqnarray}
\label{eq:Ranomaly}
[D\chi^i] [D\chi^{\bar{i}\dag}] [D \lambda] [D \lambda^\dag] = [D\chi^{'i}] [D\chi^{'\bar{i}\dag}] [D \lambda'] [D \lambda^{'\dag}] e^{i\Delta S_{\rm ano}} ,\nonumber \\
\Delta S_{\rm ano} =  \alpha\int {\rm d}^4 x\frac{1}{32\pi^2} \epsilon^{abcd}F_{ab}^AF_{cd}^A
\times3\left(T_G -\sum_i T_i\right),
\end{eqnarray}
where $T_G$ and $T_i$ are the Dynkin index of the adjoint
representation and the chiral multiplet $Q^i$, respectively.
In the superspace, Eq.~(\ref{eq:Ranomaly}) is expressed as
\begin{eqnarray}
\label{eq:superanomaly}
 \Delta  S_{\rm ano} = i \alpha \int {\rm d}^4 x {\rm d}^2 \Theta  2 {\cal
  E} \frac{1}{128\pi^2}
  W^{A\alpha} W^A_\alpha (3T_G - 3 \sum_i T_i) + {\rm h.c.}.
\end{eqnarray}
This anomaly is cancelled by adding a counter term
\begin{eqnarray}
\label{eq:local counter term}
 S_{\rm c.t.} = - \frac{1}{256\pi^2}\int {\rm d}^4 x {\rm d}^2 \Theta  2 {\cal
  E} {\rm ln}C
  W^{A\alpha} W^A_\alpha (3T_G - 3\sum_i T_i) + {\rm h.c.},
\end{eqnarray}
whose variation under the $U(1)_R$ symmetry cancels with
Eq.~(\ref{eq:superanomaly}).%
\footnote{This counter term is induced by the anomaly of the super-Weyl transformation performed to introduce the super-Weyl compensator.}

Finally, let us separate the chiral multiplets from the super-Weyl compensator by a
transformation
\begin{eqnarray}
 Q_s^i = C Q^i.
\end{eqnarray}
This transformation is anomalous,
\begin{eqnarray}
\label{eq:cano anomaly}
[D Q^i] [D Q^{i \dag}] = [D Q_s^i] [D Q_s^{i \dag}] e^{i \Delta S_c }, \nonumber \\
\Delta S_s =  \frac{1}{128\pi^2}\int {\rm d}^4 x{\rm d}^2 \Theta  2 {\cal
  E} {\rm ln}C
  W^{A\alpha} W^A_\alpha \sum_i T_i + {\rm h.c.}.
\end{eqnarray}
Adding Eqs.~(\ref{eq:local counter term}) and (\ref{eq:cano anomaly}),
we obtain
\begin{eqnarray}
 \label{eq:compensator}
 \Delta S = \frac{1}{256\pi^2}\int {\rm d}^4x{\rm d}^2 \Theta  2 {\cal
  E} {\rm ln}C
  W^{A\alpha} W^A_\alpha (3T_G - \sum_i T_i) + {\rm h.c.}.
\end{eqnarray}

From Eq.~(\ref{eq:compensator F}) and (\ref{eq:compensator}),
the gaugino mass from the $F$ term of the super-Weyl compensator is given by
\begin{eqnarray}
\label{eq:gaugino1}
 (M_{\lambda}/ g^2)_{\rm SW} = \frac{1}{16\pi^2}(\sum_i T_i -3 T_G) \left[
m_{3/2} + \frac{1}{3}K_iF^i
\right] .
\end{eqnarray}
When SUSY is broken by $F$ terms of singlet chiral fields, $\vev{K_i F^i} =O(m_{3/2})$ in general.

\subsection{Anomaly mediation from SUSY breaking field}

Let us consider the coupling of $F$ terms of SUSY breaking fields with $Q_s^i$ in the Kahler potential.
We denote the SUSY breaking fields and their $F$ terms as $Z^I$ and $F^I$.
From Eq.~(\ref{eq:SW classical action2}), couplings between the $F$ terms and $Q_s^i$ are given by
\begin{eqnarray}
\int {\rm d}^2 \theta {\rm d}^2 \theta^\dag Q_s^i Q_s^{\bar{j}\dag}
\left[
K_{i\bar{j}}\left( 1  - \frac{1}{3} K_I F^I \theta^2  - \frac{1}{3} K_{\bar{I} }F^{\bar{I}\dag} \theta^{\dag2}\right)
 + K_{i\bar{j} I} F^I \theta^2  + K_{i\bar{j} \bar{I}} F^{\bar{I}\dag} \theta^{\dag2}
\right].
\end{eqnarray}
Here, $Q^i$ is a chiral multiplet while $K_{i\bar{j}}$, $K_I$ and $K_{i\bar{j} I}$ are lowest components of the corresponding superfields.
Then canonically normalized fields are given by
\begin{eqnarray}
Q_c^l &=& Q_s^i {U_i}^k c_k^{1/2}\left[
\left( 1 - \frac{1}{3}K_I F^I \theta^2\right) {\delta_k}^l 
+ c_k^{-1/2} {U_k^{\dag}}^i K_{i\bar{j}I} F^I U^{\bar{j}\bar{l}} c_l^{-1/2} \theta^2\right],\nonumber\\
K_{i\bar{j}} &\equiv& {U_i}^k c_k \delta_{k\bar{l}}{U^{\dag \bar{l}}}_{\bar{j}},
\end{eqnarray}
where $U$ is the unitary matrix and $c_k$ are positive real constant.
Canonicallization induces the $F$ term of the gauge kinetic function,
\begin{eqnarray}
\prod_i[D Q_s^i] &=& \prod_i [D Q_c^i] \times e^{i \Delta S_c}, \\
\Delta S_c &=& \frac{1}{128\pi^2}\int {\rm d}^4x{\rm d}^2 \Theta  2 {\cal
  E} \left[
  - \frac{1}{3} K_I F^I \sum_i T_i \theta^2  + \sum_iT_i  ({\rm ln~det} K_{i\bar{j}})_I F^I \theta^2
  \right]
  W^{A\alpha} W^A_\alpha+ {\rm h.c.}.\nonumber
\end{eqnarray}
Here, we have used the identity
\begin{eqnarray}
{\rm ln}~{\rm det}_{kl} \left[ {\delta_k}^l +c_k^{-1/2} {U_k^{\dag}}^i K_{i\bar{j}I} F^I U^{\bar{j}\bar{l}} c_l^{-1/2} \theta^2 \right]
=
{\rm tr}{\rm ln}_{kl} \left[ {\delta_k}^l +c_k^{-1/2} {U_k^{\dag}}^i K_{i\bar{j}I} F^I U^{\bar{j}\bar{l}} c_l^{-1/2} \theta^2  \right]
\nonumber \\
 =
 {\rm tr}_{kl} \left[ c_k^{-1/2} {U_k^{\dag}}^i K_{i\bar{j}I} F^I U^{\bar{j}\bar{l}} c_l^{-1/2} \theta^2  \right]=
({\rm tr}~{\rm ln } K_{i\bar{j}})_I F^I \theta^2  = ({\rm ln }~{\rm det} K_{i\bar{j}})_I F^I \theta^2.
\end{eqnarray}
Then the gaugino mass is given by
\begin{eqnarray}
\label{eq:gaugino2}
(M_\lambda / g^2)_{\rm SUSY} =  \frac{1}{8\pi^2}\frac{1}{3} K_I F^I \sum_i T_i -  \frac{1}{8\pi^2} \sum_iT_i  ({\rm ln~det} K_{i\bar{j}})_I F^I
\end{eqnarray}
\\

We note that it is crucial to start from the path-integral measure $[D Q^i]$.
If one starts from measures $[DQ^i f(Z)]$ with a generic function $f$, one obtains different results from Eq.~(\ref{eq:gaugino2}).
Assuming the measure $[D Q^i]$, adding Eqs.~(\ref{eq:gaugino1}) and (\ref{eq:gaugino2}), we obtain
\begin{eqnarray}
M_\lambda /  g^2 =
\frac{1}{16\pi^2} (\sum_i T_i -3 T_G)m_{3/2} + 
\frac{1}{16\pi^2} (\sum_i T_i - T_G )  K_I F^I -
\frac{1}{8\pi^2} \sum_iT_i  ({\rm ln~det} K_{i\bar{j}})_I F^I,\nonumber\\
\end{eqnarray}
which is consistent with Ref.~\cite{Bagger:1999rd}.

\section{Annihilation of gluino}
\label{sec:gluino ani}

In this section, we derive the leading order annihilation cross section of the gluino used in Sec.~\ref{subsec: bino-gluino}.
We formulate the Sommerfeld effect with the non-relativistic effective theory of the gluino.

\subsection{Non-relativistic effective action}
Let us first derive a non-relativistic effective actin of the gluino.
We assume that squarks are heavy enough that they do not affect the interaction of the gluino except for its decay.
The interaction of the gluino $\widetilde{G}^a$ with the gluon $G_\mu^a$ is described by the action,
\begin{equation}
S=\int {\rm d}^4x\left[-\frac{1}{4}F_{\mu\nu}^{a}F^{\mu\nu}_{a}-\frac{1}{2}(\partial G^a)^2+\frac{1}{2}\overline{\widetilde{G}}^a(i\slashed{\partial}-M)\widetilde{G}^a+\frac{i}{2}g_{s}f^{abc}\overline{\widetilde{G}}^a\slashed{G}^b\widetilde{G}^c\right],
\end{equation}
where
\begin{equation}
F_{\mu\nu}^{a}=\partial_{\mu}G_{\nu}^a-\partial_{\nu}G_{\mu}^a+g_{s}f^{abc}G_{\mu}^bG_{\nu}^c,
\end{equation}
and $M$ and $g_s$ are the mass of the gluino and the $SU(3)$ gauge coupling constant, respectively.
Here, we have taken the Feynman - 't Hooft gauge and the gluino is expressed as a Majorana field.
Ghost fields are irrelevant for the leading order calculation.

In the calculation of the annihilation of the gluino, the time-like degree of freedom of the gluino field
and the soft degree of freedom of the gluon field do not appear in external lines.
By integrating them out, we obtain
\begin{eqnarray}
\label{eq:Raction}
S_{\rm eff} & = & \int {\rm d}^4x\left[
-\frac{1}{4}F_{\mu\nu}^{a}F^{\mu\nu}_{a}
-\frac{1}{2}(\partial G^a)^2
+\frac{1}{2}\overline{\widetilde{G}}^a(i\slashed{\partial}-M)\widetilde{G}^a
+\frac{i}{2}g_{s}f^{abc}\overline{\widetilde{G}}^a\slashed{G}^b \widetilde{G}^c\right] \nonumber \\
& + & \int {\rm d}^4x{\rm d}^4y\Big[
-\frac{i}{2}g_s^2f^{abc}f^{ade}\overline{\widetilde{G}}^d(y)\slashed{G}^e(y)S_{F}(y-x)\slashed{A}^b(x)\widetilde{G}^c(x) \nonumber \\
& &
-\frac{i}{8}g_s^2f^{abc}f^{ade}\overline{\widetilde{G}}^d(y)\gamma_{\mu}\widetilde{G}^e(y)D_{F}^{\mu\nu}(y-x)\overline{\widetilde{G}}^b(x)\gamma_{\nu}\widetilde{G}^c(x) \Big],
\end{eqnarray}
where
\begin{eqnarray}
S_F(y-x) & = & \int_\text{time-like} {\rm d}^4pe^{-ip(y-x)}\frac{i(\slashed{p}+M)}{p^2-M^2+i\epsilon}\\
D_F^{\mu\nu}(y-x) & = & \int_\text{soft} {\rm d}^4pe^{-ip(y-x)}\frac{-ig_{\mu\nu}}{p^2+i\epsilon}.
\end{eqnarray}

In the Dirac representation, the non-relativistic gluino field $\chi$ is defined by
\begin{equation}
\label{eq:NRgluino}
\tilde{G}^a=
\left(
\begin{array}{c}
e^{-imt}\chi^a+ie^{imt}\frac{{\bf \nabla}\cdot {\bf \sigma}}{2M}(\chi^a)^c \\
e^{imt}(\chi^a)^c-ie^{-imt}\frac{{\bf \nabla}\cdot {\bf \sigma}}{2M}\chi^a
\end{array} 
\right),
\end{equation}
where the charge conjugation is defined by
\begin{equation}
(\chi^a)^c=i\sigma^2(\chi^a)^{\ast}.
\end{equation}
By substituting Eq.~(\ref{eq:NRgluino}) into Eq.~(\ref{eq:Raction}), we obtain
\begin{eqnarray}
S_{\rm eff} & = & \int {\rm d}^4x\Big[
-\frac{1}{4}F_{\mu\nu}^{a}F^{\mu\nu}_{a}
-\frac{1}{2}(\partial G^a)^2
+\chi^{a\dag}\left(i\frac{\partial}{\partial t}
+\frac{\nabla^2}{2M}
\right)\chi^a
\nonumber \\
             &   & +\frac{i}{2}g_sf^{abc}G^{ai}\left(e^{2iMt}\chi^{b\dag}\sigma^i(\chi^c)^c+e^{-2iMt}(\chi^{b\dag})^c\sigma^i\chi^c\right)\Big] \nonumber \\
             & + & \int {\rm d}t{\rm d}^3{\bf x}{\rm d}^3{\bf y}\frac{-\alpha_3({\bf x}-{\bf y})}{\| {\bf x}-{\bf y}\|^3}e^{-M{\| {\bf x}-{\bf y} \|}}\left(1+M{\| {\bf x}-{\bf y} \|}\right)\frac{i}{2}f^{abc}f^{ade}G^{ei}({\bf x},t)G^{bj}({\bf y},t)\nonumber \\
             &   & \times\Big[i\epsilon^{ikj}e^{2imt}\chi^{d\dag}({\bf x},t)(\chi^c)^c({\bf y},t)+i\epsilon^{ikj}e^{-2imt}(\chi^d)^{c\dag}({\bf x},t)\chi^c({\bf y},t) \nonumber \\
             &   & +e^{2imt}\chi^{d\dag}({\bf x},t)(\delta^{kj}\sigma^i+\delta^{ki}\sigma^j-\delta^{ij}\sigma^k)(\chi^c)^c({\bf y},t) \nonumber \\
             &   & +e^{-2imt}(\chi^d)^{c\dag}({\bf x},t)(\delta^{kj}\sigma^i+\delta^{ki}\sigma^j-\delta^{ij}\sigma^k)\chi^c({\bf y},t)\Big] \nonumber \\
             & + & \int {\rm d}t{\rm d}^3{\bf x}{\rm d}^3{\bf y}\frac{1}{4}f^{abc}f^{ade}\frac{\alpha_3}{\| {\bf x}-{\bf y} \|}\\
             &   & \Big[\chi^{b\dag}({\bf x},t)(\chi^d)^c({\bf y},t)(\chi^c)^{c\dag}({\bf x},t)\chi^e({\bf y},t) 
              -\chi^{b\dag}({\bf x},t)\sigma^i(\chi^d)^c({\bf y},t)(\chi^c)^{c\dag}({\bf x},t)\sigma^i\chi^e({\bf y},t)\Big].\nonumber
\end{eqnarray}

Let us replace a pair of gluinos with a two body state field.  
We introduce auxiliary fields by inserting the identity,
\begin{eqnarray}
1= & \int [D\phi_R^{\mu\alpha_R}][Ds_R^{\mu\alpha_R\dag}]{\rm exp}\Big[\frac{i}{2}\int {\rm d}t{\rm d}^3{\bf x}{\rm d}^3{\bf y}\phi_R^{\mu\alpha_R}({\bf x},{\bf y},t) \nonumber \\
   & \times\left(s_R^{\mu\alpha_R\dag}({\bf y},{\bf x},t)-C^{\alpha_R}_{ab}\chi^{a\dag}({\bf x},t)\sigma^{\mu}(\chi^b)^c({\bf y},t)\right)\Big],
\end{eqnarray}
and its conjugate.  Here, $R$ indicates representations of
$SU(3)$ and $\alpha_R$ denotes indices of the representation
$R$.  $C^{\alpha_R}_{ab}$ are the Clebsch-Gordan coefficient for a decomposition
${\rm Ad} \otimes {\rm Ad} \rightarrow 1\oplus 8_A \oplus 8_S \oplus 10\oplus
\bar{10} \oplus 27$.  By integrating out $\chi, s$, and their
conjugate, we obtain the effective action,
\begin{eqnarray}
\label{eq:NRaction}
S_{\rm eff}^{\rm NR} = &
\int {\rm d}^4x\left[-\frac{1}{4}F_{\mu\nu}^{a}F^{\mu\nu}_{a}-\frac{1}{2}(\partial G^a)^2\right] \nonumber \\
               &
+ \sum_{R,\alpha_R,\mu}\int {\rm d}^3{\bf r}{\rm d}^4x\Big[
\phi_R^{\mu\alpha_R\dag}({\bf r},x)(i\partial_{t}+\frac{{\bf \nabla}^2_x}{4M}
+\frac{{\bf \nabla}^2_r}{M}
-V_R({\bf r})
)\phi_R^{\mu\alpha_R}({\bf r},x) \nonumber \\
               & + \phi_R^{\mu\alpha_R\dag}({\bf r},x)D_R^{\mu\alpha}({\bf r},x) + \phi_R^{\mu\alpha_R}({\bf r},x)D_R^{\mu\alpha\dag}({\bf r},x)\Big],
\end{eqnarray}
where
\begin{eqnarray}
V_R({\bf r})            & = & c_R\frac{\alpha_3}{r}, \\
\label{eq:potential coeff}
c_R                     & = &
 \frac{1}{2}\left(C_2(R)-C_2(Ad)-C_2(Ad)\right)=
\begin{cases}
-3 & R=1\\
-\frac{3}{2} & R=8\\
0 & R=10\\
+1 & R=27,
\end{cases}
 \\
D_R^{0\alpha_R}({\bf r},x) & = & \alpha_3\frac{r^k}{r^3}e^{-Mr}(1+Mr)e^{2iMt}c_RC^{\alpha_R}_{ab}\epsilon^{ikj}G^{ai}({\bf x}+\frac{{\bf r}}{2},t)G^{bj}({\bf x}-\frac{{\bf r}}{2},t), \\
\label{eq:spinone}
D_R^{i\alpha_R}({\bf r},x)& = & \delta_{R8_A}\delta_{\alpha_Ra}\delta^3({\bf r})ig_s\sqrt{3}e^{2iMt}G^{ai}({\bf x}+\frac{{\bf r}}{2},t)  \\
                        &   & -i\alpha_3\frac{1}{r^3}e^{-Mr}(1+Mr)e^{2iMt}c_RC^{\alpha_R}_{ab}[r^jG^{ai}({\bf x}+\frac{{\bf r}}{2},t)G^{bj}({\bf x}-\frac{{\bf r}}{2},t) \nonumber \\
                        &   & +r^jG^{aj}({\bf x}+\frac{{\bf r}}{2},t)G^{bi}({\bf x}-\frac{{\bf r}}{2},t)-r^i{\bf G}^{a}({\bf x}+\frac{{\bf r}}{2},t)\cdot {\bf G}^{b}({\bf x}-\frac{{\bf r}}{2},t)] .\nonumber 
\end{eqnarray}
$\phi^0$ and $\phi^i$ correspond to a pair of gluinos with a spin $0$ and $1$, respectively.
Accodring to Landau-Yang's theorem~\cite{Landau:1948kw,Yang:1950rg}, a spin-one state cannot decay into two gluons.
Therefore, the matrix element calculated from the second term in Eq.~(\ref{eq:spinone}) should vanish if the two gluons are on-shell.

Let us take the annihilation of the gluino into account.
Since we consider non-relativistic processes, the annihilation is dominated by s-waves and hence the annihilation is expressed by a local four Fermi term,
\begin{eqnarray}
\label{eq:fermi}
S_{\rm eff}^{\rm ani} = i \Gamma_{\rm ani}^{\mu,R}
\sum_{R,\alpha_R,\mu}\frac{i}{N}\int {\rm d}^4 x
C^{\alpha_R}_{ab}\chi^{a\dag}(x)\sigma^{\mu}(\chi^b)^c(x)
C^{\alpha_R*}_{cd}(\chi^d)^{c\dag}(x)\sigma^{\mu}\chi^c(x),
\end{eqnarray}
where $N=2$.%
\footnote{If the two body state is composed of Dirac particles, $N=1$ and the bilinear should be replaced as
$
C^{\alpha_R}_{ab}\chi^{a\dag}(x)\sigma^{\mu}(\chi^b)^c(x)
\rightarrow
C^{\alpha_R}_{ab}\chi^{a\dag}(x)\sigma^{\mu}\eta^b(x)
$,
where $\eta$ is the field which creates an anti-particle.
}
With two body state fields, the Fermi term is expressed as
\begin{eqnarray}
\sum_{R,\alpha_R,\mu}
\int {\rm d}^3 {\bf r}{\rm d}^4 x \phi_R^{\mu \alpha_R\dag}({\bf r},x)
2i \Gamma_{\rm ani}^{\mu,R}\delta({\bf r})\phi_R^{\mu\alpha_R}({\bf r},x)
\end{eqnarray}

Let us calculate $\Gamma_{\rm ahi}^{\mu,R}$ by matching tree level calculations of the annihilation cross section of non-relativistic gluinos with forward scattering amplitudes calculated from Eq.~(\ref{eq:fermi}), with an aid of the optical theorem.

For spin-0 states, annihilation cross sections are given by
\begin{eqnarray}
\sigma v (R,\mu=0\rightarrow GG) =
\begin{cases}
\frac{4\pi}{M^2} \alpha_3^2 c_R^2&(R=1,8_S,27)\\
0& (R=8_A, 10,\bar{10}),
\end{cases}
\end{eqnarray}
while those of spin-1 states are given by
\begin{eqnarray}
\sigma v (R,\mu=i\rightarrow u\bar{u},d\bar{d},\cdots) =
\begin{cases}
6\times\frac{2\pi}{M^2} \alpha_3^2& (R=8_A)\\
0& (\text{otherwise}).
\end{cases}
\end{eqnarray}
Note that spin-1 states do not annihilate into two gluinos, which is consistent with the Landau-Yang's theorem.

Annihilation cross sections are related with forward scattering amplitudes by the optical theorem for two body scatterings:
\begin{eqnarray}
\label{eq:optical_th}
{\rm Im}{\cal M}(\widetilde{G}\widetilde{G}\rightarrow \widetilde{G}\widetilde{G}) = \frac{1}{2} s \sigma v \simeq 2M^2 \sigma v,
\end{eqnarray}
where $s\simeq4M^2$ is the center of mass energy.
Here, ${\cal M}$ is the invariant matrix element normalized by
\begin{eqnarray}
\label{eq:matrix element}
<f|iT|i> = (2\pi)^4 \delta^{(4)}(p_i-p_f)i {\cal M},
\end{eqnarray}
with the Lorentz invariant normalization of one-particle states,
\begin{eqnarray}
\label{eq:normalization}
<{\bf q}|{\bf p}> = 2 E_{p} (2\pi)^3 \delta^{(3)}({\bf p}-{\bf q}).
\end{eqnarray}

With this normalization, the invariant amplitude calculated by the action in Eq.~(\ref{eq:fermi}) is given by
\begin{eqnarray}
{\rm Im}{\cal M}(R,\mu\rightarrow R,\mu) = 8N M^2 \Gamma_{\rm ani}^{\mu,R}.
\end{eqnarray}
Hence, $\Gamma_{\rm ahi}^{\mu,R}$ is given by%
\footnote{Note the difference of the factor of 4 from the formula given in Refs.~\cite{Hisano:2002fk,Hisano:2003ec}.
This is because we consider the annihilation cross section for a particular initial spin state while that in Refs.~\cite{Hisano:2002fk,Hisano:2003ec} is the averaged one. Our treatment would be useful when one handles various color and spin states.
}
\begin{eqnarray}
\Gamma_{\rm ahi}^{\mu,R} = \frac{1}{4N}\sigma v (R,\mu \rightarrow GG~{\rm or}~q\bar{q}).
\end{eqnarray}
Specifically,
\begin{eqnarray}
\Gamma_{\rm ani}^{0,1} = \frac{9\pi}{2M^2}\alpha_3^2,~~
\Gamma_{\rm ani}^{0,8_S} = \frac{9\pi}{8M^2}\alpha_3^2,~~
\Gamma_{\rm ani}^{0,27} = \frac{\pi}{2M^2}\alpha_3^2,~~
\Gamma_{\rm ani}^{i,8_A} = \frac{3\pi}{2M^2}\alpha_3^2.
\end{eqnarray}

\subsection{Extraction of s-wave component}

Non-relativistic annihilation processes are dominated by s-waves.
Thus, it is useful to extract an s-wave component of a given initial state.
For that purpose, we insert a complete set of asymptotic two body s-wave states,
\begin{eqnarray}
{\bf 1} &\sim& \sum_{\lambda}\int \frac{{\rm d}^3{\bf P}}{(2\pi)^3}\frac{{\rm d}k}{2\pi}|{\bf P},k,\lambda ><{\bf P},k,\lambda |,\\
\label{eq:normalization}
<{\bf P}',k',\lambda'|{\bf P},k,\lambda>&=&(2\pi)^4\delta^{(3)} ({\bf P}-{\bf P}')\delta (k-k')\delta_{\lambda\lambda'},
\end{eqnarray}
where ${\bf P}$ and $k$ denote the total momentum and the relative momentum of the two body state.
$\lambda$ labels the discrete quantum number of the two body state, such as the color and the spin.
$k$ is related with the relative velocity $v$ by $k= Mv/2$.
In the following,
we omit the index $\lambda$ to simplify notation.
By inserting the complete set,
the forward scattering amplitude of a state $|i>$ by its s-wave component is given by
\begin{eqnarray}
<i|iT|i>|_s = \int \frac{{\rm d}^3 {\bf P}}{(2\pi)^2}\frac{{\rm d}k}{2\pi}|<{\bf P},k|i>|^2 i {\cal M}({\bf P},k),\\
\label{eq:def amp}
<{\bf P'},k'|iT|{\bf P},k> \equiv (2\pi)^4 \delta^{(3)}({\bf P}-{\bf P}')\delta (k-k') i{\cal M}({\bf P},k).
\end{eqnarray}

The state $|{\bf P},k>$ is constructed by
\begin{eqnarray}
|{\bf P},k> = \frac{k}{\sqrt{2\pi}} \int \frac{{\rm d}\Omega_k}{4\pi}
c^\lambda_{\alpha\beta}a^\dag_\alpha (\frac{{\bf P}}{2}+{\bf k})a^\dag_\beta (\frac{{\bf P}}{2}-{\bf k})|0>,
\end{eqnarray}
where $a^\dag_\alpha ({\bf p})$ is the creation operator of a particle with a momentum ${\bf p}$ and a quantum number $\alpha$. $c^\lambda_{\alpha \beta}$ is a coefficient to construct the quantum number $\lambda$, which is normalized by $c^\lambda_{\alpha\beta}c^{\lambda'}_{\alpha \beta} = \delta^{\lambda\lambda'}$. With a normalization
\begin{eqnarray}
\{a_\alpha({\bf p}),a_\beta({\bf p}')^\dag\} = (2\pi)^3 \delta^{(3)}({\bf p}-{\bf p}') \delta_{\alpha\beta},
\end{eqnarray}
it can be shown that the normalization in Eq.~(\ref{eq:normalization}) holds.

Let us consider the two-body state in the center-of-momentum system;
\begin{eqnarray}
|i>=|{\bf 0},{\bf k}> = 2 E_k c^\lambda_{\alpha\beta}a^\dag_\alpha ({\bf k})a^\dag_\beta (-{\bf k})|0>,
\end{eqnarray}
where $E_k = \sqrt{k^2+m^2} = \sqrt{s}/2\simeq M$.
The state is normalized in the Lorentz invariant way
and has a overlap with the state $|{\bf P}',k>$,
\begin{eqnarray}
<{\bf P}',k'|i> = \frac{2M}{(2\pi)^{3/2}}(2\pi)^4\delta^{(3)}({\bf P}')\delta(k-k').
\end{eqnarray}

By remembering that the total energy of the initial state is given by $P^0\simeq 2M + k^2/M$, we obtain
\begin{eqnarray}
 \delta (k-k') \simeq \frac{2k}{M}\delta(P^0-P^{0'}).
\end{eqnarray}
Hence, the forward scattering amplitude of the state $|i>$ is given by
\begin{eqnarray}
<i|iT|i>|_s = 16\pi \frac{M}{k} i{\cal M}({\bf 0},k) \times VT,
\end{eqnarray}
where $VT = (2\pi)^4 \delta^{(4)}(P-P)$.

\subsection{Forward scattering amplitude and the Green's function}

Next, let us relate ${\cal M}$ with the Green's function of the two body state fields,
\begin{eqnarray}
G(E;{\bf r},{\bf r'})\equiv i\int {\rm d}^4x e^{iEx^0}<T\phi({\bf r},x)\phi^{\dag}({\bf r'},0)>.
\end{eqnarray}
The relation is given by the Lehmann-Symanzik-Zimmermann (LSZ) reduction formula~\cite{Lehmann:1954rq} we derive in this subsection.

We start from the effective Lagrangian,
\begin{eqnarray}
\label{eq:action phi}
S_{\rm eff} = &
\int {\rm d}^3{\bf r}{\rm d}^4x
\phi^{\dag}({\bf r},x)\left(i\partial_{t}+\frac{{\bf \nabla}^2_x}{4M}
+\frac{{\bf \nabla}^2_r}{M}
-V(r)
+2i \delta({\bf r})\Gamma_{\rm ani}
\right)\phi({\bf r},x).
\end{eqnarray}
As an asymptotic initial state, we consider the state with two free gluinos.
Therefore, to derive the LSZ reduction formula, we first expand $\phi$ by ignoring the potential and the annihilation terms.
Further, since we consider only the s-wave state, we reduce $\phi$ as
\begin{eqnarray}
\label{eq:reduction}
\phi({\bf r},x) \rightarrow \int \frac{{\rm d}k}{2\pi} \frac{k}{\sqrt{\pi}}\frac{{\rm sin}kr}{kr}\phi_k (x).
\end{eqnarray}
With this reduction, the effective Lagrangian is given by
\begin{eqnarray}
S_{\rm eff} = \int {\rm d}^4 x \frac{dk}{2\pi}\phi^\dag_k (x) (i \partial_t + \frac{{\bf \nabla}^2_x}{4M} - \frac{k^2}{M})\phi_k(x)
\end{eqnarray}

We define the annihilation operator by
\begin{eqnarray}
\phi_k = \int \frac{{\rm d}^3 {\bf P}}{(2\pi)^3}e^{-iE_{{\bf P},k} t + i {\bf P}\cdot {\bf x}} a({\bf P},k),~~E_{{\bf P},k} = \frac{{\bf P}^2}{4M} + \frac{k^2}{M}.
\end{eqnarray}
The conjugate momentum and the canonical quantization condition are given by
\begin{eqnarray}
\pi_k (x) = \frac{\delta S_{\rm eff}}{\delta \dot{\phi}_k (x) } = i \phi^\dag_k(x) = i  \int \frac{{\rm d}^3 {\bf P}}{(2\pi)^3}e^{iE_{{\bf P},k} t - i {\bf P}\cdot {\bf x}} a^\dag({\bf P},k),\nonumber\\
\left[ \phi_k (x),\pi_{k'} (y) \right] |
_{x^0=y^0} = i 2\pi \delta (k-k') \delta^{(3)}({\bf x}-{\bf y}),
\end{eqnarray}
which requires
\begin{eqnarray}
\left[
a ({\bf P},k), a^\dag({\bf P}',k)
\right]
= (2\pi)^4 \delta^{(3)}({\bf P}-{\bf P}')\delta (k-k').
\end{eqnarray}
It can be easily shown that the state,
\begin{eqnarray}
|{\bf P},k>\equiv a^\dag ({\bf P},k)|0>,
\end{eqnarray}
satisfies the normalization given in Eq.~(\ref{eq:normalization}).
Then the matrix element of the field $\phi_k (x)$ between the vacuum and the asymptotic two gluon free state is given by
\begin{eqnarray}
<0| \phi _{k'} (x) |{\bf P}, k> = 2\pi \delta (k-k') e^{-i E_{{\bf P},k} t + i {\bf P}\cdot {\bf x}}.
\end{eqnarray}

With the exactly  same procedure as the derivation of the LSZ formula for relativistic field theories, one can derive a formula,
\begin{eqnarray}
\int {\rm d}^4 x e^{i E t - i {\bf P}\cdot {\bf x}}\int {\rm d}^4 y e^{-i E' t' + i {\bf P'}\cdot {\bf x'}}
<T \phi_k (x) \phi_{k'}^\dag (x')>\nonumber \\
\rightarrow 
\frac{i}{E-E_{{\bf P},k}+ i\epsilon}\frac{i}{E'-E_{{\bf P}',k'}+i\epsilon}
<{\bf P},k|S |{\bf P}',k'>~~
({\rm for}~E\rightarrow E_{{\bf P},k},~E'\rightarrow E_{{\bf P}',k'}),
\end{eqnarray}
which yields
\begin{eqnarray}
 <{\bf 0},k|S| {\bf 0},k> &=&
 - {\rm lim}_{E\rightarrow k^2/M+i\epsilon}(E-\frac{k^2}{M})^2 \int {\rm d}^4 x e^{iEt} <T \phi_k(x)\phi^\dag_k(0)>\nonumber\\
 &&\times(2\pi)^4 \delta^{(3)}({\bf P}-{\bf P}) \delta (k-k) 
\end{eqnarray}

By remembering the inverse of the reduction in Eq.~(\ref{eq:reduction}),
\begin{eqnarray}
\phi_k(x) = \int {\rm d}^3 {\bf r} \frac{k}{\sqrt{\pi}}\frac{{\rm sin}kr}{kr} \phi ({\bf r},x),
\end{eqnarray}
we obtain,
\begin{eqnarray}
\label{eq:LSZ formula}
<{\bf 0},k|S|{\bf 0},k> &=& 8\pi i \frac{M}{k}{\rm lim}_{E\rightarrow k^2/M+i\epsilon}
\left(E-\frac{k^2}{M}\right)^2 
\int {\rm d}r{\rm d}r' rr'
{\rm sin}(kr){\rm sin}(kr')
G(E;{\bf r},{\bf r}')\times\nonumber\\
&&(2\pi)^4 \delta^{(3)} ({\bf P}-{\bf P})\delta(k-k),\nonumber\\
G(E;{\bf r},{\bf r'})&\equiv& i\int {\rm d}^4x e^{iEx^0}<T\phi({\bf r},x)\phi^{\dag}({\bf r'},0)>.
\end{eqnarray}

From Eq.~(\ref{eq:def amp}) and (\ref{eq:LSZ formula}), ${\cal M}({\bf 0},k)$ is given by
\begin{eqnarray}
i{\cal M}({\bf 0},k) = 8\pi i \frac{M}{k}{\rm lim}_{E\rightarrow k^2/M+i\epsilon}
\left(E-\frac{k^2}{M}\right)^2 
\int {\rm d}r{\rm d}r' rr'
{\rm sin}(kr){\rm sin}(kr')
G(E;{\bf r},{\bf r}'),
\end{eqnarray}
with the contribution from the free propagation subtracted.

Finally, with the optical theorem, we obtain
\begin{eqnarray}
\label{eq:Xsection}
\sigma v (\mu R \rightarrow \text{anything})&=&
\frac{64\pi^2}{k^2} \text{lim}_{E\rightarrow k^2/M+i\epsilon}\left(E-\frac{k^2}{M}\right)^2 \times\nonumber\\
&&\int {\rm d}r{\rm d}r' rr'
{\rm sin}(kr){\rm sin}(kr')
{\rm Im}\Bigl[
G^{\mu R}(E;{\bf r},{\bf r}')
\Bigr],\nonumber\\
G^{\mu R}(E;{\bf r},{\bf r}')&=&
i\int {\rm d}^4 x e^{iEx^0}<T \phi_R^{\mu \alpha_R}({\bf r},x)\phi_R^{\mu \alpha_R}({\bf r}',0)^\dag>,
\end{eqnarray}
where we have denoted the $\lambda =\{\mu,\alpha_R\}$ dependence explicitly.%
\footnote{
Again, note the difference of factor 4 from the formula given in Ref.~\cite{Hisano:2002fk,Hisano:2003ec}.
}

\subsection{Solution of Green's function}

Let us calculate the Green's function of the gluino pairs:
\begin{eqnarray}
G(E;{\bf r},{\bf r'})=i\int {\rm d}^4x e^{iEx^0}<T\phi({\bf r},x)\phi^{\dag}({\bf r'},0)>,
\end{eqnarray}
where we  have again omitted indices $\mu,~R$ and $\alpha_R$.
From the action of $\phi$ given in Eq.~(\ref{eq:action phi}),
the Green's function satisfies
\begin{eqnarray}
\bigl(-\frac{\nabla_{r}^2}{M}+V(r)-E-2i\Gamma_{\rm ani} \delta ({\bf r})\bigr)G(E;{\bf r},{\bf r'})=\delta^3({\bf r}-{\bf r'}).\label{eq:sch}
\end{eqnarray}

As we have mentioned, we consider only the s-wave state since it dominantly contributes to the annihilation of the gluino.
For the s-wave, Eq.~(\ref{eq:sch}) becomes
\begin{eqnarray}
\label{eq:ssch}
\bigl(-\frac{1}{M}(\frac{{\rm d}^2}{{\rm d}r^2}+\frac{2}{r}\frac{d}{{\rm d}r})+V(r)-E-i\Gamma_{\rm ani}\frac{\delta(r)}{2\pi r^2}\bigr)G(E;r,r')=\frac{1}{4\pi r^2}\delta(r-r').
\end{eqnarray}
With $g(E;r,r')\equiv 4\pi r r' G(E;r,r')$, Eq.~(\ref{eq:ssch}) becomes
\begin{eqnarray}
\label{eq:ssch simp}
\left(
- \frac{1}{M}\frac{\partial^2}{\partial r^2 } + V(r)-E - i \Gamma_{\rm ani} \frac{\delta(r)}{2\pi r^2}
\right)
g(E;r,r') = \delta (r-r').
\end{eqnarray}

Let us first neglect $\Gamma_{\rm ani}$ and solve Eq.~(\ref{eq:ssch simp}) analytically.
We denote the solution as $g^{(0)}$.
$\Gamma_{\rm ani}$ is taken into account as a perturbation later.
For a dimensionless variable $x\equiv Mr$, Eq.~(\ref{eq:ssch simp}) with $\Gamma_{\rm ani}=0$ is
\begin{eqnarray}
\label{eq:ssch dimless}
\left(
- \frac{\partial ^2}{\partial x^2} - \frac{\alpha}{x} - \beta^2
\right) g^{(0)} = \delta (x-x'),
\end{eqnarray}
where $\beta^2 \equiv E/M$ and $\alpha\equiv - c_R \alpha_3 $.
The solution for Eq.~(\ref{eq:ssch dimless}) is expressed as~\cite{Strassler:1990nw}
\begin{eqnarray}
g^{(0)}= f_> (x) f_< (x')\theta(x-x')+ f_< (x) f_> (x')\theta(x'-x),\\
\left(
- \frac{\partial ^2}{\partial x^2} - \frac{\alpha}{x} - \beta^2
\right) f_{>,<}(x)=0,\\
f_<(0) =0,~f_<'(0)=1,~f_>(0)=1.
\end{eqnarray}
Note that the function $f_>(x)$ is not yet fixed.
Since $f_>$ corresponds to an out going wave sourced by the delta function, we put a condition that $f_>(x)$ contains only an out going modes $\propto e^{i \beta x}$ at $x\rightarrow \infty$.%
\footnote{
Technically speaking, this boundary condition is chosen by the $i\epsilon$ prescription. If an incoming mode exists at $x\rightarrow \infty$, the Green's function diverges at infinity due to the small imaginary part in $\beta$, ${\rm Im}\beta >0$.
}
Then the solution for $f_>$ is given by
\begin{eqnarray}
f_>(x) &=& \Gamma(1- i \frac{\alpha}{2\beta}) W_{i\frac{\alpha}{2\beta},\frac{1}{2}} (- 2 i \beta x),
\end{eqnarray}
where W is the Whittaker function.
For $x\rightarrow \infty$, $f_>(x)$ behaves as
\begin{eqnarray}
\label{eq:asymptotic}
f_>(x)\rightarrow e^{i\beta x}x^{i \frac{\alpha}{2\beta}} (-2 i \beta) ^{i \frac{\alpha}{2\beta}} \Gamma(1- i \frac{\alpha}{2\beta}) \equiv e^{i\beta x} x^{i \frac{\alpha}{2\beta}} A.
\end{eqnarray}
To the first order in $\Gamma_{\rm ani}$, $g$ is given by
\begin{eqnarray}
\label{eq:green perturbation}
g(E;r,r') &\simeq& g^{(0)}(E;r,r') + \int {\rm d}r'' g^{(0)}(E;r,r'') i \Gamma_{\rm ani} \frac{\delta(r'')}{2\pi r''^2}g^{(0)}(E;r'',r')\nonumber\\
&=& g^{(0)}(E;r,r') + i \Gamma_{\rm ani}\frac{M^2}{2\pi} f_>(Mr) f_>(Mr').
\end{eqnarray}

Let us note subtlety read off from the asymptotic form of the Green's function.
As can be seen from Eq.~(\ref{eq:asymptotic}), for $x\rightarrow \infty$, the Green's function have a non-trivial phase ${\rm exp}[i \alpha/ (2\beta) {\rm ln}x]$, in addition to a trivial phase ${\rm exp}(i \beta x)$ which corresponds to free particles.
This means that our assumption of free incident particles is not good.
The non-trivial phase appears due to a long-range force mediated by gluons.
In thermal bath in the early universe, however, gluons obtain their thermal masses and hence the long-range force is screened.
Thus, the non-trivial phase should vanish once thermal effects are taken into account.
In the following, we simply drop the non-trivial phase in Eq.~(\ref{eq:asymptotic}).
Then, the wave function of gluinos at infinity is simply enhanced (or declined)
by a constant $A$, in comparison with the trivial one, $f_>(x\rightarrow \infty)|_{\alpha =0} = e^{i\beta x}$.

\subsection{Annihilation cross section}

We are now at the point to calculate the annihilation cross section of the gluino.
From Eqs.~(\ref{eq:Xsection}) and (\ref{eq:green perturbation}), we obtain
\begin{eqnarray}
\sigma v (\widetilde{G} \widetilde{G}\rightarrow \text{anything}) = \frac{8}{k^2}  \Gamma_{\rm ani} \times
\text{Re}\Bigl[
\text{lim}_{E\rightarrow \frac{k^2}{M}+i\epsilon}\left(E-\frac{k^2}{M}\right)
\int {\rm d}x
{\rm sin}(\frac{k}{M}x) f_>(x)
\Bigr]^2,
\end{eqnarray}
where we have omitted indices $\mu,R$.
Here, we have subtracted a term proportional to $g^{(0)}$ which corresponds a scattering between gluinos.

Let us evaluate the factor in the parenthesis.
Note that the factor vanishes unless the integration diverges.
Thus, we may replace the integrand with its asymptotic form for $x \rightarrow \infty$ given in Eq.~(\ref{eq:asymptotic}).
Observing that
\begin{eqnarray}
\text{lim}_{E\rightarrow k^2/M+i\epsilon}\left(E-\frac{k^2}{M}\right)
\int^\infty dx {\rm sin}(\frac{k}{M}x)e^{i\beta x} = k,
\end{eqnarray}
we find
\begin{eqnarray}
\label{eq:Xsection all}
\sigma v (\widetilde{G} \widetilde{G} \rightarrow \text{anything}) = 8 \Gamma_{\rm ani} {\rm Re} [ A ^2].
\end{eqnarray}

The cross section given in Eq.~(\ref{eq:Xsection all}) includes not only the annihilation into gluons and quarks, but also includes the scattering into a pair of gluinos.
Let us  extract the former contribution in a intuitive way.
As we have mentioned, the factor $A$ expresses the enhancement (or decline) of the wave function of gluinos by a constant, $f_>(x\rightarrow \infty) = e^{i\beta x} A$.
The factor $A$ would enter the annihilation amplitude simply as a multiplication factor.
Then, the annihilation cross section into gluinos and quarks is given by
\begin{eqnarray}
\label{eq:Xsection excracted}
\sigma v (\widetilde{G} \widetilde{G} \rightarrow GG, q\bar{q}) = 8 \Gamma_{\rm ani} | A |^2.
\end{eqnarray}

In the end, we obtain
\begin{eqnarray}
\sigma v (\mu,R) = 8 \Gamma^{\mu,R}_{\rm ani}\times  |A^{\mu,R}|^2 &=& 8 \Gamma^{\mu,R}_{\rm ani}\times
\frac{2\pi c_R \alpha_{s,R}/v}
{{\rm Exp}(2\pi c_R \alpha_{s,R}/v)-1}\\
&\simeq&
\begin{cases}
8 \Gamma^{\mu,R}_{\rm ani} 2\pi (-c_R)\frac{\alpha_{s,R}}{v} & c_R<0 \\
0 & c_R>0.
\end{cases}
~~({\rm for}~v\ll 2\pi |c_R|\alpha_{s,R})\nonumber
\end{eqnarray}
Here, $\alpha_{s,R}$ is the fine structure constant of the QCD evaluated at the scale $\mu_R$ which is determined by~\cite{Nagano:1999nw}
\begin{eqnarray}
\mu_R= \frac{M}{2}|c_R|\alpha_3(\mu_R).
\end{eqnarray}

\begin{figure}[tb]
 \begin{center}
  \includegraphics[width=0.7\linewidth]{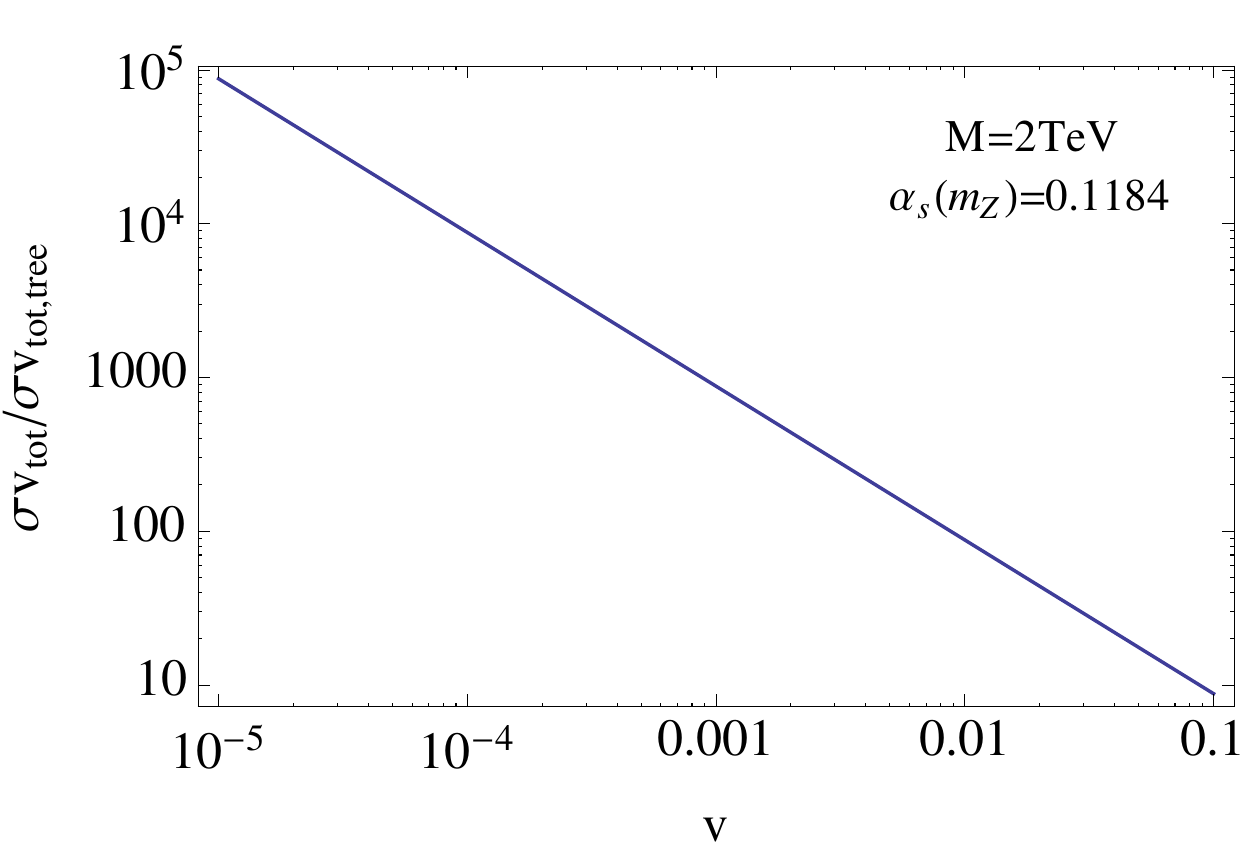}
 \end{center}
\caption{Enhancement factor $\sigma v_{\rm tot}/ \sigma v_{\rm tot,tree}$.}
\label{fig:enhancement}
\end{figure}

A spin and color averaged annihilation total annihilation cross section is given by
\begin{eqnarray}
\sigma v_{\rm tot} &=& \frac{1}{64\times4}
\left[
1\times \sigma v\left(0,1\right) +
8\times \sigma v\left(0,8_S\right) +
27\times \sigma v\left(0,27\right) +
8\times 3\times \sigma v\left(i,8_A\right)
\right]\nonumber\\
& \simeq& \frac{81\pi}{16} \frac{\alpha_{s,8}}{v} \times \frac{\pi \alpha_{s,{\rm UV}}^2}{M^2} = \frac{18\pi}{7} \frac{\alpha_{s,8}}{v} \sigma v _{\rm tot, tree},
\end{eqnarray}
where $\sigma v _{\rm tot, tree} = 63 \pi \alpha_{s,{\rm UV}}^2/(32 M^2)$.
Here, we have used the approximation that $\alpha_{s,1}\simeq \alpha_{s,8}$ and neglected the contribution from ${\bf 27}$ representation. This approximation induces only an error of one percent.
In Fig.~\ref{fig:enhancement}, we show the enhancement factor $\sigma v_{\rm tot}/ \sigma v_{\rm tot,tree}$.

\section{Wino annihilation cross sections}
\label{app: wino annihilations}
Here, we summarize annihilation cross sections of the wino.
As already mentioned in main text, there are six annihilation modes: $\widetilde{W}^0 \widetilde{W}^0$, $\widetilde{W}^+ \widetilde{W}^-$, $\widetilde{W}^0 \widetilde{W}^\pm$, and $\widetilde{W}^\pm \widetilde{W}^\pm$. Initial states of $\widetilde{W}^0 \widetilde{W}^0$ and $\widetilde{W}^\pm \widetilde{W}^\pm$ form only spin-0 states, while those of other modes form both spin-0 and spin-1 states. Below, we carefully present the cross sections in each mode .

\subsection{$\widetilde{W}^- \widetilde{W}^-$ annihilation}

Since this is the annihilation between identical particles, its initial sate forms only a spin-0 state, and it annihilates into $W^- W^-$ pair. The cross section shown here can also be applied to its conjugate case, namely $\widetilde{W}^+ \widetilde{W}^+ \to W^+ W^+$.
{\small
\begin{eqnarray}
\sigma_0 v|_{WW} =
\frac{4\pi \alpha_2^2}{M_2^2}
\left[ 1 - \frac{m_W^2}{M_2^2} \right]^{3/2}
\left[ 1 - \frac{m_W^2}{2M_2^2} \right]^{-2}.
\end{eqnarray}
}

\subsection{$\widetilde{W}^0 \widetilde{W}^-$ annihilations}

Cross sections presented here can also be applied their conjugate cases. When the initial state forms a spin-0 state, it annihilates into $W^- Z$ and $W^- \gamma$.
{\small
\begin{eqnarray}
\sigma_0 v|_{WZ} &=&
\frac{2\pi \alpha_2^2 c_W^2}{M_2^2}
\left[ 1 - \frac{m_W^2+m_Z^2}{2M_2^2}
+ \frac{\left( m_W^2 - m_Z^2 \right)^2}{16M_2^4} \right]^{3/2}
\left[ 1 - \frac{m_W^2+m_Z^2}{4M_2^2} \right]^{-2},
\\
\sigma_0 v|_{W\gamma} &=&
\frac{2\pi \alpha_2^2 s_W^2}{M_2^2}
\left[ 1 - \frac{m_W^2}{2M_2^2} + \frac{m_W^4}{16M_2^4} \right]^{3/2}
\left[ 1 - \frac{m_W^2}{4M_2^2} \right]^{-2}.
\end{eqnarray}
}

On the other hand, when the initial state forms a spin-1 state, it annihilates into $f \bar{f}^\prime$, $W^- h$, and $W^- Z$. Those cross sections are given as follows. Below, we only show the cross section of $\widetilde{W}^0 \widetilde{W}^- \to e^- \bar{\nu}$ as a representative of $\widetilde{W}^0 \widetilde{W}^- \to f \bar{f}^\prime$.
{\small
\begin{eqnarray}
\sigma_0 v|_{e\bar{\nu}} &=&
\frac{\pi \alpha_2^2}{3M_2^2}
\left[ 1 - \frac{m_W^2}{4M_2^2} \right]^{-2}
\left[ 1 - \frac{m_e^2}{4M_2^2} \right]^2
\left[ 1 + \frac{m_e^2}{8M_2^2} \right],
\\
\sigma_0 v|_{Wh} &=&
\frac{\pi \alpha_2^2}{12M_2^2}
\left[ 1 - \frac{m_W^2}{4M_2^2} \right]^{-2}
\left[ \left( 1 + \frac{m_W^2-m_h^2}{4M_2^2} \right)^2
+ 2 \frac{m_W^2}{M_2^2} \right]
\nonumber \\ && \times
\left[ 1 - \frac{m_W^2 + m_h^2}{2M_2^2}
+ \frac{ \left( m_W^2 - m_h^2 \right)^2}{16M_2^4} \right]^{1/2},
\\
\sigma_0 v|_{WZ} &=&
\frac{\pi \alpha_2^2}{12M_2^2}
\left[ 1 - \frac{m_W^2+m_Z^2}{2M_2^2}
+ \frac{ \left( m_W^2 - m_Z^2 \right)^2}{16M_2^4} \right]^{3/2}
\left[ 1 - \frac{m_W^2+m_Z^2}{4M_2^2} \right]^{-2}
\nonumber \\ && \times
\left[ 1 - \frac{m_W^2}{4M_2^2} \right]^{-2}
\left[ 1 + \frac{5}{2}\frac{m_W^2+m_Z^2}{M_2^2}
+ \frac{m_W^4 + m_Z^4 + 10 m_W^2 m_Z^2}{16M_2^4} \right].
\end{eqnarray}
}

\subsection{$\widetilde{W}^0 \widetilde{W}^0$ annihilation}

Since the neutral wino is a Majorana particle, its initial state forms only a spin-0 state. A pair of the neutral wino annihilates only into $W^+W^-$.
{\small
\begin{eqnarray}
\sigma_0 v|_{WW} =
\frac{8\pi \alpha_2^2}{M_2^2}
\left[ 1 - \frac{m_W^2}{M_2^2} \right]^{3/2}
\left[ 1 - \frac{m_W^2}{2M_2^2} \right]^{-2}.
\end{eqnarray}
}

\subsection{$\widetilde{W}^+ \widetilde{W}^-$ annihilations}

When the initial state forms a spin-0 state, it annihilates into $\gamma \gamma$, $W^+ W^-$, $Z Z$, and $Z \gamma$. Corresponding cross sections of these annihilation channels are as follows:
{\small
\begin{eqnarray}
\sigma_0 v|_{\gamma \gamma} &=&
\frac{4\pi \alpha_2^2 s_W^4}{M_2^2},
\\
\sigma_0 v|_{WW} &=&
\frac{2\pi \alpha_2^2}{M_2^2}
\left[ 1 - \frac{m_W^2}{M_2^2} \right]^{3/2}
\left[ 1 - \frac{m_W^2}{2M_2^2} \right]^{-2},
\\
\sigma_0 v|_{ZZ} &=&
\frac{4\pi \alpha_2^2 c_W^4}{M^2}
\left[ 1 - \frac{m_Z^2}{M_2^2} \right]^{3/2}
\left[ 1 - \frac{m_Z^2}{2M_2^2} \right]^{-2},
\\
\sigma_0 v|_{Z \gamma} &=&
\frac{8\pi \alpha_2^2 c_W^2 s_W^2}{M^2}
\left[ 1 - \frac{m_Z^2}{2M_2^2} + \frac{m_Z^4}{16M_2^4} \right]^{3/2}
\left[ 1 - \frac{m_Z^2}{4M_2^2} \right]^{-2},
\end{eqnarray}
}

On the other hand, when the initial state forms a spin-1 state, it annihilates into $W^+ W^-$, $Z h$, and $f \bar{f}$. Below, $Q_f$ and $I_{3f}$ denote electric charge and $SU(2)_L$ charge respectively, while $k_f$ is defined as $k_f \equiv M_2 [ 1 - m_f^2/M_2^2 ]^{1/2}$.
{\small
\begin{eqnarray}
\sigma_0 v|_{WW} &=&
\frac{\pi \alpha_2^2}{12M_2^2}
\left[ 1 - \frac{m_W^2}{M_2^2} \right]^{3/2}
\left[ 1 - \frac{m_W^2}{2M_2^2} \right]^{-2}
\left[ 1 - \frac{m_Z^2}{4M_2^2} \right]^{-2}
\nonumber \\ && \times
\left[ 1 - \frac{m_Z^2 - m_W^2}{2M_2^2} \right]^{2}
\left[ 1 + \frac{5m_W^2}{M_2^2} + \frac{3m_W^4}{4M_2^4} \right],
\\
\sigma_0 v|_{Zh} &=&
\frac{\pi \alpha_2^2}{12M_2^2}
\left[ 1 - \frac{m_Z^2}{4M_2^2} \right]^{-2}
\left[\left( 1 + \frac{m_Z^2 - m_h^2}{4M_2^2} \right)^2
+ 2 \frac{m_Z^2}{M_2^2} \right]
\nonumber \\ && \times
\left[ 1 - \frac{m_Z^2+m_h^2}{2M_2^2}
+ \frac{\left( m_Z^2 - m_h^2 \right)^2}{16M_2^4} \right]^{1/2},
\\
\sigma_0 v|_{f \bar{f}} &=&
\frac{\pi \alpha_2^2}{6M_2^2}\frac{k_f}{M_2}
\left[4 s_W^4 Q_f^2 \left( 3 - \frac{k_f^2}{M_2^2} \right)
- 3 \frac{m_f^2}{M_2^2} I_{3f}^2
\left( 1 - \frac{m_Z^2}{4M_2^2} \right)^{-2} \right.
\nonumber \\ &&
+ 4 s_W^2 Q_f \left( I_{3f}- 2 s_W^2 Q_f \right)
\left( 3 - \frac{k_f^2}{M_2^2} \right)
\left( 1 - \frac{m_Z^2}{4M_2^2} \right)^{-1}
\nonumber \\ &&
\left. + 4 \left( s_W^4 Q_f^2 - I_{3f} s_W^2 Q + I_{3f}^2/2 \right)
\left( 3 - \frac{k_f^2}{M_2^2} \right)
\left( 1 - \frac{m_Z^2}{4M_2^2} \right)^{-2} \right].
\end{eqnarray}
}

\subsection{Mixing between $\widetilde{W}^0 \widetilde{W}^0$ and $\widetilde{W}^+ \widetilde{W}^-$}

Since $\widetilde{W}^0 \widetilde{W}^0$ and the spin-0 state of $\widetilde{W}^+ \widetilde{W}^-$ have the same quantum number, they are mixed each other. In order to evaluate the Sommerfeld factor for the states, we have to calculate the imaginary part of the transition amplitude between $\widetilde{W}^0 \widetilde{W}^0$ and $\widetilde{W}^+ \widetilde{W}^-$. As can be easily understood from the interaction of the neutral wino, the intermediate state of the amplitude is the $W$ boson pair, which is evaluated as
{\small
\begin{eqnarray}
\sigma_0 v|_ {\widetilde{W}^0 \widetilde{W}^0 \leftrightarrow \widetilde{W}^+ \widetilde{W}^- \to WW} =
\frac{2\pi \alpha_2^2}{M_2^2}
\left[ 1 - \frac{m_W^2}{M_2^2} \right]^{3/2}
\left[ 1 - \frac{m_W^2}{2M_2^2} \right]^{-2}.
\end{eqnarray}
}

\newpage

\begin{center}
{\bf References}
\end{center}


\end{document}